\documentclass{lmcs}
\pdfoutput=1

\usepackage[utf8]{inputenc}
\usepackage{lastpage}
\lmcsdoi{17}{4}{25}
\lmcsheading{}{\pageref{LastPage}}{}{}%
{Jan.~22,~2018}{Dec.~23,~2021}{}


\usepackage{microtype}
\usepackage{xspace}
\usepackage{color}
\usepackage{cmll}
\usepackage{amsfonts}
\usepackage{amsmath}
\usepackage{amsthm}
\usepackage{stmaryrd}
\usepackage{mathtools}
\usepackage{url}
\usepackage{todonotes}
\usepackage{bussproofs}
\usepackage{MnSymbol}
\usepackage{caption}
\usepackage{subcaption}

\usetikzlibrary{calc}
\SetMathAlphabet{\mathtt}{bold}  {OT1}{lmtt}{b}{n} % correct a bug whenever using bold+mathtt

\newcommand\NAT{\mathtt{NAT}}           % type of integers
\newcommand\BIN{\mathtt{BIN}}           % type of binomials
\newcommand\Nat{\mathbb{N}} % natural numbers
\newcommand\Bin{\mathbb{B}} % binomial numbers
\newcommand\newop[1]{{\mathtt{#1}}}     % -
\newcommand{\bc}{\oplus}                % prob. sum
\newcommand\fixran{\newop X}      % fixpoint operator
\newcommand{\xc}{\fixran}               % alias
\newcommand\rand{\newop R}        % random integers
\newcommand{\rnd}{\rand}                % alias
\newcommand\Srand{\boldsymbol{\bar {\mathtt R}}}  % state-bounded random intergers
\newcommand\rec{\newop{rec}}       % rec operator
\newcommand\0{\boldsymbol{0}}           % constant 0
\renewcommand\S{\newop S}         % succesor operator
\newcommand\piO{\newop\pi_1}     % projections
\newcommand\piT{\newop\pi_2}     % -
\newcommand\piI{\newop\pi_i}     % -
\newcommand{\PCF}{\mathbb{PCF}}
\newcommand{\npT}{\mathbb{T}}                     % system T
\newcommand\SystT{\npT}                           % alias
\newcommand{\pT}[1]{\npT^{#1}}                    % system T with aditional operators
\newcommand\SystTot{\pT{\bc,\rnd,\xc}}            % our full probabilistic system
\newcommand\SystFix{\pT {\fixran}}                % system T with prob. fixpoint
\newcommand\SystPl{\pT {\bc}}                     % system T with boolean choice
\newcommand\SystRand{\pT{\rand}}                  % system T with random intergers
\newcommand\SystPlR{\pT {\bc,\rand}}              % system T with boolean choice and random intergers
\newcommand\SystSRand{\pT{\Srand}}                % system T with state-bounded random intergers
\newcommand\Dist[1]{\mathfrak D (#1)}              % the distributions over...  
\newcommand{\BPP}{\mathbf{BPP}}
\newcommand{\ZPP}{\mathbf{ZPP}}
\newcommand\BP{\mathbf{BPT}} 
\newcommand\BPf{\mathbf{BPT}_{\!\ge\SystT\!}}
\newcommand\LV{\mathbf{LVT}}
\newcommand\LVf{\mathbf{LVT}_{\!\ge \SystT\!}}

\newcommand\PrC{\mathbf{PT}}
\newcommand\NDC{\mathbf{NT}}
\newcommand\DT{\mathbf{DT}}

\newcommand\Eval[1]{\left\llbracket #1\right\rrbracket}                           % Evaluation
\newcommand\Evall[1]{\Eval{#1}}                           % Evaluation (long exp)
\newcommand\Evala[1]{\Eval{#1}_a}    % Accesible Evaluation
\newcommand\supp[1]{\left|#1\right|}         % Support of a distribution
\newcommand\suppR[1]{\supp{#1}_R}      % Reducible support of a distribution
\newcommand\suppV[1]{\supp{#1}_V}      % Value support of a distribution
\newcommand\Succ{Succ}%{\scalerel*{\mathbb\Sigma}{\sum}}            % probability of normalisation
\newcommand\AvLength[1]{[#1]}            % probability of normalisation
\newcommand\Red[1]{Red_{#1}}      % reducibility candidates
\newcommand\norm[1]{||#1||}

  \newcommand\llc{\llparenthesis}
  \newcommand\rrc{\rrparenthesis}

\newcommand\rta{\rightarrow}

\newcommand\Rta{\Rightarrow}
\newcommand\Lra{\Leftrightarrow}
\newcommand\typ{\mathop{:}}

\newcommand\dash{\text{-}}
\newcommand\at{@}
\newcommand\ret{\newop{ret}}
\newcommand\bindo{{=}\hspace{-0.2em}{<\hspace{-0.3em}<}}

\newcommand\pmTermM{M} %meta-variable for terms
\newcommand\pmTermN{N} %meta-variable for terms
\newcommand\pmTermL{L} %meta-variable for terms
\newcommand\pmDistM{\mathcal{M}} %meta-variable for distributions over terms
\newcommand\pmDistN{\mathcal{N}} %meta-variable for distributions over terms
\newcommand\pmDistL{\mathcal{L}} %meta-variable for distributions over terms
\newcommand\pmDistP{\mathcal{P}} %meta-variable for distributions over terms
\newcommand\pmDistQ{\mathcal{Q}} %meta-variable for distributions over terms
\newcommand\pmDistX{\mathcal{X}} %meta-variable for distributions over terms
\newcommand\pmDistvU{\mathcal{U}} %meta-variable for distributions over values
\newcommand\pmDistvV{\mathcal{V}} %meta-variable for distributions over values
\newcommand\pmDistvW{\mathcal{W}} %meta-variable for distributions over values
\newcommand\pmValU{U} %meta-variable for values
\newcommand\pmValV{V} %meta-variable for values
\newcommand\pmValW{W} %meta-variable for values
\newcommand\pmTypa{A} %meta-variable for types
\newcommand\pmTypb{B} %meta-variable for types
\newcommand\pmTypc{C} %meta-variable for types

\newcommand{\tran}{\mathrm{tran}}

\newenvironment{varitemize}
{
\begin{list}{\labelitemi}
{
\setlength{\itemsep}{0pt}
 \setlength{\topsep}{0pt}
 \setlength{\parsep}{0pt}
 \setlength{\partopsep}{0pt}
 \setlength{\leftmargin}{15pt}
 \setlength{\rightmargin}{0pt}
 \setlength{\itemindent}{0pt}
 \setlength{\labelsep}{5pt}
 \setlength{\labelwidth}{10pt}}}
{
 \end{list}
}

\newcommand{\church}{\textsc{Church}}
\newcommand{\anglican}{\textsc{Anglican}}

\newcommand{\midd}{\; \; \mbox{\Large{$\mid$}}\;\;} 

\newcommand{\tpttype}[1]{(\hspace{-0.2em}(#1)\hspace{-0.2em})}
\newcommand{\tptnumber}[1]{\tpttype{#1}_{\!\#}}
\newcommand{\tptbranch}[1]{\tpttype{#1}_{\!\downarrow}}
\newcommand{\tptvals}[1]{\tpttype{#1}_{\boldsymbol{V}}}
\newcommand{\NF}[1]{\mathtt{N\!F}(#1)}
\newcommand{\EOE}{\hfill\Large$\boxtimes$}

\newcommand{\sumd}[3]{\sum_{#1\mapsto #2}#3}

\begin{document}

\title{On Higher-Order Probabilistic Subrecursion}

\author[F. Breuvart]{Flavien Breuvart}[a]
\address{LIPN, Universit\'e Paris Nord, France}
\email{flavien.breuvart@lipn.univ-paris13.fr}

\author[U. Dal Lago]{Ugo Dal Lago}[b]
\address{DISI, Universit\`a di Bologna, Italy}
\email{ugo.dallago@unibo.it}

\author[A. Herrou]{Agathe Herrou}[c]
\address{LIRIS, Universit\'e Claude Bernard 1, France}
\email{agathe.herrou@liris.cnrs.fr}

\begin{abstract}
  We study the expressive power of subrecursive probabilistic
  higher-order calculi. More specifically, we show that endowing a
  very expressive deterministic calculus like G\"odel's $\npT$ with
  various forms of probabilistic choice operators may result in
  calculi which are \emph{not} equivalent as for the class of
  distributions they give rise to, although they all guarantee
  \emph{almost-sure} termination. Along the way, we introduce a
  probabilistic variation of the classic reducibility technique, and
  we prove that the simplest form of probabilistic choice leaves the
  expressive power of $\npT$ essentially unaltered. The paper ends
  with some observations about the functional expressive power: 
  expectedly,
  all the considered calculi capture the functions which
  $\SystT$ itself represents, at least when standard notions of
  observations are considered.
\end{abstract}

\maketitle

%%%%%%%%%%%%%%%%%%%%%%
\section*{Introduction}
%%%%%%%%%%%%%%%%%%%%%%
Probabilistic models are more and more pervasive in computer science
and are among the most powerful modeling tools in many areas like computer
vision \cite{prince2012}, machine learning \cite{pearl1988} and
natural language processing \cite{manning1999}. Since the early times
of computation theory~\cite{de1956computability}, the very concept of
an algorithm has been itself generalized from a purely deterministic
process to one in which certain elementary computation steps can have
a probabilistic outcome. This has further stimulated research in
computation and complexity theory \cite{gill1974}, but also in
programming languages \cite{SahebDjahromi,kozen1981}.

Endowing programs with probabilistic primitives ({\em e.g.}, an 
operator
which models sampling from a distribution) poses a challenge to
programming language semantics. Already for a minimal, imperative
probabilistic programming language, giving a denotational semantics is
nontrivial~\cite{kozen1981}. When languages also have higher-order
constructs, everything becomes even harder \cite{jungtix1998} to the
point of disrupting much of the beautiful theory known in the
deterministic case~\cite{barendregt1984}. This has stimulated research
about the denotational semantics of higher-order probabilistic programming
languages, with some surprising positive results coming out recently
({\em e.g.}, 
\cite{ehrhardtassonpagani2014,dallagocrubille2014,DLFVY17,HKSY2017}).

Not much is known about the expressive power of \emph{probabilistic}
higher-order calculi, as opposed to the extensive literature on
\emph{deterministic} such calculi (see,
{\em e.g.} \cite{Statman1979,Sorensen2006}). What happens to the class of
representable functions if one enriches, say, a deterministic
$\lambda$-calculus $\mathbb{X}$ with certain probabilistic choice
primitives? Are the expressive power and/or the good properties 
of
$\mathbb{X}$ somehow preserved?  These questions have been given
answers in the case in which $\mathbb{X}$ is the pure, untyped,
$\lambda$-calculus~\cite{dallagozorzi2012}: in that case, the calculus
stays universal, mimicking what happens in Turing machines
\cite{Santos69}. But what if $\mathbb{X}$ is one of the many typed
$\lambda$-calculi ensuring strong normalization for typed
terms~\cite{proofsandtypes}?

Let us do a step back, first: when should a higher-order probabilistic
program be considered terminating in the first place?  The question
can be given a satisfactory answer being inspired by, {\em e.g.}, recent
work on probabilistic termination in imperative languages and term
rewrite systems \cite{mcivermorgan,bournez2005}: one could ask the
probability of divergence to be $0$, called the \emph{almost sure
  termination} property, or the stronger \emph{positive almost sure
  termination} property, in which one requires the average 
  number of evaluation
steps to be finite. That termination is desirable property, even in a
probabilistic setting, can be seen, {\em e.g.} in the field of languages
like \church\ and \anglican, in which programs are often assumed to be
almost surely terminating, {\em e.g.} when doing inference by MH algorithms
\cite{gmrbt2008}: if they are not, inference is bound to fail
miserably.

In this paper, we initiate a study on the expressive power of
terminating higher-order calculi, in particular those obtained by
endowing G\"odel's $\npT$ with various forms of probabilistic choice
operators. In particular, three operators will be analyzed in this
paper:
\begin{varitemize}
\item
  A binary probabilistic operator $\bc$ such that for every pair of
  terms $M,N$, the term $M\bc N$ evaluates to either $M$ or $N$,
  each with probability $\frac{1}{2}$. This is a rather minimal
  option, which, however, guarantees universality if applied to
  the untyped $\lambda$-calculus \cite{dallagozorzi2012} (and, more
  generally, to universal models of computation \cite{Santos69}).
\item
  A combinator $\rnd$, which evaluates to any natural number $n\geq 0$
  with probability $\frac{1}{2^{n+1}}$. This is the natural
  generalization of $\bc$ to sampling from the \emph{geometric distribution}
  which has \emph{countable} rather than \emph{finite} support. This
  apparently harmless generalization (which is absolutely
  non-problematic in the context of a Turing-complete computational
  model) has dramatic consequences in a subrecursive scenario, as we
  will discover soon.
\item
  A combinator $\xc$ such that for every pair of values $V,W$,
  the term $\xc\langle V,W\rangle$ evaluates to either $W$
  or $V (\xc\langle V,W\rangle)$, each with probability $\frac{1}{2}$. The
  operator $\xc$ can be seen as a probabilistic variation on $\PCF$'s
  fixpoint combinator. As such, $\xc$ is potentially problematic
  for termination, giving rise to infinite reduction trees.
\end{varitemize}
This way, various calculi can be obtained, {\em e.g.}, $\pT{\bc}$, 
namely a
minimal extension of $\npT$, or the full calculus $\pT{\bc,\rnd,\xc}$,
in which the three operators are all available. In principle, the only
obvious fact about the expressive power of the above mentioned
operators is that both~$\rnd$ and~$\xc$ are at least as expressive as
$\bc$: binary choice can be easily expressed by either $\rnd$ or
$\xc$. Less obvious, but still easy to prove, is the equivalence
between $\rnd$ and $\xc$ in presence of an operator for primitive
recursion (see
Section \ref{sect:roadmap}). But how about, say, $\pT{\bc}$ vs. $\pT{\rnd}$?
%: we will discuss the encoding in more details in Section~\ref{sec:fullCalculus}. 

Traditionally, the expressive power of such languages are 
compared by
looking at the set of functions $f:\Nat\rta\Nat$ defined by typable
programs $\pmTermM:\NAT\rta\NAT$. However, in a probabilistic setting,
programs $\pmTermM:\NAT\rta\NAT$ computes functions from natural
numbers to \emph{distributions} of natural numbers. In order to
evaluate the impact of the underlying probabilistic choice
operator(s), we thus need to fix a notion of observation.  There are at
least two relevant notions of observations, corresponding to two
randomized programming paradigms, namely the so-called \emph{Las
  Vegas} and \emph{Monte Carlo}
observations~\cite{motwani1995,AroraBarak2009}. The main problem,
then, consists in understanding how the obtained classes relate to
each other, and to the class of $\SystT$-representable functions,
which is well-known to comprise precisely the provably total
functions of Peano's arithmetic~\cite{proofsandtypes}. Along the way,
however, we manage to understand how to compare the expressive power
of probabilistic calculi \emph{per se}. Summing up, this paper's main
contributions are the following ones:
\begin{varitemize}
\item
  We first take a look at the full calculus $\pT{\bc,\rnd,\xc}$, and
  prove that it enforces almost-sure termination, namely that the
  probability of termination of any typable term is $1$. This is done
  by appropriately adapting the well-known reducibility technique
  \cite{proofsandtypes} to a probabilistic operational semantics. We
  then observe that while $\pT{\bc,\rnd,\xc}$ cannot be
  \emph{positively} almost surely terminating, $\pT{\bc}$ indeed
  is. This already shows that there must be a gap in
  expressive power. This is done in Section \ref{sec:fullCalculus}.
\item
  In Section \ref{sec:fragmentBinChoice}, we look more closely
  at the expressive power of $\pT{\bc}$, proving that the mere
  presence of probabilistic choice does not add much to the expressive
  power of $\npT$: in a sense, probabilistic choice can be ``lifted
  up'' to the ambient deterministic calculus.
\item
  We look at other fragments of $\pT{\bc,\rnd,\xc}$ and at their
  expressive power. More specifically, we will prove that (the
  equiexpressive) $\pT{\rnd}$ and $\pT{\xc}$ represent precisely
  what $\pT{\bc}$ can do \emph{at the limit}, in a sense which will be
  made precise in Section \ref{sec:fullCalculus}.  This part, which turns
  out to be the most challenging, is in Section \ref{sec:fragmentRand}.
\item 
  Section~\ref{sec:subrec} is devoted to proving that both for
  \emph{Monte Carlo} and for \emph{Las Vegas} observations, the class
  of functions representable in $\pT{\rnd}$ coincides with the
  $\SystT$-representable ones, the only exception being classes
  obtained by observing the most likely outcome, which are much
  larger and of questionable interest.
\end{varitemize}

\section{Probabilistic Choice Operators, Informally}
    \label{sec:2ndSec}
    Any term of G\"odel's $\npT$ can be seen as a purely deterministic
computational object whose dynamics is finitary, due to the well-known
strong normalization theorem (see, e.g., \cite{proofsandtypes}). In
particular, the non-determinism due to multiple redex occurrences is
completely harmless because of confluence. Confluence is
well-known \emph{not} to hold in a probabilistic
scenario \cite{dallagozorzi2012}, but in this paper we neglect this
problem, and work with a fixed reduction strategy, namely weak
call-by-value reduction (keeping in mind that all what we say here
also holds when call-by-name is the underlying notion of
reduction). Evaluation of a $\npT$-term $M$ of type $\NAT$ can be seen
as a finite sequence of terms ending in the normal form
$\boldsymbol{n}$ of $M$ (see Figure~\ref{fig:exSystT}). More
generally, the unique normal form of any $\npT$ term $M$ will be
denoted as~$\Eval{M}$.  Noticeably, $\npT$ is computationally very
powerful. In particular, the $\npT$-representable functions on
$\Nat$ coincide with the functions which are provably total
in Peano's arithmetic~\cite{proofsandtypes}.

As we already mentioned, the most natural way to enrich deterministic
calculi and turn them into probabilistic ones consists in endowing
their syntax with one or more probabilistic choice
operators. Operationally, each of them models the process of sampling from a 
distribution and proceeding
depending on the outcome.  Of course, one has many options here as
for \emph{which one(s)} of the various operators to grab. The aim of this
work is precisely the one of studying to which extent this choice have
consequences on the overall expressive power of the underlying
calculus.

Suppose, for example, that $\npT$ is endowed with the binary probabilistic
choice operator~$\bc$ described in the Introduction, whose evaluation
corresponds to tossing a fair coin and choosing one of the two
arguments accordingly. The presence of $\bc$ has indeed an impact on
the dynamics of the underlying calculus: the evaluation of any term $M$
is not deterministic anymore, but can be modeled as a finitely branching tree (see,
e.g. Figure \ref{fig:exSystT+} for such a tree when $M$ is
$(\boldsymbol 3\bc\boldsymbol 4)\bc\boldsymbol 2\:$). The fact that all
branches of this tree have finite height (and the tree is thus finite)
is intuitive, and a proof of it can be given by adapting the
well-known reducibility proof of termination for $\npT$. In this
paper, we in fact prove much more, and establish that $\pT{\bc}$ can be
embedded into $\npT$.

If $\bc$ is replaced by $\rnd$, the underlying tree is not finitely
branching anymore, but, again, there is no infinitely long branch,
since each of them can somehow be seen as a $\npT$ computation (see
Figure \ref{fig:exSystTR} for an example). What happens to the
expressive power of the obtained calculus? Intuition tells us that the
calculus should not be too expressive viz. $\pT{\bc}$. If $\bc$ is
replaced by $\xc$, on the other hand, the underlying tree \emph{is}
finitely branching, but its height can well be infinite (see
Figure~\ref{fig:exSystTX}). Actually, $\xc$ and $\rnd$ are easily
shown to be equiexpressive in presence of higher-order recursion, as
we show in Section~\ref{sect:roadmap}. On the other hand, for $\rnd$
and $\bc$, no such result is possible. Nonetheless, $\pT{\rnd}$ can
still be somehow encoded into $\npT$, as we will detail in
Section \ref{sec:fragmentRand}.  From this embedding, we can show that
neither Monte Carlo nor Las Vegas algorithms on $\pT{\bc,\xc,\rnd}$
add any expressive power to $\npT$.  This is done in
Section~\ref{sec:subrec}.

\begin{figure}
  \begin{center}
    \fbox{
      \begin{minipage}{.97\textwidth}
      	\begin{center}
        \begin{minipage}[c]{.328\textwidth}
          \begin{subfigure}[c]{.99\textwidth}
          $$\pmTermM\rta\cdots\rta\boldsymbol{n}$$
          \caption{$\SystT$}\label{fig:exSystT}	
          \end{subfigure}
          \\
          \begin{subfigure}[c]{.99\linewidth}
          \begin{tikzpicture}[scale=0.95]
            \node (v) at (1.5,3.5) {$\rand$};
            \node (v1) at (0.5,2) {$\boldsymbol0$};
            \node (v2) at (1.2,2) {$\boldsymbol1$};
            \node (v3) at (1.9,2) {$\boldsymbol2$};
            \node (v4) at (2.6,2) {$\boldsymbol3$};
            \node (v5) at (3.2,2) {$\boldsymbol4$};
            \node (v6) at (4,2) {$\vphantom 1$};
            \node (v7) at (4,2.2) {$\vphantom 1$};
            \node (v8) at (4,2.37) {$\vphantom 1$};
            \node (v9) at (4,2.5) {$\vphantom 1$};
            \node (v10) at (4,2.6) {$\vphantom 1$};
            \draw[->] ($(v.south)+(-0.2,0)$) -- node [below left] {{\scriptsize $\frac 1 2\ \ \ \phantom.$}} (v1.north);
            \draw[->] ($(v.south)+(-0.1,0)$) -- node [below left] {{\scriptsize $\frac 1 {4_{\vphantom 1}}$}} (v2.north);
            \draw[->] ($(v.south)+(0,0)$) -- node [below] {{\scriptsize $\frac 1 8 \phantom.$}} (v3.north);
            \draw[->] ($(v.south)+(0.05,0)$) -- node [below] {{\scriptsize $\ \frac 1 {16}$}} (v4.north);
            \draw[->] ($(v.south)+(0.1,0)$) -- node [below right] {} (v5.north);
            \draw[dashed,->] ($(v.south)+(0.15,0)$) -- node [right] {{}} (v6.north west);
            \draw[dashed,->] ($(v.south)+(0.2,0)$) -- node [right] {{}} (v7.north);
            \draw[dashed,->] ($(v.south)+(0.23,0)$) -- node [right] {{}} (v8.north);
            \draw[dashed,->] ($(v.south)+(0.26,0)$) -- node [right] {{}} (v9.north);
            \draw[dashed,->] ($(v.south)+(0.3,0)$) -- node [right] {{}} (v10.north);
          \end{tikzpicture}
          \caption{$\SystRand$}
          \label{fig:exSystTR}
          \end{subfigure}
        \end{minipage}
        \begin{minipage}[c]{.328\textwidth}
          \begin{subfigure}[c]{.99\textwidth}
          \begin{tikzpicture}[scale=0.95]
            \node (t) at (2,3) {$(\boldsymbol{3}\bc\boldsymbol{4})\bc\boldsymbol{2}$};
            \node (t1) at (1,1.5) {$\boldsymbol{3}\bc\boldsymbol{4}$};
            \node (t2) at (3,1.5) {$\boldsymbol{2}$};
            \node (t11) at (0,0) {$\boldsymbol{3}$};
            \node (t12) at (2,0) {$\boldsymbol{4}$};
            \draw[->] ($(t.south)+(-0.4,0)$) -- node [right] {{\scriptsize $\frac 1 2$}} (t1.north);
            \draw[->] ($(t.south)+(0.4,0)$) -- node [left] {{\scriptsize $\frac 1 2$}} (t2.north);
            \draw[->] ($(t1.south)+(-0.2,0)$) -- node [right] {{\scriptsize $\frac 1 2$}} (t11.north);
            \draw[->] ($(t1.south)+(0.2,0)$) -- node [left] {{\scriptsize $\frac 1 2$}} (t12.north);
          \end{tikzpicture}
          \caption{$\SystPl$}
          \label{fig:exSystT+}
          \end{subfigure}
        \end{minipage}
        \begin{minipage}[c]{.328\textwidth}
          \begin{subfigure}[c]{.99\textwidth}
          \begin{tikzpicture}[scale=0.95]
            \node (u) at (0.7,3) {$\fixran\ \S\ \boldsymbol 3$};
            \node (u1) at (0,2) {$\boldsymbol 3$};
            \node (u2) at (1.4,2) {$\S(\fixran\ \S\ \boldsymbol 3)$};
            \node (u21) at (0.7,1) {$\boldsymbol 4$};
            \node (u22) at (2.1,1) {$\S\S(\fixran\ \S\ \boldsymbol 3)$};
            \node (u221) at (1.4,0) {$\boldsymbol 5$};
            \node (u222) at (3,0.1) {$\ddots$};
            \draw[->] ($(u.south)+(-0.2,0)$) -- node [left] {{\scriptsize $\frac 1 2$}} (u1.north);
            \draw[->] ($(u.south)+(0.2,0)$) -- node [right] {{\scriptsize $\frac 1 2$}} (u2.north);
            \draw[->] ($(u2.south)+(-0.2,0)$) -- node [left] {{\scriptsize $\frac 1 2$}} (u21.north);
            \draw[->] ($(u2.south)+(0.2,0)$) -- node [right] {{\scriptsize $\frac 1 2$}} (u22.north);
            \draw[->] ($(u22.south)+(-0.2,0)$) -- node [left] {{\scriptsize $\frac 1 2$}} (u221.north);
            \draw[->] ($(u22.south)+(0.2,0)$) -- node [right] {{\scriptsize $\frac 1 2$}} ($(u222.north west)+(0.1,-0.3)$);
          \end{tikzpicture}
          \caption{$\SystFix$}\label{fig:exSystTX}
          \end{subfigure}
        \end{minipage}
        \end{center}
      \end{minipage}}
  \end{center}
  \caption{Reduction in Fragments of $\SystTot$}
\end{figure}

\section{The Full Calculus $\pT{\bc,\mathtt R,\mathtt X}$}
    \label{sec:fullCalculus}
    All along this paper, we work with a calculus $\SystTot$ whose
\emph{terms} are the ones generated by the following grammar:
\begin{align*}
\pmTermM,\pmTermN,\pmTermL\ ::=\ &
      x \midd \lambda x.\pmTermM \midd \pmTermM\ \pmTermN \midd 
      \langle\pmTermM,\pmTermN\rangle \midd \piO \midd \piT 
      \midd\rec \midd \0 \midd \S \midd \pmTermM \oplus \pmTermN \midd \rand \midd \fixran.
\end{align*}
Please observe the presence of the usual constructs from the
untyped $\lambda$-calculus, but also of primitive recursion,
constants for natural numbers, pairs, and the three choice operators
we have described in the previous sections.

As usual, terms are taken modulo $\alpha$-equivalence. Terms in which
no variable occurs free are, as usual, dubbed \emph{closed}, and are
collected in the set $\SystTot_C$. A \emph{value} is simply a closed
term from the following grammar
\begin{equation}\label{equ:grammarvalues}
 \pmValU,\pmValV\ ::=\ \lambda
 x.\pmTermM \midd \langle \pmValU,\pmValV\rangle \midd
 \piO\midd \piT \midd \rec \midd \0 \midd \S \midd \S\ \pmValV \midd \fixran,
\end{equation}
and the set of all values is $\SystTot_{V}$. Closed terms that are not values are
called \emph{reducible} and their set is denoted $\SystTot_R$. A
\emph{context} is a term with a unique hole:
$$
  C := \llc\cdot\rrc \midd \lambda x.C\midd C\ \pmTermM \midd \pmTermM\ C \midd
  \langle C,\pmTermM\rangle \midd \langle\pmTermM, C\rangle \midd
  C\oplus \pmTermM \midd \pmTermM\oplus C 
$$
We write $\SystTot_{\llc\cdot\rrc}$ for the set of all such contexts.
The expression $C\llc\pmTermM\rrc$ indicates the term obtained by
substituting $\llc\cdot\rrc$ with $\pmTermM$ inside $C$.

Termination of G\"odel's $\npT$ is guaranteed by the presence of
types, which we also need here. \emph{Types} are expressions generated
by the following grammar
$$
\pmTypa,\pmTypb ::= \NAT \midd \pmTypa\rta \pmTypb \midd \pmTypa\times \pmTypb.
$$
\emph{Environmental contexts} are expressions of the form $\Gamma=x_1\typ
\pmTypa_1,\ldots,x_n\typ \pmTypa_n$, while \emph{typing judgments} are of
the form $\Gamma\vdash \pmTermM\typ \pmTypa$. \emph{Typing rules} are
given in Figure~\ref{fig:TySysTot}. From now on, only typable terms
will be considered. We denote as $\SystTot_C(\pmTypa)$ the set of
closed terms of type $\pmTypa$, similarly for $\SystTot_V(\pmTypa)$.
\begin{figure}
  \fbox{
  \begin{minipage}{0.976\textwidth}
  \vspace*{1em}
  \begin{small}
  \begin{center}
    \AxiomC{$\vphantom{\Gamma}$}
    \UnaryInfC{$\Gamma,x\typ \pmTypa\vdash x\typ \pmTypa$}
    \DisplayProof \hspace{3.5em}
    \AxiomC{$\Gamma,x\typ \pmTypa\vdash \pmTermM: \pmTypb$}
    \UnaryInfC{$\Gamma \vdash \lambda x.\pmTermM : \pmTypa\rta \pmTypb$}
    \DisplayProof \hspace{3.5em}
    \AxiomC{$\Gamma\vdash \pmTermM:\pmTypa\rta \pmTypb$}
    \AxiomC{$\Gamma\vdash \pmTermN:\pmTypa$}
    \BinaryInfC{$\Gamma \vdash \pmTermM\ \pmTermN: \pmTypb$}
    \DisplayProof \\[1.5em]
    
    \AxiomC{}
    \UnaryInfC{$\Gamma\vdash \0:\NAT$}
    \DisplayProof \hspace{2.5em}
    \AxiomC{}
    \UnaryInfC{$\Gamma\vdash \S:\NAT\rta\NAT$}
    \DisplayProof \hspace{2.5em}
    \AxiomC{}
    \UnaryInfC{$\Gamma\vdash \rec:(\pmTypa\times(\NAT\rta\pmTypa\rta\pmTypa)\times\NAT)\rta\pmTypa$}
    \DisplayProof \\[1.5em]
    
    \AxiomC{$\Gamma\vdash \pmTermM:\pmTypa$}
    \AxiomC{$\Gamma\vdash \pmTermN:\pmTypb$}
    \BinaryInfC{$\Gamma \vdash \langle \pmTermM, \pmTermN\rangle: \pmTypa\times\pmTypb$}
    \DisplayProof \hspace{2.5em}
    \AxiomC{$\vphantom\Gamma$}
    \UnaryInfC{$\Gamma\vdash \piO: (\pmTypa\times\pmTypb)\rta\pmTypa$}
    \DisplayProof \hspace{2.5em}
    \AxiomC{$\vphantom\Gamma$}
    \UnaryInfC{$\Gamma\vdash \piT: (\pmTypa\times\pmTypb)\rta\pmTypb$}
    \DisplayProof \\[1.5em]
    
    \AxiomC{$\Gamma\vdash \pmTermM:\pmTypa$}
    \AxiomC{$\Gamma\vdash \pmTermN:\pmTypa$}
    \BinaryInfC{$\Gamma\vdash \pmTermM\oplus\pmTermN:\pmTypa$}
    \DisplayProof \hspace{2.5em}
    \AxiomC{$\vphantom\Gamma$}
    \UnaryInfC{$\Gamma\vdash \rand: \NAT$}
    \DisplayProof \hspace{2.5em}
    \AxiomC{$\vphantom\Gamma$}
    \UnaryInfC{$\Gamma\vdash \fixran: (\pmTypa\rta\pmTypa)\times\pmTypa\rta\pmTypa$}
    \DisplayProof %\hspace{2.5em}
  \end{center}
  \end{small}
  \end{minipage}}
  \caption{Typing Rules}\label{fig:TySysTot}
\end{figure}
Given $n\in\Nat$, we use the shortcut $\boldsymbol{n}$ for the
corresponding value of type $\NAT$: $\boldsymbol{0}$ is
already part of the language of terms, while $\boldsymbol {n+1}$ is simply
$\S\ \boldsymbol {n}$. For simplicity of notations, we also write $\S\S$ for
$\lambda x. \S(\S\ x)$; similarly for $\S\S\ldots\S$ ($n$ times).

%%%%%%%%%%%%%%%%%%%%%%%%%%%%%%%%%
\subsection{Operational Semantics}
%%%%%%%%%%%%%%%%%%%%%%%%%%%%%%%%%
While evaluating terms in a deterministic calculus ends up in
a \emph{value}, the same process leads to a \emph{distribution} of
values when applied to terms in a probabilistic calculus~\cite{dallagozorzi2012}.  The
reader should note that \emph{countable} distributions are perfectly
sufficient to give semantics to $\SystTot$, and that we will focus
our attention on them here. Formalizing all this requires some care,
but can be done following one of the many definitions from the
literature (e.g.,~\cite{dallagozorzi2012}).

Given a countable set $X$, a \emph{distribution $\pmDistL$ on $X$} is
a function mapping elements of $X$ to elements of the interval
$[0,1]$:
$$
\pmDistL,\pmDistM,\pmDistN \in \Dist{X} = \Bigl\{f:X\rta[0,1] \Bigm| \sum_{x\in X} f(x) \le 1\Bigr\}
$$
Observe that we take distributions as functions summing to
a real in $[0,1]$ rather than to $1$. This is an implicit way
to handle computations with nonzero probability
of divergence as distributions whose sum is strictly below $1$.

If $\pmDistL$ is a distribution on $X$ and $x\in X$, then $\pmDistL(x)$
is the real number which $\pmDistL$ put in correspondence to $x$.
We will use the pointwise order $\le$ on distributions, which turns
them into an $\omega\mathbf{CPO}$. The \emph{support} of a distribution $\pmDistM\in\Dist{X}$,
namely the subset of $X$ to which $\pmDistM$ assigns strictly positive
probability, is indicated as~$\supp\pmDistM$.
For any $\pmDistM\in\Dist{X}$ and $Y\subseteq X$, 
$\pmDistM^Y$ is another distribution in $\Dist{X}$ such
that $\pmDistM^Y(x)=\pmDistM(x)$ if $x\in Y$ and $\pmDistM^Y(x)=0$
otherwise. Another useful operation on distributions is
scalar multiplication: if $\pmDistL\in\Dist{X}$ and $p\in[0,1]$
then $p\cdot\pmDistL$ is the distribution assigning to $x$
the probability $p\pmDistL(x)$. 

We are especially interested in distributions over \emph{terms} here. In
particular, a \emph{distribution of type $\pmTypa$} is simply an
element of $\Dist{\SystTot_C(\pmTypa)}$ and, as such, it is a function
from $\SystTot_C(\pmTypa)$ to $[0,1]$ whose sum is itself less or equal to $1$.
The set $\Dist{\SystTot_{V}}$ is the set of of distribution over values
and will thus be used to assign a meaning to close terms; it
is ranged over by metavariables like $\pmDistvU,\pmDistvV,\pmDistvW$.
We use the following notation for Dirac's distributions over terms: 
$$
\{\pmTermM\}:=
\left\{\begin{matrix}\pmTermM\mapsto 1 \phantom{\text{ if }
  \pmTermM\neq\pmTermN}\\ \pmTermN\mapsto 0\text{ if }
\pmTermM\neq\pmTermN \end{matrix}\right\}.
$$
We define the \emph{reducible} and \emph{value supports} of a
distribution $\pmDistM$ on terms as
$\suppR\pmDistM:=\supp\pmDistM\cap\SystTot_R$
and~$\suppV\pmDistM:=\supp\pmDistM\cap\SystTot_V$, respectively. This
way, notations like $\pmDistM^R$ and $\pmDistM^V$ have an obvious and
natural meaning.

As syntactic sugar, we use the following convenient notation to manipulate
distributions, {\em i.e.}, for any family of distributions
$(\pmDistN_\pmTermM)_{\pmTermM\in\SystTot}$ such
that $\pmDistN_\pmTermM$ is in $\Dist{\SystTot}$ for
every $\pmTermM\in\SystTot$,
the expression $\sumd{\pmTermM}{\pmDistM}{\pmDistN_\pmTermM}$ stands 
for
$$
\sumd{\pmTermM}{\pmDistM}{\pmDistN_\pmTermM}
:=\ \sum_{\pmTermM\in\SystTot}\pmDistM(\pmTermM)\cdot\pmDistN_{\pmTermM}.
$$
By a slight abuse of notation, we may define $\pmDistN_\pmTermM$ only for
$\pmTermM\in|\pmDistM|$, since the other elements of the family
$(\pmDistN_\pmTermM)_{\pmTermM\in\SystTot}$ do not anyway contribute to the sum.
The sum notation as we use it here can be easily generalized, e.g., to families 
of real numbers
$(p_\pmTermM)_{\pmTermM\in\SystTot}$ and to other kinds of distributions.
\begin{exa}\label{ex:Distributions}
  Suppose that $\pmDistM=\left\{\boldsymbol n\mapsto \frac 1 {2^{n+1}}\mid 
n\in \Nat\right\}\in\Dist{\SystTot(\NAT)}$, which is the so-called exponential distribution 
over the natural numbers. Suppose, moreover, that for any $n$, 
$\pmDistN_{\boldsymbol n}=\left\{\begin{matrix} \boldsymbol {n+1} &\mapsto 
\frac 12\\ \0&\mapsto \frac 12 \end{matrix}\right\}$ is another
distribution in $\Dist{\SystTot(\NAT)}$. Then 
\begin{align*}
  \sumd{\pmTermM}{\pmDistM}{\pmDistN_\pmTermM} 
  &= \sum_n\frac 1 {2^{n+1}}\cdot \left\{
    \begin{matrix} 
      \boldsymbol {n+1} &\mapsto \frac 12\\[0.1em] \0&\mapsto \frac 12 
    \end{matrix}\right\} 
  = \left\{\begin{matrix}
      \boldsymbol {n+1} &\mapsto& \frac 1{2^{n+1}}\cdot\frac 12\\[0.1em] 
      \0 &\mapsto& \sum_m\frac 1{2^{m+1}} 
    \end{matrix}\right\}\\
  &=\left\{\boldsymbol n\mapsto \frac 1 {2^{n+1}}\mid n\in \Nat\right\},
\end{align*}
 which is exactly the same as $\pmDistM$.\EOE
\end{exa}
We indicate as $C\llc \pmDistM\rrc$ the
push-forward distribution $\sumd{\pmTermM}{\pmDistM}{\{C\llc\pmTermM\rrc]\}}$
induced by a context $C$, and as
$\norm\pmDistM$ the norm~$\sumd{\pmTermM}{\pmDistM}{1}$ of $\pmDistM$.
Remark, finally, that we have the useful equality
$\pmDistM=\sumd{\pmTermM}{\pmDistM}{\{\pmTermM\}}$. More generally,
the sum notation
as defined here satisfies some familiar identities, e.g.,
\begin{equation}\label{eq:IntInt}
   \sumd{\pmTermN}{\left(\sumd{\pmTermM}{\pmDistM}{\pmDistN_\pmTermM}\right)}{\pmDistL_\pmTermN} =
   \sumd{\pmTermM}{\pmDistM}{\left(\sumd{\pmTermN}{\pmDistN_\pmTermM}{\pmDistL_\pmTermN}\right)}.
\end{equation}
\begin{exa}\label{ex:Contexts}
If $\pmDistM$ is the exponential distribution from Example~\ref{ex:Distributions}
above, and $C$ is the context $(\lambda x.x)\llc\cdot\rrc$,
then the push-forward distribution $C\llc\pmDistM\rrc$ is the distribution
assigning probability $\frac{1}{2^{n+1}}$ to any term in the form
$(\lambda x.x)\boldsymbol{n}$, and probability $0$ to any other term.\EOE
\end{exa}
In the following we will often, by abuse of notation, write
push-forward distributions like the distribution
$C\llc\pmDistM\rrc$ from Example~\ref{ex:Contexts}
as $(\lambda x.x)\pmDistM$. We even go beyond that, and
write expressions like, e.g., $\pmDistM\pmDistN$, which stands
for the following distribution:
$$
(\pmDistM\pmDistN)(\pmTermM)=
\left\{
\begin{array}{ll}
\pmDistM(\pmTermN)\pmDistN(\pmTermL) & \mbox{if $\pmTermM=\pmTermN\pmTermL$};\\
0 & \mbox{otherwise}.
\end{array}
\right.
$$

Reduction rules of $\SystTot$ are given by Figure \ref{fig:OSTot}.
\begin{figure}
  \fbox{
  \begin{minipage}{.976\textwidth}
  \vspace*{1em}
  \begin{small}
  \begin{center}
    \AxiomC{\vphantom{M}}
    \RightLabel{{\scriptsize$(r\protect\dash\beta)$}}
    \UnaryInfC{$(\lambda x.\pmTermM)\ \pmValV\rta \Bigl\{\pmTermM[\pmValV/x]\Bigr\}$}
    \DisplayProof\hspace{5em}
    \AxiomC{$\pmTermM\rta \pmDistM$}
    \RightLabel{{\scriptsize$(r\protect\dash\protect\at L)$}}
    \UnaryInfC{$\pmTermM\ \pmValV \rta \pmDistM\ \pmValV \vphantom{\Bigl\{}$}
    \DisplayProof\\[1.5em]
    \AxiomC{$\pmTermN\rta \pmDistN$}
    \RightLabel{{\scriptsize$(r\protect\dash\protect\at R)$}}
    \UnaryInfC{$\pmTermM\ \pmTermN \rta \pmTermM\ \pmDistN \vphantom{\Bigl\{}$}
    \DisplayProof\hspace{3em}
    \AxiomC{$\pmTermM\rta \pmDistM$}
    \RightLabel{{\scriptsize$(r\protect\dash\langle\cdot\rangle L)$}}
    \UnaryInfC{$\langle\pmTermM,\pmTermN\rangle \rta \langle\pmDistM,\pmTermN \rangle \vphantom{\Bigl\{}$}
    \DisplayProof\hspace{3em}
    \AxiomC{$\pmTermM\rta \pmDistM$}
    \RightLabel{{\scriptsize$(r\protect\dash\langle\cdot\rangle R)$}}
    \UnaryInfC{$\langle\pmValV,\pmTermM\rangle \rta \langle\pmValV,\pmDistM\rangle \vphantom{\Bigl\{}$}
    \DisplayProof\\[1.5em] 
    \AxiomC{$\vphantom{M}$}
    \RightLabel{{\scriptsize$(r\protect\dash \protect\rec\protect \0)$}}
    \UnaryInfC{$\rec\langle\pmValU,\pmValV,\0\rangle \rta \Bigl\{\pmValU\Bigr\}$}
    \DisplayProof\hspace{3em}
    \AxiomC{$\vphantom{M}$}
    \RightLabel{{\scriptsize$(r\protect\dash \protect\rec\protect \S)$}}
    \UnaryInfC{$\rec\langle\pmValU,\pmValV,\S\ \boldsymbol n\rangle\rta \Bigl\{\pmValV\ \boldsymbol n\ (\rec\langle\pmValU,\pmValV,\boldsymbol n\rangle)\Bigr\}$}
    \DisplayProof\\[1.5em]
    \AxiomC{$\vphantom{M}$}
    \RightLabel{{\scriptsize$(r\protect\dash \protect\piO$)}}
    \UnaryInfC{$\piO\ \langle\pmValV,\pmValU\rangle \rta \Bigl\{\pmValV\Bigr\}$}
    \DisplayProof\hspace{3em}
    \AxiomC{$\vphantom{M}$}
    \RightLabel{{(\scriptsize$r\protect\dash \protect\piT)$}}
    \UnaryInfC{$\piT\ \langle\pmValV,\pmValU\rangle \rta \Bigl\{\pmValU\Bigr\}$}
    \DisplayProof\hspace{3em}
    \AxiomC{$\vphantom{M}$}
    \RightLabel{{\scriptsize$(r\protect\dash \rand)$}}
    \UnaryInfC{$\rand \rta \Bigl\{\boldsymbol{n}\mapsto \frac{1}{2^{n+1}}\Bigr\}_{n\in\Nat\hspace{-0.5em}}\!\!\!\!\!$}
    \DisplayProof\\[1.5em]
    \AxiomC{$\vphantom{M}$}
    \RightLabel{{\scriptsize$(r\protect\dash \bc)$}}
    \UnaryInfC{$\pmTermM \bc\pmTermN \rta \left\{\begin{matrix} \pmTermM &\mapsto &\frac 1 2 \\ \pmTermN &\mapsto &\frac 1 2 \end{matrix}\right\}$}
    \DisplayProof\hspace{3em} 
    \AxiomC{$\vphantom{M}$}
    \RightLabel{{\scriptsize$(r\protect\dash \fixran)$}}
    %\UnaryInfC{$\fixran\ \pmTermM\ \pmTermN \rta \frac 1 2\left\{\pmTermM\ (\fixran\ \pmTermM\ \pmTermN)\right\}+\frac 1 2 \left\{\pmTermN\right\}$}
    \UnaryInfC{$\fixran\langle\pmValV,\pmValW\rangle \rta \left\{\begin{matrix} \pmValV\ (\fixran\;\langle\pmValV,\pmValW\rangle) &\mapsto &\frac 1 2 \\ \pmValW &\mapsto &\frac 1 2 \end{matrix}\right\}$}
    \DisplayProof\\[1.8em]
    \AxiomC{$\forall M\in \suppR\pmDistM,\ \pmTermM\rta \pmDistN_{\pmTermM}$}
    \AxiomC{$\forall \pmValV\in \suppV\pmDistM,\ \pmDistN_{\pmValV}=\{\pmValV\}$}
    \RightLabel{{\scriptsize$(r\protect\dash{\in})$}}
    \BinaryInfC{$\pmDistM \rta  \sumd{\pmTermM}{\pmDistM}{\pmDistN_\pmTermM}$}
    \DisplayProof
  \end{center}
  \end{small}
  \smallskip
  \end{minipage}}
  \caption{Operational Semantics.}\label{fig:OSTot}
\end{figure}
For simplicity, we use the notation $\pmTermM\rta^?\pmDistM$ for
$\{\pmTermM\}\rta\pmDistM$ ({\em i.e.}, $\pmTermM\rta\pmDistM$)
whenever $\pmTermM$ is reducible and $\pmDistM=\{\pmTermM\}$ whenever
$\pmTermM$ is a value. The sum notation allows us to rewrite rule
$(r\dash{\in})$ as a form of monadic lifting:
\begin{center}
    \AxiomC{$\forall M\in \supp\pmDistM,\ \pmTermM\rta^? \pmDistN_{\pmTermM}$}
    \RightLabel{{\scriptsize$(r\protect\dash{\in})$}}
    \UnaryInfC{$\pmDistM \rta  \sumd{\pmTermM}{\pmDistM}{\pmDistN_\pmTermM}$}
    \DisplayProof
\end{center}

\begin{exa}\label{example:Expo}
  Notice that we can faithfully simulate system $\SystT$ inside
  $\SystTot$. For example, consider the following term:
  $$
  \boldsymbol{\mathtt{Expo}}\ :=\ \lambda n. \rec\ \langle \boldsymbol 1 \:,\: \lambda xy.\rec\ \langle \0 , \lambda x.\S\S , y \rangle \:,\: \S n\rangle \, :\ \NAT\rta\NAT
  $$
  where $\S\S$ is a shortcut for the term $\lambda z.\S(\S z)$
  computing the double-successor of its argument. This term computes
  the function $n\mapsto 2^{n+1}$ in time $O(2^{n})$. Indeed, when
  applied to a natural number~$\boldsymbol n$, it will get the
  following reduction where we denote $E_k:= \rec\ \langle \boldsymbol
  1 \:,\: \lambda xy.\rec\ \langle \0 , \lambda x.\S\S ,
  y \rangle \:,\: \boldsymbol k\rangle$:
  \begin{align*}
    \left(\lambda n. \rec\ \langle \boldsymbol 1 \:,\: \lambda xy.\rec\ \langle \0 , \lambda x.\S\S , y \rangle \:,\: \S n\rangle\right)\ \boldsymbol n
    \rta\ & \Bigl\{E_{n+1} \Bigr\}
    \rta\ \Bigl\{ (\lambda xy.\rec\ \langle \0, \lambda x.\S\S, y\rangle)\ \boldsymbol n\ E_{n} \Bigr\}\\
    \rta\ &\ldots\rta\Bigl\{ \rec\ \langle\0, \lambda x.\S\S, \boldsymbol{2 ^{n}}\rangle \Bigr\}\\
    \rta\ &\ldots \rta\ \Bigl\{ \boldsymbol{2^{n+1}}\Bigr\}.
  \end{align*}  
  Notice that we are only considering Dirac distributions since
  reduction is deterministic: no probabilistic choice operator occurs
  in $\boldsymbol{\mathtt{Expo}}$.\EOE
\end{exa}

\begin{exa}\label{ex:fixran}
  As a second example, we are presenting the term
  $\fixran\langle\S,\0\rangle$, whose reduction
  is essentially probabilistic:
\begin{align*}
    \fixran\langle\S,\0\rangle 
    &\rta  \left\{\begin{matrix} \S\ (\fixran\;\langle\S,\0\rangle) &\mapsto &\frac 1 2 \\[0.1em] \0 &\mapsto &\frac 1 2 \end{matrix}\right\} 
    &\rta  \left\{\begin{matrix} \S\S\ (\fixran\;\langle\S,\0\rangle) &\mapsto &\frac 1 4 \\[0.1em] \boldsymbol 1 &\mapsto &\frac 1 4 \\ \0 &\mapsto &\frac 1 2 \end{matrix}\right\} 
    &\rta  \left\{\begin{matrix} \S\S\S\ (\fixran\;\langle\S,\0\rangle) &\mapsto &\frac 1 8 \\[0.1em] \boldsymbol 2 &\mapsto &\frac 1 8 \\ \boldsymbol 1 &\mapsto &\frac 1 4 \\ \0 &\mapsto &\frac 1 2 \end{matrix}\right\} 
    &\rta \cdots
  \end{align*}
  Notice that, after $n+1$ reduction steps, the support of the underlying
  distribution has precisely $n$ elements.\EOE
\end{exa}

We can easily define $\rta^n$ as the $n^{th}$ iteration of $\rta$ and
$\rta^*$ as the reflexive and transitive closure of $\rta$. If
$\pmDistM\rta^n\pmDistN$ and $\pmDistvU=\pmDistN^V$, then
we write $\pmDistM\rta^{\leq n}\pmDistvU$. In other words, $\pmDistvU$
is the distribution on values to which $\pmDistM$ reduces in at
most $n$ steps. For every $\pmDistM$ and for every natural number
$n$, there are unique $\pmDistN$ and $\pmDistvU$ such that
$\pmDistM\rta^n\pmDistN$ and $\pmDistM\rta^{\leq n}\pmDistvU$.

In probabilistic systems, we might want to consider
infinite reduction sequences such as the ones induced by $\fixran\langle(\lambda
x.x),\0\rangle$, which reduces to $\{\0\}$, but only after an infinite number of
steps. Please note that for any value $\pmValV$,
and whenever $\pmDistM\rta\pmDistN$, it holds that
$\pmDistM(\pmValV)\le\pmDistN(\pmValV)$. As a consequence, we can
finally give the following definition, one of the most crucial ones
in this paper:
\begin{defi}\label{def:sem}
 Let $\pmTermM$ be a term and let $(\pmDistM_n)_{n\in\Nat}$ be the unique distribution family
 such that~$\pmTermM\rta^{\le n} \pmDistM_n$.
 The \emph{evaluation} of $\pmTermM$ is the value distribution
 $$
 \Eval{\pmTermM}:= \{\pmValV\mapsto\lim_{n\rta\infty}\pmDistM_n(\pmValV)\}\in\Dist{\SystTot_V}.
 $$
 The \emph{success} of $\pmDistM$ is its probability of
 termination, which is formally defined as the norm of its evaluation, i.e.,
 $\Succ(\pmTermM):= \norm{\Eval\pmTermM}$. 
 %$\pmDistM^{\Delta V}_n$ stands for $\{\pmValV\mapsto \pmDistM_n(\pmValV)-\pmDistM_{n-1}(\pmValV)\}$,
 %namely the distribution of values reachable in exactly $n$ steps from $\pmTermM$.
 The \emph{average reduction length} from $\pmTermM$ is
 $$
 \AvLength{\pmTermM}\ :=\ \sum_{n}\norm{\pmDistM^{R}_n}\ \in \ \Nat\cup\{+\infty\}
 $$
The operator $\Eval{\cdot}$ can be easily generalized to one
on \emph{distributions of terms}, since the sequence of distributions
$\pmDistM_n$ is anyway unique. Whenever $\Eval{\pmTermM}=\{\pmTermN\}$,
it makes sense to consider $\pmTermN$ as the normal form of $\pmTermM$; indeed,
we write $\pmTermN=\NF{\pmTermM}$ in all these cases, e.g. when $\pmTermM$
is a term from $\mathbb{T}$.
\end{defi}

\begin{exa}\label{ex:fixran2}
  Take as an example the term $\fixran\langle\S,\0\rangle$ from Example~\ref{ex:fixran}. We have that, for all $n$:
  $$ \fixran\langle\S,\0\rangle \rta^n \pmDistM_n = \left\{\begin{matrix} \S^n\ (\fixran\;\langle\S,\0\rangle) 
      &\mapsto &\frac 1 {2^n} \\[0.1em] \boldsymbol m &\mapsto &\frac 1 
      {2^{m+1}} &\text{for } m< n \end{matrix}\right\}\ , $$
  so that $\pmDistM_n(\boldsymbol m)=\0$ if $m\ge n$ and $\pmDistM_n(\boldsymbol m)=\frac 1 {2^{m+1}}$ otherwise. Thus:
  \begin{align*}
    \Eval{\fixran\langle\S,\0\rangle} 
    &= \left\{\boldsymbol m\mapsto\lim_{n\rta\infty}\pmDistM_n(\boldsymbol m)\ \middle|\ m\in \Nat\right\} \\
    &= \left\{\boldsymbol m\mapsto \frac 1 {2^{m+1}}\ \middle|\ m\in 
    \Nat\right\}.
      \tag*{\EOE}
  \end{align*}
\end{exa}
\noindent
Notice that, by Rule $(r\dash\in)$, evaluation is
continuous: 
$$
\Eval\pmDistM=\sumd{\pmTermM}{\pmDistM}{\Eval\pmTermM}
$$

Which kind of mathematical object does any term $\pmTermM$ of type
$\NAT\rta\NAT$ \emph{compute}? Following, e.g., \cite{dallagozuppiroli2014}, we
can say that any such term \emph{represents} a function $g:\Nat\rta\Dist{\Nat}$
iff for every $m,n$ it holds that~$g(n)(m)=\Eval{\pmTermM\;\boldsymbol{n}}(\boldsymbol{m})$.
This will be a key notion not only to \emph{evaluate} the expressive power of various
fragments of $\SystTot$, but also to \emph{compare} them.

%%%%%%%%%%%%%%%%%%%%%%%%%%%%%%%%%%%%%%%%%%%%%%%%%%%%%%%
\subsection{On the Continuity of the Operational Semantics}\label{sect:lemmas}
%%%%%%%%%%%%%%%%%%%%%%%%%%%%%%%%%%%%%%%%%%%%%%%%%%%%%%%

In this section, we will ask ourselves whether the semantics
$\Eval{MN}$ of an application $MN$ can somehow be given \emph{from} the
semantics of $\Eval{M}$ and $\Eval{N}$. Lemma~\ref{lm:continuity}
below gives a positive answer to this question, but 
some auxiliary lemmas are necessary beforehand.

The first such lemma states that the (one-step) reduction of a sum
is the sum of the one-step reducts of its addends. Of
course, the different reductions of the addends can have
interpolations which makes the decomposition nontrivial:

\begin{lem}\label{lm:int1step}
  For every $\pmDistM$ and $(\pmDistN_{\pmTermM})$,
  $\sumd{\pmTermM}{\pmDistM}{\pmDistN_\pmTermM}\rta\pmDistL$ iff $\pmDistL$ can be
  written as the sum $\sumd{\pmTermM}{\pmDistM}{\pmDistL_\pmTermM}$,
  where $\pmDistN_\pmTermM\rta\pmDistL_\pmTermM$ for any $\pmTermM\in\supp\pmDistM$.
\end{lem}
\begin{proof}
Let us analyse the two implications separately:
  \begin{varitemize}
  \item
    Let $(\pmDistL_\pmTermM)_{\pmTermM\in\supp\pmDistM}$ such that
  $\pmDistN_\pmTermM\rta\pmDistL_\pmTermM$ for any
  $\pmTermM\in\supp\pmDistM$. Then in order to derive
  $\pmDistN_\pmTermM\rta\pmDistL_\pmTermM$, the only rule we can apply
  is $(r\dash{\in})$ so that there is
  $(\pmDistQ_{\pmTermM,\pmTermN})_{\pmTermM\in\supp\pmDistM,\pmTermN\in\supp{\pmDistN_\pmTermM}}$
  such that
  $\pmDistL_\pmTermM=\sumd{\pmTermN}{\pmDistN_\pmTermM}{\pmDistQ_{\pmTermM,\pmTermN}}$
  and $\pmTermN\rta^?\pmDistQ_{\pmTermM,\pmTermN}$. Notice that for
  $\pmTermN\in\supp{\pmDistN_\pmTermM}\cap\supp{\pmDistN_{\pmTermL}}$,
  the determinism of $\rta^?$ gives that
  $\pmDistQ_{\pmTermM,\pmTermN}=\pmDistQ_{\pmTermL,\pmTermN}$, thus
  we can define $\pmDistQ_\pmTermN$ without ambiguity for any
  $\pmTermN\in\bigcup_{\pmTermM\in\supp\pmDistM}\supp{\pmDistN_\pmTermM}=\supp{\sumd{\pmTermM}{\pmDistM}{\pmDistN_\pmTermM}}$.
  Then we have $\pmTermN\rta^?\pmDistQ_{\pmTermN}$ and, thanks
  to Equation~\eqref{eq:IntInt},
  $$
  \sumd{\pmTermN}{\left(\sumd{\pmTermM}{\pmDistM}{\pmDistN_\pmTermM}\right)}{\pmDistQ_\pmTermN}=
  \sumd{\pmTermM}{\pmDistM}{\left(\sumd{\pmTermM}{\pmDistN_\pmTermM}{\pmDistQ_\pmTermN}\right)}=
  \sumd{\pmTermM}{\pmDistM}{\pmDistL_\pmTermM}.
  $$
  By rule
  $(r\dash{\in})$ we get $\sumd{\pmTermM}{\pmDistM}{\pmDistN_\pmTermM}
  \rta\sumd{\pmTermM}{\pmDistM}{\pmDistL_\pmTermM}$.
\item
  Conversely, assume that $\sumd{\pmTermM}{\pmDistM}{\pmDistN_\pmTermM}
  \rta\pmDistL$.  In order to derive it, the only rule we can
  apply is $(r\dash{\in})$ so that there is
  $(\pmDistQ_{\pmTermN})_{\pmTermN\in\supp{\sumd{\pmTermM}{\pmDistM}{\pmDistN_\pmTermM}}}$ such that
  $\pmDistL=\sumd{\pmTermN}{\left(\sumd{\pmTermM}{\pmDistM}{\pmDistN_\pmTermM}
  \right)}{\pmDistQ_\pmTermN}
  = \sumd{\pmTermM}{\pmDistM}{\sumd{\pmTermM}{\pmDistN_\pmTermM}{\pmDistQ_\pmTermN}}$ and
  $\pmTermN\rta^?\pmDistQ_\pmTermN$. We conclude by setting
  $\pmDistL_\pmTermM:=\sumd{\pmTermN}{\pmDistN_\pmTermM}{\pmDistQ_\pmTermN}$.
 \qedhere
  \end{varitemize}
\end{proof}
\noindent The decomposition can of course be iterated to reductions of any length:
\begin{lem}\label{lm:intmultstep}
  For every $\pmDistM$ and $(\pmDistN_{\pmTermM})$,
  $\sumd{\pmTermM}{\pmDistM}{\pmDistN_\pmTermM}\rta^n\pmDistL$, iff
  $\pmDistL$ can be written as $\sumd{\pmTermM}{\pmDistM}{\pmDistL_\pmTermM}$,
  where $\pmDistN_\pmTermM\rta^n\pmDistL_\pmTermM$ for any
  $\pmTermM\in\supp\pmDistM$.
\end{lem}
\begin{proof}
  By induction on $n$:
  \begin{varitemize}
  \item If $n=0$ then $\pmDistL_\pmTermM=\pmDistN$. 
  \item
  Otherwise, $\sumd{\pmTermM}{\pmDistM}{\pmDistN_\pmTermM}
  \rta\pmDistQ\rta^{n-1}\pmDistL$. By
  Lemma~\ref{lm:int1step}, the first step is possible iff
  $\pmDistQ=\sumd{\pmTermM}{\pmDistM}{\pmDistQ_\pmTermM}$ with
  $\pmDistN_\pmTermM\rta\pmDistQ_\pmTermM$ for any
  $\pmTermM\in\supp\pmDistM$. By IH, the remaining steps are then
  possible iff $\pmDistL=\sumd{\pmTermM}{\pmDistM}{\pmDistL_\pmTermM}$
  such that $\pmDistQ_\pmTermM\rta^{n-1}\pmDistL_\pmTermM$ for any
  $\pmTermM\in\supp\pmDistM$.
  \qedhere
  \end{varitemize}
\end{proof}
\noindent 
Thanks to Lemma~\ref{lm:intmultstep},
it is possible to trace back values obtained by reducing an application,
as stated in the following, intermediate, lemma:
\begin{lem}\label{lmaux1:continuity}
  If $(\pmTermM\ \pmTermN)\rta^n \pmDistL \ge \pmDistvW$ for some
  $\pmDistvW\in\Dist{\SystTot_V}$, then there are distributions
  $\pmDistM,\pmDistN,\pmDistP\in\Dist{\SystTot}$ and
  $\pmDistvU,\pmDistvV\in\Dist{\SystTot_V}$ such that
  $\pmTermM \rta^* \pmDistM \ge \pmDistvU$,
  $\pmTermN \rta^* \pmDistN \ge \pmDistvV$, and
  $(\pmDistvU\ \pmDistvV) \rta^* \pmDistP \ge \pmDistvW$.
\end{lem}
  \begin{proof}
    This is again an induction on $n$. The base case is trivial.  If,
    instead, $n\ge 1$, we proceed differently depending whether
    $\pmTermM$ and $\pmTermN$ are values or not.
    \begin{varitemize}
      \item
        If $\pmTermN$ is not a value then $\pmTermN\rta \pmDistN'$ and
        $(\pmTermM\ \pmTermN)\rta(\pmTermM\ \pmDistN')\rta^{n-1} \pmDistL \ge \pmDistvW$.
        Notice that
        $(\pmTermM\ \pmDistN'):=\sumd{\pmTermN'}{\pmDistN'}{\{\pmTermM\ \pmTermN'\}}$.
        Using Lemma~\ref{lm:intmultstep}, we can decompose
        $\pmDistvW=\sumd{\pmTermN'}{\pmDistN'}{\pmDistvW_{\pmTermN'}}$
        and~$\pmDistL=\sumd{\pmTermN'}{\pmDistN'}{\pmDistL_{\pmTermN'}}$
        in such a way that for any
        $\pmTermN'\in\supp{\pmDistN'}$,
        $(\pmTermM\ \pmTermN')\rta^{\le
        n-1} \pmDistL_{\pmTermN'} \ge \pmDistvW_{\pmTermN'}$. By
        induction we have
        $\pmTermN'\rta^*\pmDistN_{\pmTermN'}\ge\pmDistvV_{\pmTermN'}$
        and
        $\pmTermM\rta^*\pmDistM_{\pmTermN'}\ge\pmDistvU_{\pmTermN'}$
        with
        $(\pmDistvU_{\pmTermN'}\ \pmDistvV_{\pmTermN'})\rta^*\pmDistP_{\pmTermN'}\ge\pmDistvW_{\pmTermN'}$
        for all $\pmTermN'\in\supp{\pmDistN'}$. Summing up we have
        $\pmDistM:=\sumd{\pmTermN'}{\pmDistN'}{\pmDistM_{\pmTermN'}}$,
        $\pmDistvU:=\sumd{\pmTermN'}{\pmDistN'}{\pmDistvU_{\pmTermN'}}$,
        $\pmDistN:=\sumd{\pmTermN'}{\pmDistN'}{\pmDistN_{\pmTermN'}}$
        and
        $\pmDistvV:=\sumd{\pmTermN'}{\pmDistN'}{\pmDistvV_{\pmTermN'}}$.
      \item
        If $\pmTermN$ is a value and $\pmTermM\rta \pmDistM'$ then
        $(\pmTermM\ \pmTermN)\rta(\pmDistM'\ \pmTermN)\rta^{n-1}\pmDistL\ge\pmDistvW$. Thus
        we can decompose the equation similarly along $\pmDistM'$ and
        apply our IH.
      \item
        If both $\pmTermM$ and $\pmTermN$ are values it is trivial
      since $\pmDistvU=\{\pmTermM\}$ and
      $\pmDistvV=\{\pmTermN\}$.
      \qedhere
    \end{varitemize}
\end{proof}
\noindent We now have all the necessary ingredients for at least proving a
      restricted form of our desired continuity result:
\begin{lem}\label{lmaux2:continuity}
  For every $m,n\in\Nat$ and every
  $\pmDistM,\pmDistN\in\Dist{\SystTot}$, whenever
  $\pmTermM\rta^m\pmDistM$ and $\pmTermN\rta^n\pmDistN\ge\pmDistvV$,
  we have $\Eval{\pmTermM\ \pmTermN}\ge\Eval{\pmDistM\ \pmDistvV}$.
  In particular, if $\pmDistM\ge\pmDistvU$, then
  $\Eval{\pmTermM\ \pmTermN}\ge\Eval{\pmDistvU\ \pmDistvV}$.
\end{lem}
  \begin{proof}
    By induction on $m+n$. The base case $m+n=0$ is trivial.
    If $n\ge 1$, we proceed differently depending whether $\pmTermM$ and $\pmTermN$ are values or not.
    \begin{varitemize}
    \item
    If $\pmTermN\rta \pmDistL\rta^{n-1}\pmDistN\ge\pmDistvV$ then
      $\pmTermN$ is not a value. By Lemma~\ref{lm:intmultstep}, we
      can decompose
      $\pmDistN=\sumd{\pmTermL}{\pmDistL}{\pmDistN_{\pmTermL}}$ and
      $\pmDistvV=\sumd{\pmTermL}{\pmDistL}{\pmDistvV_{\pmTermL}}$ in
      such a way that for any $\pmTermL\in\supp{\pmDistL}$,
      $\pmTermL\rta^{\le
      n-1} \pmDistN_{\pmTermL}\ge\pmDistvV_{\pmTermL} $ (with
      $\rta^{\le n-1}$ standing for either $=$ or $\rta^{n-1}$ depending
      whether $\pmTermL$ is a value or not).  Then we get
      $$
      \Eval{\pmTermM\ \pmTermN} = \Eval{\pmTermM\ \pmDistL}
      = \sumd{\pmTermL}{\pmDistL}{\Eval{\pmTermM\ \pmTermL}}\ge \sumd{\pmTermL}{\pmDistL}{\Eval{\pmDistM\ \pmDistvV_{\pmTermL}}}
      = \Eval{\pmDistM\ \left(\sumd{\pmTermL}{\pmDistL}{\pmDistvV_{\pmTermL}}\right)}
      = \Eval{\pmDistM\ \pmDistvV}\ .
      $$
\item
    If $\pmDistvV$ is the empty (sub)distribution, this is trivial.  \item
    If $n=0$ and $\pmDistvV$ is not empty, then
    $\pmDistvV=\{\pmTermN\}$, thus $\pmTermN$ is a value and
    $\pmTermM\rta \pmDistL\rta^{m-1} \pmDistM$, then we can decompose
    the equation similarly along $\pmDistL$ and apply our
    IH.
    \qedhere
    \end{varitemize}
\end{proof}
\noindent
The following is a crucial intermediate step towards Theorem \ref{thm:ASTofFullCalc}, the
main result of this section.
\begin{lem}\label{lm:continuity}
  For any $\pmTermM,\pmTermN$ it holds that 
  $\Eval{\pmTermM\ \pmTermN}= \Eval{\Eval{\pmTermM}\ \Eval{\pmTermN}}$.
   In particular, if the application
   $\pmTermM\ \pmTermN$ is almost-surely terminating, so are $\pmTermM$ and  $\pmTermN$.
\end{lem}
\begin{proof}
  $(\le)\ $ There is $(\pmDistL_n,\pmDistvW_n)_{n\ge 1}$ such that
  $(\pmTermM\ \pmTermN)\rta^n \pmDistL_n \ge \pmDistvW_n$ for all $n$
  and such that~$\Eval{\pmTermM\ \pmTermN}
  = \lim_n(\pmDistvW_n)$. Applying Lemma~\ref{lmaux1:continuity} gives
  $\pmDistM_n,\:\pmDistN_n,\:\pmDistQ_n\in\Dist{\SystTot}$ and
  $\pmDistvU_n,\pmDistvV_n\:{\in}\:\Dist{\SystTot_V}$ such that
  $\pmTermM\rta^*\pmDistM_n\ge\pmDistvU_n$,
  $\pmTermN\rta^*\pmDistN_n\ge\pmDistvV_n$ and
  $(\pmDistvU_n\ \pmDistvV_n)\rta^*\pmDistQ_n\ge\pmDistvW_n$. Thus
  $\lim_n\pmDistvU_n\le \Eval{\pmTermM}$ and
  $\lim_n\pmDistvV_n\le \Eval{\pmTermN}$ with
  $\pmDistvW_n\le \Eval{\pmDistvU_n\ \pmDistvV_n}$ leading to the
  required inequality:
  $$ \Eval{M\ N} =\lim_n\pmDistvW_n \le \lim_n\Eval{\pmDistvU_n\ \pmDistvV_n} 
   \le \Eval{\lim_n \left(\pmDistvU_n\ \pmDistvV_n\right)}
   \le \Eval{\left(\lim_n\pmDistvU_n\right)\ \left(\lim_n\pmDistvV_n\right)}
   \le \Eval{\Eval{\pmTermM}\ \Eval{\pmTermN}}\ .
  $$

  $(\ge)\ $ There is
  $(\pmDistM_n,\pmDistN_n,\pmDistvU_n,\pmDistvV_n)_{n\ge 1}$ such that
  $\pmTermM\rta^n \pmDistM_n \ge \pmDistvU_n$ and
  $\pmTermN\rta^n \pmDistN_n \ge \pmDistvV_n$ for all $n$ and such that
  $\Eval{\pmTermM} = \lim_n(\pmDistvU_n)$ and $\Eval{\pmTermN}
  = \lim_n(\pmDistvV_n)$. This leads to the 
  equality~$\Eval{\Eval{\pmTermM}\ \Eval{\pmTermN}}\:{=}\:\lim_{m,n}\Eval{\pmDistvU_m\ \pmDistvV_n}$.
  Finally, by Lemma~\ref{lmaux2:continuity}, for any $m,n$, each approximant
  of~$\Eval{\pmDistvU_m\ \pmDistvV_n}$ is below
  $\Eval{\pmTermM\ \pmTermN}$, so is their sup. 
\end{proof}

%%%%%%%%%%%%%%%%%%%%%%%%%%%%%%%%%%%%
\subsection{Almost-Sure Termination}\label{sect:ast}
%%%%%%%%%%%%%%%%%%%%%%%%%%%%%%%%%%%%
We now have all the necessary ingredients to specify a quite powerful
notion of probabilistic computation. When, precisely, should such a
process be considered \emph{terminating}? Do all probabilistic
branches (see figures \ref{fig:exSystT}-\ref{fig:exSystTX}) need to be
finite?  Should we stay more liberal? The literature on the subject
is unanimously pointing to a notion called
\emph{almost-sure termination}: a probabilistic computation should be
considered terminating if the set of infinite computation branches, although
not necessarily empty, has null probability \cite{mcivermorgan,fioritihermanns2015,kaminskikatoen}. This has
the following incarnation in our setting:
\begin{defi}
A term $\pmTermM$ is said to be \emph{almost-surely terminating} (AST) iff
its probability of convergence is maximal, namely if $\Succ(\pmTermM)=1$.
\end{defi}
This section is concerned with proving that $\SystTot$ indeed guarantees
almost-sure termination. This will be done by adapting Girard-Tait's
reducibility technique to a probabilistic operational semantics.
\begin{thm}\label{thm:ASTofFullCalc}
  The full system $\SystTot$ is almost-surely terminating (AST), {\em i.e.},
  for every $\pmTermM\in \SystTot$ it holds that $\Succ(\pmTermM) = 1$.
\end{thm}
\begin{proof}
  The proof is is based on the the notion of a \emph{reducible} term,
  which is given as follows by induction on the structure of types:
  \begin{align*}
    \Red\NAT &:= \ \left\{\pmTermM\in\SystTot_C(\NAT)\mid \pmTermM \text{ is AST}\right\}\\
    \Red{\pmTypa\rta\pmTypb} &:= \ \{\pmTermM\in\in\SystTot_C(\pmTypa\rta\pmTypb)\mid \forall \pmValV\in\Red\pmTypa\cap\SystTot_V,\ \pmTermM\ \pmValV\in\Red\pmTypb\}\\
    \Red{\pmTypa\times\pmTypb} &:= \ \{\pmTermM\in\in\SystTot_C(\pmTypa\times\pmTypb)\mid (\piO\ \pmTermM)\in \Red\pmTypa,\ (\piT\ \pmTermM)\in \Red\pmTypb\}
  \end{align*}
  Then we can observe that:
  \begin{enumerate}
  \item
    {\em The reducibility candidates over $\Red\pmTypa$ are
    $\rta$-saturated}.  By an induction on $\pmTypa$ we can indeed
    show that if $\pmTermM\rta\pmDistM$ then
    $\supp\pmDistM \subseteq \Red\pmTypa$ iff
    $\pmTermM\in\Red\pmTypa$.
    \begin{varitemize}
    \item
       Trivial for $\pmTypa=\NAT$.
   \item
     If $\pmTypa={\pmTypb\rta\pmTypc}$, then for all values
     $\pmValV\in\Red\pmTypb$, $(\pmTermM\ \pmValV)\rta
     (\pmDistM\ \pmValV)$, thus by IH
     $(\pmTermM\ \pmValV)\in\Red\pmTypc$ iff
     $\supp{\pmDistM\ \pmValV}=\{\pmTermN\ \pmValV\mid \pmTermN\in\supp\pmDistM\}\subseteq \Red\pmTypc$;
     which exactly means that
     $\supp\pmDistM \subseteq \Red{\pmTypb\rta\pmTypc}$ iff
     $\pmTermM\in\Red{\pmTypb\rta\pmTypc}$.
   \item
     If $\pmTypa={\pmTypa_1\times\pmTypa_2}$: then for $i\in\{1,2\}$,
     $(\piI\ \pmTermM)\rta (\piI\ \pmDistM)$ so that by IH,
     $(\piI\ \pmTermM)\in\Red{\pmTypa_i}$ iff
     $(\piI\ \pmDistM)\subseteq\Red{\pmTypa_i}$; which exactly means
     that $\supp\pmDistM \subseteq \Red{\pmTypa_1\times\pmTypa_2}$ iff
     $\pmTermM\in\Red{\pmTypa_1\times\pmTypa_2}$.
\end{varitemize}
  \item
    {\em The reducibility candidates over $\Red\pmTypa$ are precisely the AST terms $\pmTermM$
    such that $\Eval{\pmTermM}\subseteq\supp{\Red\pmTypa}$}.
    This goes by induction on $\pmTypa$. 
    \begin{varitemize} 
    \item 
      Trivial when $\pmTypa=\NAT$. 
    \item Suppose $\pmTypa=\pmTypb\rta\pmTypc$.
      Let $\pmTermM\in \Red{\pmTypb\rta\pmTypc}$. 
      Remark that there is a value $\pmValV\in\Red\pmTypb$, thus
      $(\pmTermM\ \pmValV)\in\Red\pmTypc$ and $(\pmTermM\ \pmValV)$ is
      AST by IH; using Lemma~\ref{lm:continuity} we get $\pmTermM$ AST
      and it is easy to see that if $\pmValU\in\supp{\Eval{\pmTermM}}$
      then $\pmValU\in\supp\pmDistM$ for some $\pmTermM\rta^*\pmDistM$
      so that $\pmValU\in\Red{\pmTypb\rta\pmTypc}$ by
      saturation.
      Conversely, let $\pmTermM$ be AST with
      $\supp{\Eval{\pmTermM}}\subseteq\Red{\pmTypb\rta\pmTypc}$ and let
      $\pmValV\in\Red\pmTypb$ be a value.
      By IH, for any
      $\pmValU\in\supp{\Eval{\pmTermM}}\subseteq\Red{\pmTypb\rta\pmTypc}$
      we have $(\pmValU\ \pmValV)$ AST with an evaluation supported by
      elements of $\Red\pmTypc$; by Lemma~\ref{lm:continuity}
      $\Eval{\pmTermM\ \pmValV}=\Eval{\Eval{\pmTermM}\ \pmValV}$ meaning
      that $(\pmTermM\ \pmValV)$ is AST and has an evaluation supported
      by elements of $\Red\pmTypc$, so that we can conclude by
      IH. 
    \item Suppose $\pmTypa=\pmTypa_1\times\pmTypa_2$.
      Let $\pmTermM\in \Red{\pmTypa_1\times\pmTypa_2}$. 
      then $(\piO\ \pmTermM)\in\Red{\pmTypa_1}$ and $(\piO\ \pmTermM)$
      is AST by IH; using Lemma~\ref{lm:continuity} we get $\pmTermM$
      AST and it is easy to see that if
      $\pmValU\in\supp{\Eval{\pmTermM}}$ then $\pmValU\in\supp\pmDistM$
      for some $\pmTermM\rta^*\pmDistM$ so that
      $\pmValU\in\Red{\pmTypa_1\rta\pmTypa_2}$ by saturation.
      Conversely, let $\pmTermM$ be AST with
      $\supp{\Eval{\pmTermM}}\subseteq\Red{\pmTypa_1\rta\pmTypa_2}$ and
      let $i\in\{1,2\}$. By IH, for any
      $\pmValU\in\supp{\Eval{\pmTermM}}\subseteq\Red{\pmTypa_1\rta\pmTypa_2}$
      we have $(\piI \pmValU)$ AST with an evaluation supported by
      elements of $\Red{\pmTypa_i}$; by Lemma~\ref{lm:continuity}
      $\Eval{\piI\ \pmTermM}=\Eval{\piI\ \Eval{\pmTermM}}$ meaning that
      $(\piI\ \pmTermM)$ is AST and has an evaluation supported by
      elements of $\Red{\pmTypa_i}$, so that we can conclude by
      IH.  
    \end{varitemize} 
  \item 
    {\em Every term $\pmTermM$ such that
      $x_1:\pmTypa_1,\ldots,x_n:\pmTypa_n\vdash \pmTermM:\pmTypb$ is a
      candidate in the sense that if $\pmValV_i\in\Red{\pmTypa_i}$ for
      every $1\leq i\leq n$, then
      $\pmTermM[\pmValV_1/x_1,\ldots,\pmValV_n/x_n]\in\Red\pmTypb$}.
    This one goes by induction on the type derivation. The only difficult cases
    are the recursion, the application, the binary choice $\oplus$ and 
    the denumerable choice $\fixran$:
    \begin{itemize} 
    \item 
      For the operator $\rec$: We have to show that if
      $\pmValU\in\Red\pmTypa$ and
      $\pmValV\in\Red{\NAT\rta\pmTypa\rta\pmTypa}$ then for all
      $n\in\Nat$, $(\rec\ \langle\pmValU, \pmValV, \boldsymbol
      n\rangle)\in\Red\pmTypa$. We proceed by induction on
      $n$: 
      \begin{varitemize} 
      \item   
        If $n=0$: $\rec\ \langle\pmValU, \pmValV, \boldsymbol
        0\rangle\rta\{\pmValU\}\subseteq\Red\pmTypa$ and we conclude by
        saturation.  
      \item
        Otherwise: $\rec\ \langle\pmValU, \pmValV, \boldsymbol
        {(n+1)}\rangle\rta \pmValV\ \boldsymbol n\
        (\rec\ \langle\pmValU, \pmValV, \boldsymbol n\rangle)\in\Red\pmTypa$ since
        $(\rec\ \langle\pmValU, \pmValV, \boldsymbol n\rangle)\in\Red\pmTypa$ by IH and
        since $\boldsymbol n\in\Red\Nat$ and
        $\pmValV\in\Red{\Nat\rta\pmTypa\rta\pmTypa}$, we conclude by
        saturation.      
      \end{varitemize} 
    \item 
      For the application: we have
      to show that if $\pmTermM\in \Red {\pmTypa\rta\pmTypb}$ and
      $\pmTermN\in \Red {\pmTypa}$ then $(\pmTermM\ \pmTermN)\in \Red
      {\pmTypb}$. But since $\pmTermN\in \Red {\pmTypa}$, this means
      that it is AST and for every $\pmValV\in\supp{\Eval{\pmTermN}}$,
      $(\pmTermM\ \pmValV)\in\Red\pmTypb$. In particular, by
      Lemma~\ref{lm:continuity}, we have
      $\Eval{\pmTermM\ \pmTermN}= \Eval{\pmTermM\ \Eval{\pmTermN}}$ so
      that $(\pmTermM\ \pmTermN)$ is AST and
      $\supp{\Eval{\pmTermM\ \pmTermN}}\subseteq\bigcup_{\pmValV\in\supp{\Eval{\pmTermN}}}\supp{\Eval{\pmTermM\ \pmValV}}\subseteq \Red\pmTypb$.  
    \item For the operator $\oplus$: 
      If $\pmTermM,\pmTermN\in\Red\pmTypa$ then 
      $\supp{\left\{\begin{matrix}\pmTermM\mapsto \frac12\\[0.2em]\pmTermN\mapsto\frac12\end{matrix}\right\}}\subseteq\Red\pmTypa$,
      and, by $\rta$-saturation,  $(\pmTermM\oplus\pmTermN)\in\Red\pmTypa$.
    \item
      For the operator $\fixran$: we have to show that for any value
      $\pmValU{\in} \Red {\pmTypa\rta\pmTypa}$ and $\pmValV{\in} \Red
      {\pmTypa}$ if holds that $(\fixran\ \pmValU\ \pmValV)\in \Red
      {\pmTypa}$.\\ By an easy induction on $n$,
      $(\pmValU^n\ \pmValV)\in\Red\pmTypa$ since
      $\pmValU^0\ \pmValV=\pmValV\in\Red\pmTypb$ and
      $\pmValU^{n+1}\ \pmValV=\pmValU\
      (\pmValU^n\ \pmValV)\in\Red\pmTypb$ whenever
      $\pmValU^n\ \pmValV\in\Red\pmTypb$ and
      $\pmValU\in\Red{\pmTypb\rta\pmTypb}$.
      Moreover, by an easy
      induction on $n$ we have 
      $$\Evall{\fixran\ \pmValU\ \pmValV}=\frac
      1{2^{n+1}}\Evall{\pmValU^n\
        (\fixran\ \pmValU\ \pmValV)}+ \sum_{i\le n}\frac
      1{2^{i+1}}\Eval{\pmValU^i\ \pmValV}\ .$$ 
      \begin{varitemize} 
      \item Trivial for $n=0$. 
      \item We know that
        $\Eval{\fixran\ \pmValU\ \pmValV}=\frac 1 2 \Eval{\pmValV}+\frac 1
        2 \Eval{\pmValV (\fixran\ \pmValU\ \pmValV)}$ and we get
        $\Eval{\pmValV
          (\fixran\ \pmValU\ \pmValV)}=\Eval{\pmValV \Eval{\fixran\ \pmValU\ \pmValV}}=\frac
        1{2^{n+1}}\Evall{\pmValU^{n+1}
          (\fixran\ \pmValU\ \pmValV)}+ \sum_{i\le n}\frac
        1{2^{i+1}}\Eval{\pmValU^{i+1} \pmValV}$ by
        Lemma~\ref{lm:continuity} and~IH, which is sufficient to
        conclude.  
      \end{varitemize} 
      At the limit, we get 
      $\Evall{\fixran\ \pmValU\ \pmValV}=\sum_{i\in\Nat}\frac
      1{2^{i+1}}\Eval{\pmValU^i\ \pmValV}$. We can then conclude that
      $(\fixran\ \pmValU\ \pmValV)$ is AST (since each of the
      $(\pmValU^i\ \pmValV)\in\Red\pmTypb$ are AST and $\sum_{i}\frac
      1{2^{i+1}}=1$) and that
      $\supp{\Eval{\pmTermM\ \pmTermN}}=\bigcup_i\supp{\Eval{\pmValU^i\ \pmValV}}\subseteq\Red\pmTypa$.  
    \end{itemize}
  \end{enumerate}
  Points 2 and 3 above together leads to the thesis: if $\vdash\pmTermM:\pmTypa$,
  then Point 3 implies that $M\in\Red{\pmTypa}$ which, by Point 2, implies that $\pmTermM$ is AST.       
\end{proof}
Almost-sure termination could however be seen as too weak a property:
there is no guarantee about the average computation length, which can well
be infinite even if the probability of termination is finite. For this
reason, a stronger notion is often considered,
namely \emph{positive} almost-sure termination:
\begin{defi}\label{def:past}
  A term $\pmTermM$ is said to be \emph{positively almost-surely
terminating} (or \emph{PAST}) iff the average reduction
length~$\AvLength{\pmTermM}$ is finite.
\end{defi}
G\"odel's $\npT$, when paired with $\rand$, is combinatorially too
powerful to guarantee positive almost sure termination already. This comes
from the possibility to describe programs with exponential
reduction time such as the term $\boldsymbol{\mathtt{Expo}}$ from
Example~\ref{example:Expo}, which is computing the function $n\mapsto 2^{n+1}$ 
in time $\Theta(2^{n})$.

\begin{thm}\label{thm:cePASN}
  $\SystTot$ is not positively almost-surely terminating.
\end{thm}
\begin{proof}
  The term
  $(\boldsymbol{\mathtt{Expo}}\ \rand):\NAT$ is computing, with
  probability $\frac 1 {2^{n+1}}$ the number $2^{n+1}$ in time
  $\Theta(2^{n})$; the average reduction length is thus
  \[\AvLength{\boldsymbol{\mathtt{Expo}}\ \rand}\ =\ \sum_n \frac{\Theta(2^{n})}{2^{n+1}}\ =\ +\infty\ .
  \tag*{\qedhere}
  \]
\end{proof}
Please observe that the counterexample to positive almost-sure termination
for $\SystTot$ has been obtained by applying $\mathtt{Expo}$ to $\rand$,
and both these terms are positively almost surely terminating when
considered \emph{in isolation}. In other words, positive almost sure
termination is \emph{not} a compositional property.
%%%%%%%%%%%%%%%%%%%%%%%%%%%%%%%%%%%%%%%%%%%%%%%%%%%%%%%%%%%%%%%%%%%% 
\subsection{On Fragments of $\SystTot$: a Roadmap}\label{sect:roadmap}
%%%%%%%%%%%%%%%%%%%%%%%%%%%%%%%%%%%%%%%%%%%%%%%%%%%%%%%%%%%%%%%%%%%%
The calculus $\SystTot$ contains at least four fragments, namely
G\"odel's $\SystT$ and the three fragments $\SystPl$, $\SystRand$ and
$\SystFix$ corresponding to the three probabilistic choice operators
we consider. It is then natural to ask ourselves how these fragments
relate to each other as for their respective expressive power. At the
end of this paper, we will have a very clear picture in front of us.

The first such result is the equivalence between the two
fragments $\SystRand$ and~$\SystFix$. The embeddings are in fact quite
simple: getting $\fixran$ from $\rand$ only requires ``guessing''
the number of iterations via $\rand$ and then use $\rec$ to execute
them; capturing $\rand$ from $\fixran$ is even easier: 
it corresponds to counting the total number of iterations performed
by $\fixran$:
\begin{prop}\label{prop:fixvsrand}
  $\SystRand$ and $\SystFix$ are both equiexpressive with $\SystTot$.
\end{prop}
\begin{proof}
  The calculus $\SystRand$ embeds the full system $\SystTot$ via the
  encoding:\footnote{Notice that the dummy abstractions on $z$ and the
  $\0$ at the end ensure the correct reduction order by making
  $\lambda z.\pmTermN$ a value.}
  $$
    \pmTermM\bc_\rand\pmTermN :=\ \rec\ \langle\lambda z.\pmTermN,\lambda xyz.\pmTermM,\rand\rangle\ \0; 
    \qquad\qquad
    \fixran_\rand\ :=\ \lambda x.\rec\ \langle \pi_2 x,\lambda z.\pi_1 x,\rand\rangle.
  $$
  The fragment $\SystFix$ embeds the full system $\SystTot$ via the encoding:
  $$
    \pmTermM\bc_\fixran\pmTermN :=\ \fixran\ \langle\lambda xy. \pmTermM,\lambda y.\pmTermN\rangle\ \0;
    \qquad\qquad
    \rand_\fixran\ :=\ \fixran\ \langle\S,\0\rangle.
  $$
  In both cases, the embedding is compositional and preserves types. 
    We have to prove the correctness of the two embeddings:
  \begin{itemize}
  \item For any $\pmTermM$ and $\pmTermN$:
    $$ \Eval{\pmTermM\bc_\rand\pmTermN}\ =\  \Eval{\pmTermM\bc_\fixran\pmTermN}\ =\  \Eval{\pmTermM\bc\pmTermN}\ =\ \frac12\Eval\pmTermM +\frac12\Eval\pmTermN\ . $$
    Indeed, we only have to perform a few reductions:
    \begin{align*}
      \Eval{\pmTermM\bc_\rand\pmTermN} &= \sum_{n\ge 0} \frac 1 {2^{n+1}} \Eval{\rec\ \langle\lambda z.\pmTermN,\lambda xyz.\pmTermM,\boldsymbol n\rangle\ \0} \\
      &=  \frac 1 2 \Eval{\rec\ \langle\lambda z.\pmTermN,\lambda xyz.\pmTermM,\0\rangle\ \0} + \sum_{n\ge 0} \frac 1 {2^{n+2}} \Eval{\rec\ \langle\lambda z.\pmTermN,\lambda xyz.\pmTermM,\S\boldsymbol n\rangle\ \0}\\
      &=  \frac 1 2 \Eval{(\lambda z.\pmTermN)\ \0} + \sum_{n\ge 0} \frac 1 {2^{n+2}} \Eval{(\lambda xyz.\pmTermM)\ \boldsymbol n\ (\rec\ \langle\lambda z.\pmTermN,\lambda xyz.\pmTermM,\boldsymbol n\rangle)\ \0}\\
      &=  \frac 1 2 \Eval{\pmTermN} + \sum_{n\ge 0} \frac 1 {2^{n+2}} \Eval{\pmTermM}\\
      &=  \frac 1 2 \Eval{\pmTermN} + \frac 1 2 \Eval{\pmTermM}\\
      \Eval{\pmTermM\bc_\fixran\pmTermN} &= \frac12 \Eval{(\lambda y.\pmTermN)\ \0} + \frac12\Eval{(\lambda xy. \pmTermM)\ \rand_\fixran\ \0} \\
      &= \frac12 \Eval{(\lambda y.\pmTermN)\ \0} + \frac12\Eval{(\lambda xy. \pmTermM)\ \rand_\fixran\ \0} \\
      &=  \frac 1 2 \Eval{\pmTermN} + \frac 1 2 \Eval{\pmTermM}
    \end{align*}
  \item For any $\pmValU$ and $\pmValV$:
    $$ \Eval{\fixran_\rand \langle \pmValU,\pmValV\rangle}\ =\ \Eval{\fixran \langle \pmValU,\pmValV\rangle} $$
    Indeed, both of them are the unique fixpoint of the following contractive 
    function:
    $$ f(\pmDistX)\ :=\ \frac 12 \Eval\pmValU + \frac12\Eval{\pmValV\ \pmDistX}\ .$$
    That $\Eval{\fixran \langle \pmValU,\pmValV\rangle}=f(\Eval{\fixran \langle \pmValU,\pmValV\rangle})$ is
    immediate after a reduction, as for the other, we have:
    \begin{align*}
      \Eval{\fixran_\rand \langle \pmValU,\pmValV\rangle} 
      &= \sum_{n\ge 0}\frac 1{2^{n+1}}\Eval{\rec\ \langle \pmValU,\lambda z.\pmValV,\boldsymbol n\rangle} \\
      &= \frac12\Eval{U} + \frac12\sum_{n\ge 0}\frac 1{2^{n+1}}\Eval{\pmValU\ (\rec\ \langle \pmValU,\lambda z.\pmValV,\boldsymbol n\rangle)} \\
      &= \frac12\Eval{U} + \frac12\sum_{n\ge 0}\frac 1{2^{n+1}}\Eval{\pmValU\ \Eval{\rec\ \langle \pmValU,\lambda z.\pmValV,\boldsymbol n\rangle}} 
        &\text{by Lemma~\ref{lm:continuity}}\\
      &= \frac12\Eval{U} + \frac12\Eval{\pmValU\ \left(\sum_{n\ge 0}\frac 1{2^{n+1}}\Eval{\rec\ \langle \pmValU,\lambda z.\pmValV,\boldsymbol n\rangle}\right)} \\
      &= \frac12\Eval{U} + \frac12\Eval{\pmValU\ \Eval{\fixran_\rand \langle \pmValU,\pmValV\rangle}}
    \end{align*}
  \item Finally:
    $$ \Eval{\rand_\fixran}\ =\ \Eval{\rand}\ =\ \left\{\boldsymbol n \mapsto \frac 1 {2^{n+1}} \middle| n\ge 0\right\}\ .$$
    That $\Eval{\rand} = \left\{\boldsymbol n \mapsto \frac 1 {2^{n+1}} \middle| n\ge 0\right\}$ is just one step of reduction, while 
    $ \Eval{\rand_\fixran}= \left\{\boldsymbol n \mapsto \frac 1 {2^{n+1}} \middle| n\ge 0\right\}$ was shown in Example~\ref{ex:fixran2}.
    \qedhere
  \end{itemize}
\end{proof}
\noindent
Notice how simulating $\fixran$ by $\rand$ requires the presence of
recursion, while the converse is not true. The implications of this
fact are intriguing, but lie outside the scope of this work.
In the following, we will no longer consider $\SystFix$ nor $\SystTot$
but only $\SystRand$, keeping in mind that all these are equiexpressive due
to Proposition \ref{prop:fixvsrand}. The rest of this paper, thus, will be concerned
with understanding the relative expressive power of the three fragments
$\SystT$, $\SystPl$, and $\SystRand$. Can any of the (obvious) strict \emph{syntactical}
inclusions between them be turned into a strict \emph{semantic} inclusion? Are the
three systems equiexpressive?

In order to compare probabilistic and deterministic calculi,
several options are available.  The most common one is to consider
notions of observations over the probabilistic outputs; this will be
pursued in Section~\ref{sec:subrec}, where we 
look at whether Monte Carlo or Las Vegas algorithms
on $\SystPl$ or $\SystRand$
result in deterministically $\SystT$-definable functions or not.
Notice that neither Monte Carlo nor Las Vegas algorithms are \emph{natively}
definable inside $\SystTot$. Indeed, those algorithms are based on
restrictions on the resulting distribution, which cannot
be described in the calculus. For example, a Las Vegas algorithm
is captured by, say, a term $\pmTermM:\NAT\rta\NAT$ such that 
$\Eval{\pmTermM\ \boldsymbol n}(0) \le\frac 1 2$ for any $n$.
In the next two sections, instead, we look at how $\SystTot$ can
be seen as a way to compute functions returning \emph{distributions} over the
base type $\Nat$ rather than \emph{elements} of if.

%Since those algorithms are not representable, we have to perform an external
%study that can be quite heavy. Hopefully, we where able to define an internal
%property, namely the (parameterized) \emph{functional representability},
%that is sufficient to collapse the results of both algorithms into the
%deterministic system $\SystT$.

We say that the distribution $\pmDistM\in\Dist{\Nat}$ is 
\emph{finitely represented} by\footnote{Here we denote $\Bin$ for binomial numbers 
$\frac m {2^n}$ (where $m,n\in\Nat$) and $\BIN$ for their representation in system $\SystT$
encoded by pairs $\langle m,n\rangle$ of natural numbers.} $f:\Nat\rightarrow\Bin$,
if there exists a natural number $q$ such
that for every $k\ge q$ it holds that~$f(k)=0$ and
$\pmDistM=\left\{\boldsymbol k\mapsto f(k)\right\}$.
Moreover, the definition can be extended to families of distributions
$(\pmDistM_n)_n$ by requiring the
existence of functions $f:\Nat\times\Nat\rightarrow\Bin$ and $q:\Nat\rta\Nat$ 
such that for every $k\ge q(n)$ it holds that $f(n,k)=0$ and for every $n$
it holds that
$ \pmDistM_n =\ \left\{\boldsymbol k\mapsto f(n,k)\right\}$.
In this case, we say that the representation is \emph{parametric}.
We will see in Section~\ref{sec:fragmentBinChoice} that the distributions computed 
by $\SystPl$ are exactly the ones (parametrically) finitely representable by $\SystT$
terms. Concretely, this means that for any $\pmTermM\in \SystPl(\NAT)$ or any
$\pmTermM\in \SystPl(\NAT\rta\NAT)$, the distributions $\Eval\pmTermM$ and
$(\Eval{\pmTermN \boldsymbol n})_n$ are (parametrically) finitely representable.

In $\SystRand$, however, distributions are more complex, having infinite support
and giving rise to non-rational probabilities.
That is why only a characterization in terms of approximations is possible.
More specifically, a distribution $\pmDistM\in\Dist{\Nat}$ is said to be
\emph{approximated} by two functions
$f:\Nat\times\Nat\rta\Bin$ and $g:\Nat\rta\Nat$ iff for every~$n\in\Nat$
and for every $k\ge g(n)$ it holds that $f(n,k)=0$ and 
$$
\sum_{k\in\Nat}\ \Bigl|\ \pmDistM(\boldsymbol k)\ -\ f(n,k)\ \Bigr|\ \le\ \frac{1}{n}.
$$
In other words, the distribution $\pmDistM$ can be approximated
arbitrarily well, and uniformly, by finitely representable
ones. Similarly, we can define a parametric version of this
definition at first order.
In Section~\ref{sec:fragmentRand} , we show that distributions
generated by $\SystRand$ terms are indeed uniform limits over those of
$\SystPl$; our result on $\SystPl$ thus induces their (parametric)
approximation in~$\SystT$.

\section{Binary Probabilistic Choice}
    \label{sec:fragmentBinChoice}
    This section is concerned with two theoretical results on the
expressive power of $\SystPl$. Taken together, they tell us that
this fragment is not far from $\npT$.
%%%%%%%%%%%%%%%%%%%%%%%%%%%%%%%%%%%%%%%%%%%%%
\subsection{Positive Almost-Sure Termination}
%%%%%%%%%%%%%%%%%%%%%%%%%%%%%%%%%%%%%%%%%%%%%
As we already observed, the average number of steps to normal form can be infinite for terms of $\SystTot$.
We will prove that, on the contrary, $\SystPl$ is \emph{positive}
almost-surely terminating. This will be done by adapting (and strengthening!)
the reducibility-based result from Section \ref{sect:ast}.
To this end, we will first give a formalization of the notion of 
execution tree discussed in Section~\ref{sec:2ndSec} in the form of
a multistep reduction procedure. Then, we will formally show that this 
tree is finite. We will see later that the multistep reduction is nothing more
than $\rta^*$ for $\SystPl$, but that this is
not the case in richer fragments of $\SystTot$.

\begin{defi}\label{def:BSS}
  The multistep reduction relation $\Rta$ is defined by induction in
  Figure~\ref{fig:BSTrand}. Due to the (potentially) countably many 
  preconditions of the rule $(R\dash{\in})$, the derivation tree of a multistep
  reduction $\Rta$ can be infinitely wide and even of unbounded height, but
  each path have to be finite.
\end{defi}
\begin{figure*}
  \fbox{
    \begin{minipage}{.97\textwidth}
      \vspace*{1em}
      \begin{small}
        \begin{center}
          \AxiomC{$\phantom {\pmDistM}$}
          \RightLabel{{\scriptsize $(R\dash 0)$}}
          \UnaryInfC{$\!\pmTermM\Rta \Bigl\{\pmTermM\Bigr\}\!$}
          \DisplayProof \hspace{2.5em}
          %% 
          % \AxiomC{$\pmTermM \rta \pmDistM$}
          % \RightLabel{{\scriptsize $(R\dash r)$}}
          % \UnaryInfC{$\pmTermM\Rta \pmDistM\vphantom{\Bigr\}}$}
          % \DisplayProof\hskip 20pt
          % 
          \AxiomC{$\!\pmTermM\rta\pmDistM$}
          \AxiomC{$\!\!\!\!\!\!\pmDistM\Rta\pmDistN\!$}
          \RightLabel{{\scriptsize $(R\dash \tran)$}}
          \BinaryInfC{$\pmTermM\Rta\pmDistN\vphantom{\Bigr\}}$}
          \DisplayProof \hspace{2.5em}
          \AxiomC{$\!\forall M\in |\pmDistM|,\ \pmTermM\Rta\pmDistN_{\pmTermM}\!$}
          \RightLabel{{\scriptsize $(R\dash{\in})\!\!$}}
          \UnaryInfC{$\pmDistM \Rta\sumd{\pmTermM}{\pmDistM}{\pmDistN_\pmTermM}$}
          \DisplayProof
        \end{center}
      \end{small}
    \end{minipage}}
      \caption{Multistep Reduction.}\label{fig:BSTrand}
\end{figure*}

The infiniteness of the width and the fact that the height is unbounded is 
an essential tool for analising $\SystRand$. In fact, most
theorems in this section will be given for both $\SystPl$ and $\SystRand$.
But, for now, we will focus on $\SystPl$ and finite derivations, while
$\SystRand$ and transfinite derivations will be analyzed in details in 
Section~\ref{sec:multStepRand}.

\begin{exa}\label{ex:doubleflip}
  Consider the already considered example term
  $\rec \langle \0,\lambda xy.y \oplus (\S y),\boldsymbol 2\rangle$;
  the execution tree is the following, where $U_n$ stands
  for $\rec \langle \0,\lambda xy.y \oplus (\S y),\boldsymbol n\rangle$:
  \begin{center}
    \begin{tikzpicture}[scale=0.95]
      \node (t0) at (0,2) {$U_2$};
      \node (t1) at (0.8,2) {$\bullet$};
      \node (t2) at (1.5,2) {$\bullet$};
      \node (t3) at (3,2) {$U_1 \oplus \S U_1$};
      \node (l0) at (5.2,3) {$U_1 $};
      \node (r0) at (5.2,1) {$\S U_1 $};
      \node (l1) at (6,3) {$\bullet $};
      \node (l2) at (6.7,3) {$\bullet $};
      \node (l3) at (8.5,3) {$U_0\oplus \S U_0$};
      \node (ll0) at (11,3.5) {$U_0 $};
      \node (ll1) at (12.5,3.5) {$\0 $};
s      \node (lr0) at (11,2.5) {$\S U_0 $};
      \node (lr1) at (12.5,2.5) {$\boldsymbol 1 $};
      \node (r1) at (6,1) {$\bullet $};
      \node (r2) at (6.7,1) {$\bullet $};
      \node (r3) at (8.5,1) {$\S(U_0\oplus \S U_0)$};
      \node (rl0) at (11,1.5) {$\S U_0 $};
      \node (rl1) at (12.5,1.5) {$\boldsymbol 1 $};
      \node (rr0) at (11,0.5) {$\S\S U_0 $};
      \node (rr1) at (12.5,0.5) {$\boldsymbol 2 $};
      \draw[->] (t0.east) -- node [right] {} ($(t1.west)+(0.15,0)$);
      \draw[->] ($(t1.east)+(-0.1,0)$) -- node [right] {} ($(t2.west)+(0.15,0)$);
      \draw[->] ($(t2.east)+(-0.1,0)$) -- node [right] {} (t3.west);
      \draw[->] (l0.east) -- node [right] {} ($(l1.west)+(0.15,0)$);
      \draw[->] ($(l1.east)+(-0.1,0)$) -- node [right] {} ($(l2.west)+(0.15,0)$);
      \draw[->] ($(l2.east)+(-0.1,0)$) -- node [right] {} (l3.west);
      \draw[->] (r0.east) -- node [right] {} ($(r1.west)+(0.15,0)$);
      \draw[->] ($(r1.east)+(-0.1,0)$) -- node [right] {} ($(r2.west)+(0.15,0)$);
      \draw[->] ($(r2.east)+(-0.1,0)$) -- node [right] {} (r3.west);
      \draw[->] (ll0.east) -- node [right] {} (ll1.west);
      \draw[->] (lr0.east) -- node [right] {} (lr1.west);
      \draw[->] (rl0.east) -- node [right] {} (rl1.west);
      \draw[->] (rr0.east) -- node [right] {} (rr1.west);
      \draw[->] ($(t3.east)+(0,0.2)$) -- node [right] {{\scriptsize $\frac 1 2$}} (l0.west);
      \draw[->] ($(t3.east)+(0,-0.2)$) -- node [right] {{\scriptsize $\frac 1 2$}} (r0.west);
      \draw[->] ($(l3.east)+(0,0.1)$) -- node [right] {{\scriptsize $\frac 1 2$}} (ll0.west);
      \draw[->] ($(l3.east)+(0,-0.1)$) -- node [right] {{\scriptsize $\frac 1 2$}} (lr0.west);
      \draw[->] ($(r3.east)+(0,0.1)$) -- node [right] {{\scriptsize $\frac 1 2$}} (rl0.west);
      \draw[->] ($(r3.east)+(0,-0.1)$) -- node [right] {{\scriptsize $\frac 1 2$}} (rr0.west);
    \end{tikzpicture}\vspace{-1em}
  \end{center}
  This tree is subsumed by the (more complex) derivation of $U_2 \Rta\ \frac 
  14\{\0\}+\frac12\{\boldsymbol 1\}+\frac14\{\boldsymbol 2\}$, with every arrow 
  of the exectution tree replaced by a $(R\dash \tran)$ rule followed by a 
  $(R\dash{\in})$ rule:
  \begin{center}
    \scriptsize
      \begin{minipage}[c]{\linewidth}
      	\centering
        %\AxiomC{\hspace{-4em}}
        \AxiomC{\hspace{-4em}}
		\AxiomC{\hspace{5em}}
        \AxiomC{\hspace{1em}}
        \AxiomC{\hspace{-10em}}
        \AxiomC{\hspace{-4em}}	
        \AxiomC{\hspace{3em}}
        \AxiomC{\hspace{0.5em}}
        \AxiomC{\hspace{-0.5em}}	
        %\AxiomC{\hspace{-12.5em}}

        %\AxiomC{\hspace{-4em}}
        %\AxiomC{\hspace{3.5em}}
        %\AxiomC{\hspace{2em}}
        %\AxiomC{\hspace{0.5em}}

        \AxiomC{\hspace{0em}}
        \AxiomC{$\phantom {\pmDistM}$}
        \RightLabel{{\scriptsize \rlap{$(R\dash 0)$}}}
        \UnaryInfC{$\boldsymbol 1\Rta\{\boldsymbol 1\}$}
        \RightLabel{{\scriptsize $(R\dash{\in})\!\!$}}
        \UnaryInfC{\llap{$ U_0\rta\: $}$\{\boldsymbol 1\}\Rta\{\boldsymbol 1\}$}
        \RightLabel{{\scriptsize \rlap{$(R\dash\tran)$}}}
        \BinaryInfC{$ U_0\Rta \{\boldsymbol 1\}$}
        \AxiomC{\hspace{2em}}
        \AxiomC{\hspace{0em}}
        \AxiomC{$\phantom {\pmDistM}$}
        \RightLabel{{\scriptsize \rlap{$(R\dash 0)$}}}
        \UnaryInfC{$\boldsymbol 2\Rta\{\boldsymbol 2\}$}
        \RightLabel{{\scriptsize \rlap{$(R\dash{\in})$}}}
        \UnaryInfC{\llap{$\S U_0\rta\: $}$\{\boldsymbol 2\}\Rta\{\boldsymbol 
        2\}$}
        \RightLabel{{\scriptsize \rlap{$(R\dash \tran)$}}}
        \BinaryInfC{$\S U_0\Rta \{\boldsymbol 2\}$}
        \RightLabel{{\scriptsize \rlap{$(R\dash{\in})$}}}
        \TrinaryInfC{\llap{$U_0 \oplus (\S U_0)\rta\:$}$\frac12\{ 
        U_0\}+\frac12\{\S U_0\}\Rta\frac12\{\boldsymbol 
        1\}+\frac12\{\boldsymbol 2\}$}
        \RightLabel{{\scriptsize \rlap{$(R\dash \tran)$}}}
        \BinaryInfC{$U_0 \oplus (\S U_0)\Rta\frac12\{\boldsymbol 
        1\}+\frac12\{\boldsymbol 2\}$}
        \RightLabel{{\scriptsize \rlap{$(R\dash{\in})$}}}
        \UnaryInfC{\llap{$ (\lambda y.y \oplus (\S y)) U_0\rta\:$}$\{U_0 \oplus 
        (\S U_0) \}\Rta\frac12\{\boldsymbol 1\}+\frac12\{\boldsymbol 2\}$}
        \RightLabel{{\scriptsize \rlap{$(R\dash \tran)$}}}
        \BinaryInfC{$ (\lambda y.y \oplus (\S y)) U_0\Rta\frac12\{\boldsymbol 
        1\}+\frac12\{\boldsymbol 2\}$}
        \UnaryInfC{\llap{$ (\lambda xy.y \oplus (\S y)\boldsymbol 1 
        U_0)\rta\:$}$\{ (\lambda y.y \oplus (\S y)) 
        U_0\}\Rta\frac12\{\boldsymbol 1\}+\frac12\{\boldsymbol 2\}$}
        \BinaryInfC{$ (\lambda xy.y \oplus (\S y))\boldsymbol 1 
        U_0\Rta\frac12\{\boldsymbol 1\}+\frac12\{\boldsymbol 2\}$}
        \RightLabel{{\scriptsize \rlap{$(R\dash{\in})$}}}
        \UnaryInfC{\llap{$ U_1\rta\:$}$\{ (\lambda xy.y \oplus (\S 
        y))\boldsymbol 1 U_0\}\Rta\frac12\{\boldsymbol 1\}+\frac12\{\boldsymbol 
        2\}$}
        \RightLabel{{\scriptsize \rlap{$(R\dash \tran)$}}}
        \BinaryInfC{$U_1\Rta\frac 12\{\0\}+\frac12\{\boldsymbol 
        1\}$}

        \AxiomC{$\xi$}
        
        \RightLabel{{\scriptsize \rlap{$(R\dash{\in})$}}}
        \BinaryInfC{\llap{$U_1 \oplus (\S 
        U_1)\rta\:$}$\frac12\{U_1\}+\frac12\{\S U_1\}\Rta\frac 
        14\{\0\}+\frac12\{\boldsymbol 1\}+\frac14\{\boldsymbol 
        2\}\hspace{-1em}$}
        \RightLabel{{\scriptsize \rlap{$(R\dash \tran)$}}}
        \BinaryInfC{$U_1 \oplus (\S U_1)\Rta\frac 14\{\0\}+\frac12\{\boldsymbol 
        1\}+\frac14\{\boldsymbol 2\}\hspace{-1em}$}
        \RightLabel{{\scriptsize \rlap{$(R\dash{\in})$}}}
        \UnaryInfC{\llap{$(\lambda y.y \oplus (\S y)) U_1\rta\:$}$\{U_1 \oplus (\S U_1)\}\Rta\frac 14\{\0\}+\frac12\{\boldsymbol 1\}+\frac14\{\boldsymbol 2\}\hspace{-1em}$}
        \RightLabel{{\scriptsize \rlap{$(R\dash \tran)$}}}
        \BinaryInfC{$(\lambda y.y \oplus (\S y)) U_1\Rta\frac 
        14\{\0\}+\frac12\{\boldsymbol 1\}+\frac14\{\boldsymbol 
        2\}\hspace{-1em}$}
        \RightLabel{{\scriptsize \rlap{$(R\dash{\in})$}}}
        \UnaryInfC{\llap{$(\lambda xy.y \oplus (\S y))\boldsymbol 1 U_1\rta\:$}$\{(\lambda y.y \oplus (\S y)) U_1\}\Rta\frac 14\{\0\}+\frac12\{\boldsymbol 1\}+\frac14\{\boldsymbol 2\}\hspace{-1em}$}
        \RightLabel{{\scriptsize \rlap{$(R\dash \tran)$}}}
        \BinaryInfC{$(\lambda xy.y \oplus (\S y))\boldsymbol 1 U_1\Rta\frac 
        14\{\0\}+\frac12\{\boldsymbol 1\}+\frac14\{\boldsymbol 
        2\}\hspace{-1em}$}
        \RightLabel{{\scriptsize \rlap{$(R\dash{\in})$}}}
        \UnaryInfC{\llap{$U_2\rta\:$}$\{(\lambda xy.y \oplus (\S y))\boldsymbol 1 U_1\}\Rta\frac 14\{\0\}+\frac12\{\boldsymbol 1\}+\frac14\{\boldsymbol 2\}\hspace{-1em}$}
        % \LeftLabel{\hspace{-10em}}
        \RightLabel{{\scriptsize \rlap{$(R\dash \tran)$}}}
        \BinaryInfC{$U_2 \Rta\ \frac 14\{\0\}+\frac12\{\boldsymbol 
        1\}+\frac14\{\boldsymbol 2\}\hspace{-1em}$}
        \DisplayProof
      \end{minipage}
    %\end{small}
  \end{center}
  where $\xi$ is the following derivation:
  \begin{center}
  	\scriptsize
  	\begin{minipage}[c]{\linewidth}
  		\centering
  		%\AxiomC{\hspace{-7.5em}}
  		
  		\AxiomC{\hspace{-4.5em}}
  		\AxiomC{\hspace{5em}}
  		\AxiomC{\hspace{2em}}
  		\AxiomC{\hspace{2em}}
  		
  		\AxiomC{\hspace{-0.5em}}
  		\AxiomC{$\phantom {\pmDistM}$}
  		\RightLabel{{\scriptsize \rlap{$(R\dash 0)$}}}
  		\UnaryInfC{$\boldsymbol 1\Rta\{\boldsymbol 1\}\hspace{-1em}$}
  		\RightLabel{{\scriptsize $(R\dash{\in})\!\!$}}
  		\UnaryInfC{\llap{$\S U_0\rta\: $}$\{\boldsymbol 1\}\Rta\{\boldsymbol 
  			1\}\hspace{-1em}$}
  		\RightLabel{{\scriptsize \rlap{$(R\dash \tran)$}}}
  		\BinaryInfC{$\S U_0\Rta \{\boldsymbol 1\}\hspace{-1em}$}
  		\AxiomC{\hspace{0.5em}}
  		\AxiomC{$\phantom {\pmDistM}$}
  		\RightLabel{{\scriptsize \rlap{$(R\dash 0)$}}}
  		\UnaryInfC{$\boldsymbol 2\Rta\{\boldsymbol 2\}\hspace{-1em}$}
  		\RightLabel{{\scriptsize \rlap{$(R\dash{\in})$}}}
  		\UnaryInfC{\llap{$\S\S U_0\rta\: $}$\{\boldsymbol 2\}\Rta\{\boldsymbol 
  			2\}\hspace{-1em}$}
  		\RightLabel{{\scriptsize \rlap{$(R\dash \tran)$}}}
  		\BinaryInfC{$\S\S U_0\Rta \{\boldsymbol 2\}\hspace{-1em}$}
  		\RightLabel{{\scriptsize \rlap{$(R\dash{\in})$}}}
  		\BinaryInfC{\llap{$\S (U_0 \oplus (\S U_0))\rta\:$}$\frac12\{\S 
  			U_0\}+\frac12\{\S\S U_0\}\Rta\frac12\{\boldsymbol 
  			1\}+\frac12\{\boldsymbol 2\}\hspace{-1em}$}
  		\RightLabel{{\scriptsize \rlap{$(R\dash \tran)$}}}
  		\BinaryInfC{$\S (U_0 \oplus (\S U_0))\Rta\frac12\{\boldsymbol 
  			1\}+\frac12\{\boldsymbol 2\}$}
  		\RightLabel{{\scriptsize \rlap{$(R\dash{\in})$}}}
  		\UnaryInfC{\llap{$\S ((\lambda y.y \oplus (\S y)) U_0)\rta\:$}$\{\S 
  			(U_0 \oplus (\S U_0))\}\Rta\frac12\{\boldsymbol 
  			1\}+\frac12\{\boldsymbol 2\}\hspace{-1em}$}
  		\RightLabel{{\scriptsize \rlap{$(R\dash \tran)$}}}
  		\BinaryInfC{$\S ((\lambda y.y \oplus (\S y)) 
  			U_0)\Rta\frac12\{\boldsymbol 1\}+\frac12\{\boldsymbol 
  			2\}\hspace{-1em}$}
  		\RightLabel{{\scriptsize \rlap{$(R\dash{\in})$}}}
  		\UnaryInfC{\llap{$\S ((\lambda xy.y \oplus (\S y))\boldsymbol 1 
  				U_0)\rta\:$}$\{\S ((\lambda y.y \oplus (\S y)) 
  			U_0)\}\Rta\frac12\{\boldsymbol 1\}+\frac12\{\boldsymbol 
  			2\}\hspace{-1em}$}
  		\RightLabel{{\scriptsize \rlap{$(R\dash \tran)$}}}
  		\BinaryInfC{$\S ((\lambda xy.y \oplus (\S y))\boldsymbol 1 
  			U_0)\Rta\frac12\{\boldsymbol 1\}+\frac12\{\boldsymbol 
  			2\}\hspace{-1em}$}
  		\RightLabel{{\scriptsize \rlap{$(R\dash{\in})$}}}
  		\UnaryInfC{\llap{$\S U_1\rta\:$}$\{\S ((\lambda xy.y \oplus (\S 
  			y))\boldsymbol 1 U_0)\}\Rta\frac12\{\boldsymbol 
  			1\}+\frac12\{\boldsymbol 2\}\hspace{-1em}$}
  		\RightLabel{{\scriptsize \rlap{$(R\dash \tran)$}}}
  		\BinaryInfC{$\S U_1\Rta\frac12\{\boldsymbol 1\}+\frac12\{\boldsymbol 
  			2\}\hspace{-1em}$}
  		\DisplayProof
  	\end{minipage}
  \end{center}
\noindent Notice that the derivation is correct and finite because the 
execution tree is finite.  
\end{exa}

\begin{lem}
  The multistep semantics $\Rta$ is confluent.
\end{lem}
\begin{proof}
  By an easy induction, we show that if 
  $\pmDistN\Leftarrow\pmTermM\Rta\pmDistL$ 
  (resp. $\pmDistN\Leftarrow\pmDistM\Rta\pmDistL$), then there is $\pmDistP$
  such that $\pmDistN\Rta\pmDistP\Leftarrow\pmDistL$. Now:
  \begin{varitemize}
  \item If both reductions are using the same rule ( either $(R\dash 0)$, 
  $(R\dash \tran)$, or
    $(R\dash{\in})$ ), then it is an immediate use of the induction hypothesis on the premises
    as those rules are determinist.
  \item If one of them use the rule $(R\dash 0)$, then it is trivial.
  \item No other case is possible as $(R\dash \tran)$ and $(R\dash{\in})$ 
  cannot apply together
    (one require a term as source and the other a distribution).
    \qedhere
  \end{varitemize}
\end{proof}

\begin{lem}\label{lmaux1:BSSred}
  If $(\pmTermM\ \pmValV)\Rta \pmDistvU\in\Dist{\SystTot_V}$ then there is $\pmDistvW\in\Dist{\SystTot_V}$ such that $\pmTermM\Rta\pmDistvW$.
\end{lem}
  \begin{proof}
    By induction on $\Rta$ we can show that if $(\pmTermM\ \pmValV)\Rta \pmDistN$ then $\pmDistN=\pmDistL+(\pmDistM\ \pmValV)$ with $\pmTermM\Rta \pmDistvW+\pmDistM$ and $(\pmDistvW\ \pmValV)\Rta \pmDistL$ (and similarly if $\pmTermM$ is a distribution):
    \begin{varitemize}
    \item The $(R\dash 0)$ and $(R\dash\in)$ are trivial,
    \item If $(\pmTermM\ \pmValV)\rta \pmDistQ\Rta\pmDistN$ then there are two 
    cases:
      \begin{varitemize}
      \item Either $\pmTermM\rta\pmDistP$ and $\pmDistQ=(\pmDistP\ 
      \pmValV)\Rta\pmDistN$ and we can conclude by induction hypothesis;
      \item Or $\pmTermM=\pmValW$ is a value and $\pmTermM\Rta\{\pmValW\}$ with 
      $(\{\pmValW\}\ \pmValV)\rta \pmDistQ\Rta\pmDistN=\pmDistL$.
      \end{varitemize}
    \end{varitemize}
    Notice that if $\pmDistN$ is a value distribution then $\pmDistM$ has to be one.
  \end{proof}

The following lemma is an alternative version of
Lemma~\ref{lm:continuity}, where the evaluation is obtained after a
finite number of steps (in both hypothesis and conclusions).

\begin{lem}\label{lmaux2:BSSred}
  If $\pmTermN\Rta \pmDistvV\in\Dist{\SystTot_V}$ and
  $\pmTermM\ \pmDistvV\Rta\pmDistvU\in\Dist{\SystTot_V}$ then
  $\pmTermM\ \pmTermN\Rta\pmDistvU$.
\end{lem}
  \begin{proof}
    By induction on the derivation of $\pmTermN\Rta\pmValV$
    (generalizing the property for any distribution $\pmDistN$ in
    place of $\pmTermN$):
  \begin{varitemize}
    \item
     If $\pmValV=\{\pmTermN\}$ this is trivial.
    \item
     If $\pmTermM\rta\pmDistN\Rta\pmDistvV$ then by IH,
     $\pmTermM\ \pmDistN\Rta \pmDistvU$ and thus
     $\pmTermM\ \pmTermN\rta \pmTermM\ \pmDistN \Rta\pmDistvU$ so that
     we can conclude by rule $(R\dash +)$.
   \item   
    If for all $\pmTermN\in \supp\pmDistN$,
    $\pmTermN\Rta\pmDistvV_\pmTermN$ with
    $\pmDistvV=\sumd{\pmTermN}{\pmDistN}{\pmDistvV_\pmTermN}$ then by
    applying the IH on each $\pmTermN\in \supp\pmDistN$, we get that
    $\pmTermM\ \pmTermN\Rta\pmDistvU_\pmTermN$ for some
    $\pmDistvU_\pmTermN$ and we conclude by rule $(R\dash\in)$ using
    $\pmDistvU = \sumd{\pmTermN}{\pmDistN}{\pmDistvU_\pmTermN}$.
  \end{varitemize}
\end{proof}

\begin{thm} \label{thm:BSSred} 
  For any term $\pmTermM\in \SystPlR$, there exists a value distribution $\Evala{\pmTermM}\in\Dist{\SystPlR_V}$ 
  such that $\pmTermM\Rta\Evala{\pmTermM}$. We call it the accessible evaluation.
\end{thm}
\begin{proof}
  When it exists, $\Evala\pmTermM$ is unique due to confluence. Thus
  we only have to prove its existence. The proof goes by
  reducibility over reducibility sets defined as follows:
  \begin{align*}
    \Red\NAT &:=\ \{\pmTermM\in\SystPl(\NAT) \mid \exists \Evala{\pmTermM}\in\Dist{\SystPl_V},\ \pmTermM \Rta \Evala{\pmTermM}\}\\
    \Red{\pmTypa\rta\pmTypb} &:= \ \{\pmTermM\in\SystPl(\pmTypa\rta\pmTypb)\mid \forall \pmValV\in\Red\pmTypa\cap\SystPl_V,\ (\pmTermM\ \pmValV)\in\Red\pmTypb\}\\
    \Red{\pmTypa\times\pmTypb} &:= \ \{\pmTermM\in\SystPl(\pmTypa\times\pmTypb)\mid (\piO\ \pmTermM)\in \Red\pmTypa,\ (\piT\ \pmTermM)\in \Red\pmTypb\}
  \end{align*}
  \begin{enumerate}
    \item {\em The reducibility candidates over $\Red\pmTypa$ are $\rta$-saturated}.
      By induction on $\pmTypa$ we can show that if $\pmTermM\rta\pmDistM$ then 
      $\supp\pmDistM \subseteq \Red\pmTypa$ iff $\pmTermM\in\Red\pmTypa$.
      \begin{varitemize}
      \item
       If $\pmTypa=\NAT$: then whenever
       $\pmTermN\Rta \Evala{\pmTermN}$ for all
       $\pmTermN\in\supp\pmDistM$ we get
       $\pmDistM\Rta \sumd{\pmTermN}{\pmDistM}{\Evala{\pmTermN}}=\Evala\pmDistM$
       by $(R\dash\in)$ and thus
       $\pmTermM\Rta\Evala\pmDistM=\Evala\pmTermM$ by $(R\dash
       trans)$. Conversely, whenever $\pmTermM\Rta\Evala\pmTermM$,
       this reduction cannot come from rule $(R\dash 0)$ since
       $\Evala\pmTermM$ is a value distribution and $\pmTermM$ is
       reducible, thus it comes from rule $(R\dash +)$ and
       $\pmDistM\Rta \Evala\pmDistM$ which itself necessarily comes
       from an application of $(R\dash\in)$ so that
       $\pmTermN\Rta\Evala{\pmTermN}$ for any
       $\pmTermN\in\supp\pmDistM$.
    \item
      If $\pmTypa={\pmTypb\rta\pmTypc}$: then for all values
      $\pmValV\in\Red\pmTypb$, $(\pmTermM\ \pmValV)\rta
      (\pmDistM\ \pmValV)$, thus by IH
      $(\pmTermM\ \pmValV)\in\Red\pmTypc$ iff
      $\supp{\pmDistM\ \pmValV}=\{\pmTermN\ \pmValV\mid 
      \pmTermN\in\supp\pmDistM\}\subseteq \Red\pmTypc$;
      which exactly means that
      $\supp\pmDistM \subseteq \Red{\pmTypb\rta\pmTypc}$ iff
      $\pmTermM\in\Red{\pmTypb\rta\pmTypc}$.
    \item
      If $\pmTypa={\pmTypa_1\times\pmTypa_2}$: then for $i\in\{1,2\}$,
      $(\piI\ \pmTermM)\rta (\piI\ \pmDistM)$ so that by IH,
      $(\piI\ \pmTermM)\in\Red{\pmTypa_i}$ iff
      $(\piI\ \pmDistM)\subseteq\Red{\pmTypa_i}$; which exactly means
      that $\supp\pmDistM \subseteq \Red{\pmTypa_1\times\pmTypa_2}$
      iff $\pmTermM\in\Red{\pmTypa_1\times\pmTypa_2}$.
    \end{varitemize}
    \item
      {\em The reducibility candidates over $\Red\pmTypa$ are
      $\Rta$-saturated}. This is a trivial induction on $\Rta$ using the
      $\rta$-saturation for the $(R\dash +)$ case.
    \item {\em $\Red\pmTypa$ is inhabited by a value}.
      By induction on $\pmTypa$: $\0\in\Red\NAT$, $\lambda x.\pmValV\in\Red{\pmTypa\rta\pmTypb}$ and $\langle \pmValU,\pmValV\rangle\in\Red{\pmTypa\times\pmTypb}$ whenever $\pmValU\in\Red\pmTypa$ and $\pmValV\in\Red\pmTypb$.
  \item {\em The reducibility candidates $\pmTermM$ over $\Red\pmTypa$ $\Rta$-reduce to $\Evala\pmTermM$}.
    This goes by induction on~$\pmTypa$:
    \begin{varitemize}
    \item Trivial for $\pmTypa=\NAT$. 
    \item Let $\pmTermM\in \Red{\pmTypb\rta\pmTypc}$, there is a value $\pmValV\in\Red\pmTypb$, thus $(\pmTermM\ \pmValV)\in\Red\pmTypc$ and $\pmTermM\ \pmValV\Rta \Evala{\pmTermM\ \pmValV}$ by IH; we can conclude using Lemma~\ref{lmaux1:BSSred}.
    %\longv{\item Let $\pmTermM\in \Red{\pmTypb\times\pmTypc}$, then $\piO\ \pmTermM\in\Red\pmTypb$ and $\piO\ \pmTermM\Rta \Evala{\piO\ \pmTermM}$ by IH; we can conclud using Lemma~\ref{lmaux2:BSSred}.\todo{small issue her}} 
    \item Similar for products.
    \end{varitemize}
  \item
    {\em Every term $x_1:\pmTypa_1,\ldots,x_n:\pmTypa_n\vdash \pmTermM:\pmTypb$ is a
    candidate in the sense that if $\pmValV_i\in\Red{\pmTypa_i}$
    then $\pmTermM[\pmValV_1/x_1,\ldots,\pmValV_n/x_n]\in\Red\pmTypb$.} By induction
    on the type derivation. The only difficult cases are applications,
    recursion, and binary probabilistic choices:
    \begin{varitemize}
     \item
        For the application, we have to show that if $\pmTermM\in \Red
        {\pmTypa\rta\pmTypb}$ and $\pmTermN\in \Red {\pmTypa}$ then
        $(\pmTermM\ \pmTermN)\in \Red {\pmTypb}$. But since
        $\pmTermN\in \Red {\pmTypa}$ we get that
        $\pmTermN\Rta\Evala{\pmTermN}$ with
        $|\Evala{\pmTermN}|\subseteq \Red {\pmTypa}$.  This means that
        $\supp{\pmTermM\ \Evala{\pmTermN}}\subseteq\Red\pmTypb$ and
        that
        $\pmTermM\ \Evala{\pmTermN}\Rta\Evala{\pmTermM\ \Evala{\pmTermN}}$
        supported into $\Red\pmTypb$. We conclude by
        Lemma~\ref{lmaux2:BSSred} that
        $\pmDistvU=\Evala{\pmTermM\ \pmTermN}$ and thus that
        $(\pmTermM\ \pmTermN)\in\Red\pmTypb$.
   \item
     For the operator $\rec$, We have to show that if
     $\pmValU\in\Red\pmTypa$ and
     $\pmValV\in\Red{\NAT\rta\pmTypa\rta\pmTypa}$ then for all
     $n\in\Nat$, $(\rec\ \langle \pmValU, \pmValV, \boldsymbol
     n\rangle)\in\Red\pmTypa$. We proceed by induction on $n$:
     \begin{varitemize}
     \item
         If $n=0$: $\rec\ \langle\pmValU, \pmValV, \boldsymbol
         0\rangle\rta\{\pmValU\}\subseteq\Red\pmTypa$ and we conclude
         by saturation.
     \item
        Otherwise: $\rec\ \langle\pmValU, \pmValV, \boldsymbol
        {(n+1)}\rangle\rta \{\pmValV\ \boldsymbol n\
        (\rec\ \langle\pmValU, \pmValV, \boldsymbol
        n\rangle)\}\subseteq\Red\pmTypa$ since
        $(\rec\ \langle\pmValU, \pmValV, \boldsymbol
        n\rangle)\in\Red\pmTypa$ by IH and since $\boldsymbol
        n\in\Red\NAT$ and
        $\pmValV\in\Red{\NAT\rta\pmTypa\rta\pmTypa}$, we conclude by
        saturation.
     \end{varitemize}
    \item
      For the operator $\oplus$, if $\pmTermM,\pmTermN\in\Red\pmTypa$
      then
      $\supp{\left\{\begin{matrix}\pmTermM\mapsto \frac12\\[0.2em]\pmTermN\mapsto\frac12\end{matrix}\right\}}\subseteq\Red\pmTypa$,
      and, by $\rta$-saturation,
      $(\pmTermM\oplus\pmTermN)\in\Red\pmTypa$.
     \end{varitemize}
  \end{enumerate}
  The thesis, as usual, can be proved as a corollary of points 4 and 5.
\end{proof}
Notice that this theorem does not apply to $\SystFix$ (and a fortiori
to $\SystTot$) because step~(5) of the proof would not hold.
\begin{thm}\label{thm:BSS-ASN}
  The accessible evaluation of a term $\pmTermM\in \SystPlR$ is its
  evaluation, i.e., $\Evala{\pmTermM}=\Eval{\pmTermM}$.
  Moreover, any term $\pmTermM$ is almost surely
  terminating.
\end{thm}
\begin{proof}
  By a trivial induction on $\Rta$, we can easily show that if
  $\pmTermM\Rta \pmDistM$ then $\Eval{\pmTermM}=\Eval{\pmDistM}$.
  As a consequence,
  $\Eval{\pmTermM}=\Eval{\Evala{\pmTermM}}=\Evala{\pmTermM}$.
  By another
  trivial induction on $\Rta$  we can moreover show that if
  $\pmDistM\Rta\pmDistN$ then $\norm\pmDistM = \norm\pmDistN$,
  which yields the thesis.
\end{proof}

\begin{cor}\label{thm:BSS-PASN}
  Any term $\pmTermM\in\SystPl$ is positively almost-surely terminating.
\end{cor}
  \begin{proof}
    By an induction on $\Rta$ we can show that if
    $\pmTermM\Rta\Eval\pmTermM$ (resp. $\pmDistM\Rta\Eval\pmDistM$ for
    $\pmDistM$ finitely supported) then $\pmTermM\rta^*\Eval\pmTermM$
    (resp. $\pmDistM\rta^*\Eval\pmDistM$):
    \begin{varitemize}
    \item
        $(R\dash 0)$ is trivial.
    \item
        $(R\dash +)$ is immediate once we remark that in $\SystPl$
        whenever $\pmTermM\rta\pmDistM$, necessarily $\pmDistM$ is
        finitely supported and we can use our induction hypothesis.
   \item
        If for all $\pmTermM\in\supp\pmDistM$,
        $\pmTermM\Rta\Eval\pmTermM$, then by IH,
        $\pmTermM\rta^{n_\pmTermM}\Eval\pmTermM$ for some
        $n_\pmTermM\in\Nat$. Moreover, since $\supp\pmDistM$ is
        finite, we can set $n= sup_{\pmTermM\in\supp\pmDistM}$ so that
        $\pmTermM\rta^{\le n}\Eval\pmTermM$ and $\pmDistM\rta^{\le
        n}\int_\pmDistM\Eval\pmTermM d\pmTermM=\Eval\pmDistM$.
\end{varitemize}
    The reduction time of a term is then bounded by $n$ such that $\pmTermM\rta^n\Eval\pmTermM$. 
  \end{proof}
Notice that Corollary~\ref{thm:BSS-PASN} does \emph{not} apply to
$\SystRand$ (and a fortiori to $\SystTot$) because the second bullet
of the proof would not be verified.

%%%%%%%%%%%%%%%%%%%%%%%%%%%%%%%
\subsection{Mapping to $\npT$} \label{sec:coding_T+}
%%%%%%%%%%%%%%%%%%%%%%%%%%%%%%%

Positive almost sure termination of terms in $\SystPl$ is not the only 
consequence of
Theorem~\ref{thm:BSS-ASN}.  In fact, the finiteness of the resulting
distribution over values allows a \emph{finite} representation of
$\SystPl$-distributions by $\SystT$-definable functions.
Indeed, we can consider an extension of classic system $\SystT$ with a
single memory cell of type $\NAT$ that we use to store (the binary
encoding of) the outcomes of the coin flips we will perform in the
future. If we denote $c$ the memory-cell, this means that the $\oplus$
can be encoded\footnote{Notice that conditionals, parity and
fractions are easily implementable in $\SystT$.}:
$$
(\pmTermM\oplus\pmTermN)^*\quad :=\quad \newop{if}\ (\newop{mod}_2\:
c)\ \newop{then}\ (c\newop{:=}\frac c 2\:
;\: \pmTermM^*)\ \newop{else}\ (c\newop{:=} \frac c 2\:
;\: \pmTermN^*)
$$

From Theorem~\ref{thm:BSSred}, we know that for any
$\pmTermM\in\SystPl(\NAT)$, there is $n\in\Nat$ such that
$\pmTermM\rta^n\Eval\pmTermM$. Since the execution is bounded by $n$,
there cannot be more than $n$ successive probabilistic choice so that:
$$ \Eval\pmTermM = \left\{k\mapsto \frac{\#\{m<2^n\mid k=\NF{c\newop{:=}\boldsymbol m\:;\:\pmTermM^*}\}}{2^n}\right\}. $$
Using a well known state-passing style transformation, we can enforce
$(c\newop{:=}\boldsymbol m\:;\:\pmTermM^*)$ into a term of $\SystT$. Then, using a simple recursive
operation on $m$, we can represent the whole $\#\{m<2^n\mid k=\NF{c\newop{:=}\boldsymbol m\:;\:\pmTermM^*}\}$ into the
result of a term $k:\Nat\vdash \pmTermN:\Nat$ so that $\lambda k.\pmTermN$ define a function that
represent the distribution $\Eval \pmTermM$.

\begin{exa}
  Take the term $\pmTermM\ =\ \rec\ \langle \0,\ \lambda xy.y\oplus \S
y,\boldsymbol 2\rangle$ from Example~\ref{ex:doubleflip}. Its
  encoding in $\SystT$ is
  $$
  \pmTermM^* = \rec\langle \0,\ \lambda xy.\:\newop{if}\:
  (\newop{mod}\: c\: \boldsymbol 2)\: \newop{then}\: (c\newop{:=}\frac
  c 2\: ;\: y)\: \newop{else}\: (c\newop{:=} \frac c 2\: ;\:\S
  y),\ \boldsymbol 2\rangle.
  $$
  By a standard state passing lifting (and a few simplifications) we obtain the
  term:
  $$
  \pmTermM^\sim\ =\ \rec\ \langle \lambda c.(\0,c),\ \lambda
  xyc.\:\newop{if}\: (\newop{mod}\: c\: \boldsymbol
  2)\: \newop{then}\: y\ (\newop{div}\: c\: \boldsymbol
  2)\: \newop{else}\: \S\ (y\ (\newop{div}\: c\: \boldsymbol
  2)),\ \boldsymbol 2 \rangle
  $$
  As we know that there are at most two choices, we can count the
  number of $c$ below $4$ which result to a certain $u$, getting:
  $$
  \pmTermM^{\$}\ :=\ \lambda u.\: \rec\ \langle\0,\ \lambda
  xy.\: \newop{if}\ (\pi_1 (\pmTermM^\sim x) == u)\ \newop{then}\ \S
  y\ \newop{else}\ y,\ \boldsymbol 4 \rangle.
  $$
  Then we have:
  \[
  \Eval\pmTermM = \left\{k\mapsto \frac{\NF{\pmTermM^{\$}\ \boldsymbol
  k}}{4}\right\}.
  \tag*{\EOE}
  \]
\end{exa}

What remains to be shown is that this encoding can be made parameteric, in the
sense that for any $\pmTermM\in\SystPl(\NAT\rta\NAT)$, we can generate
$\pmTermM_\downarrow\in\SystT(\NAT\rta\NAT\rta\NAT)$ and
$\pmTermM_\#\in\SystT(\NAT\rta\NAT)$ such that for all $n\in\Nat$:
  $$
  \Eval{\pmTermM\ \boldsymbol n} \quad = \quad \left\{\boldsymbol
k\mapsto \frac{\#\{m<2^{\NF{\pmTermM_\#\ \boldsymbol n}}\mid
k=\NF{\pmTermM_\downarrow\ \boldsymbol n}\}}
{2^{\NF{\pmTermM_\#\ \boldsymbol n}}}\right\}. $$
  The difficulty, here, comes from the bound
$\pmTermM_\#$ that have to be computed dynamically by a complex
monadic encoding. To this purpose, a translation of $\SystPl$ into
$\SystT$ has to be appropriately defined.

First of all, let us define two maps $\tpttype{\cdot}$ and
$\tptvals{\cdot}$ on types as follows:
\begin{align*}
\tpttype{\pmTypa} &:= \bigl(\NAT\rta\tptvals{\pmTypa}\bigr)\times\NAT\:;  &
\tptvals{\NAT}&:=\NAT\:;\\
\tptvals{\pmTypa\rta\pmTypb}&:=\tptvals{\pmTypa}\rta\tpttype{\pmTypb}\:;  &
\tptvals{\pmTypa\times\pmTypb}&:=\tptvals{\pmTypa}\times\tptvals{\pmTypb}\:.
\end{align*}
This can be seen as the monadic lifting of the probabilistic
monad. The maps $\tpttype{\cdot}$ and
$\tptvals{\cdot}$ can be generalized to type environments
in a natural way. Their extension to $\pT{\bc}$ terms, for $\tpttype{\cdot}$,
and to $\pT{\bc}$ extended values, for $\tptvals{\cdot}$,
is the core of the embedding, and will be defined shortly.
In the meantime, it is instructive to examine the properties
we expect from these maps. First of all, whenever $\Gamma\vdash\pmTermM\typ\pmTypa$
and $\Gamma\vdash\pmValV\typ\pmTypa$, it holds that
$$
\tptvals{\Gamma}\vdash\tpttype{\pmTermM}\typ\tpttype{\pmTypa}\:;\qquad\qquad
\tptvals{\Gamma}\vdash\tptvals{\pmValV}\typ\tptvals{\pmTypa}\:;
$$
in $\npT$. With a slight abuse of notation, we see the type $\tpttype{\pmTypa}$,
which by definition is a product type, as given through two components
$\tptbranch{\pmTypa}:=\NAT\rta\tptvals{\pmTypa}$ and
$\tptnumber{\pmTypa}:=\NAT$. Accordingly, we denote
$\pmTermM_\downarrow:=\piO\pmTermM :\tptbranch{\pmTypa}$ and $\pmTermM_\#:=\piT\pmTermM : \tptnumber{\pmTypa}$
whenever $\pmTermM:\tpttype{\pmTypa}$. Similarly, we may
directly define~$\tptbranch\pmTermM$ and $\tptnumber\pmTermM$ whenever
$\tpttype\pmTermM=\langle\tptbranch\pmTermM,\tptnumber\pmTermM\rangle$.

We can now give a precise (although laborious) definition
of the maps above. What is important for the rest of the development
is that for every natural number $n$,
$\tptvals{\boldsymbol{n}}=\boldsymbol{n}$, and that $\tptvals{\lambda
y.\pmTermM}=\lambda y.\tpttype{\pmTermM}$. The encoding $\tptvals \cdot$ of 
extended values is given by:
\begin{align*}
  \tptvals\S&:= \lambda y. \tpttype{\S\ y}\:; & 
  \tptvals{\S\ \pmValV}&:= \S\ \tptvals{\pmValV}\:; &
  \tptvals\0&:= \0\:; \\
  \tptvals{\langle\pmTermM,\pmTermN\rangle} &:=  
  \langle\tptvals\pmTermM,\tptvals\pmTermN\rangle\:; &
  \tptvals{\lambda x.M}&:= \lambda x.\tpttype \pmTermM\:; &
  \tptvals x &:= x\:;\\
  \tptvals{\pi_i} &:= \lambda x. \ret (\pi_i x)\:;\\
  \tptvals\rec &:= \rlap{$\lambda \langle u,v,w\rangle.\rec\:\langle \ret\: u 
  ,\: \lambda xy. (v x)\bindo y ,\: w\rangle\:;$}
\end{align*}
where $\ret$ and $\bindo$ are the return and the bind of the underlying 
monad\footnote{Technically, this bind is not the usual bind, but a lifted 
version.}:
\begin{align*}
  \ret &: \tptvals\pmTypa\rta\tpttype\pmTypa\:;  &
  \ret &:= \lambda x.\langle \lambda y.x ,\; \0 \rangle\:; \\
  (\cdot)\bindo(\cdot) & : 
  \tpttype{\pmTypa\rta\pmTypb}\times\tpttype\pmTypa\rta\tpttype\pmTypb\:; &
  \pmTermM\bindo \pmTermN & := \langle\pmTermM\bindo_\downarrow 
  \pmTermN,\pmTermM\bindo_\# \pmTermN\rangle\:.
\end{align*}
However, the computation of the bind is
complex. Intuitively, the right part is computing the number of
choices in $\pmTermM$ in $\pmTermN$ and in all the possibles outcomes
$(\pmValU\ \pmValV)$ for $\pmValU\in \supp{\Eval\pmTermM}$ and
$\pmValV\in\supp{\Eval\pmTermN}$. Of course we take the least upper
bound of those outcomes as they are independent:
\begin{align*}
  (\cdot)\bindo_\# (\cdot) & : 
  \tpttype{\pmTypa{\rta}\pmTypb}\times\tpttype\pmTypa\rta\NAT\:; \\
  \pmTermM\bindo_\# \pmTermN &:= \pmTermM_\# + \pmTermN_\# + 
  \mathop{\newop{max}}_{x< 
  2^{\pmTermM_\#}}\mathop{\newop{max}}_{y<2^{\pmTermN_\#}}(\pmTermM_\downarrow\ 
  x\ (\pmTermN_\downarrow\ y))_\#\:;
\end{align*}
where we use the following syntactical sugar:
\begin{align*}
  (\cdot)+(\cdot) & : \NAT\times\NAT\rta\NAT\:; &
  \pmTermM + \pmTermN &:= \rec \langle \pmTermM,\: \lambda \_x.\S x 
  ,\:\pmTermN\rangle\:;\\
  2^{\:(\cdot)}&: \NAT\rta\NAT\:;  &
  2^\pmTermM&:= \rec\langle \boldsymbol 1,\: \lambda \_y.\rec\langle y, \lambda 
  \_v.\S v, y\rangle,\: \pmTermM\rangle\:;\\
  & &
  \mathop{\newop{max}}_{x< \pmTermM}(\pmTermN)&:= \rec\langle \0,\: \lambda 
  \_y.\:\pmTermN {\vee} y,\: \pmTermM\rangle\:; \\
    (\cdot)\vee(\cdot)&:\NAT\times\NAT\rta\NAT\:; &
    \pmTermM\vee\pmTermN&:= \rec\langle \lambda u.u ,\: \lambda 
    xyu.\S(\rec\langle x, \lambda a \_. y a ,u\rangle) ,\ \pmTermM \rangle\ 
    \pmTermN\:.
\end{align*}
The first member of the bind is also complex. It is taking a stream of
probabilistic choices $s$, computes $\pmTermM$ over this choices, then
shifts the stream $s$ in order to remove the choices relative to
$\pmTermM$, this way we can compute the result from $\pmTermN$. After
all this, we obtain an object of type $\tpttype \pmTypb$, but our job
is not finished yet: we have to select the first member, to which we
give what remains of the stream (computed by shifting $x$ twice).
\begin{align*}
(\cdot)\bindo_{\downarrow\:} (\cdot)& : 
\tpttype{\pmTypa{\rta}\pmTypb}\times\tpttype\pmTypa\rta\NAT\rta 
\tptvals\pmTypb\:; \\
  \pmTermM\bindo_{\downarrow\:} \pmTermN &:= \lambda s.(\pmTermM_\downarrow\ s\ 
  (\pmTermN_\downarrow\ (\newop{shift}\ s\ \pmTermM_\#)))_\downarrow\ 
  (\newop{shift}\ s\ (\pmTermM_\#{+}\pmTermN_\#))\:;
\end{align*}
where we use the following syntactical sugar:
\begin{align*}
  \newop{shift} &: \NAT \rta \NAT\rta\NAT\:;&
  \newop{shift} &:= \lambda sy. \rec\langle s,\: \lambda u.div_2 ,\: 
  y\rangle\:;\\
  \newop{div_2} &: \NAT\rta\NAT\:;&
  \newop{div_2} &:= \lambda x. \rec \langle \lambda \_.\0 ,\:\lambda 
  \_vw.\newop{ite}\langle w,\S(v \0), v \boldsymbol 1\rangle ,\:\ x \rangle\ 
  \0\:.
\end{align*}
The encoding $\tpttype\cdot$ is given through the return and the bind operations:
\begin{align*}
  \tpttype{\pmValV} &:= \ret\ \tptvals\pmValV\:;  &
  \tpttype{\pmTermM\ \pmTermN} &:= \tpttype\pmTermM\bindo\tpttype\pmTermN\:.
\end{align*}
The binary choice is the defined as expected:
\begin{align*}
  \tptnumber{\pmTermM\oplus\pmTermN} &:= \tptnumber\pmTermM {\vee} 
  \tptnumber\pmTermN\:;\\
  \tptbranch{\pmTermM\oplus\pmTermN} &:= \lambda x. \newop{ite} \langle 
  \newop{mod_2} x,\:\tptbranch\pmTermM (\newop{div_2} x),\: \tptbranch\pmTermN 
  (\newop{div_2} x)\rangle\:;
  \end{align*}
where we use the following syntactical sugar:
\begin{align*}
    \newop{ite}&:\NAT\times\pmTypa\times\pmTypa\rta\pmTypa\:;&
    \newop{ite} &:= \lambda x. \rec\ \langle \pi_3 x,\: \lambda \_\_.\pi_2 x,\: 
    \pi_1 x \rangle\:;\\
    \newop{mod_2} &: \NAT\rta\NAT\:;&
    \newop{mod_2} &:= \lambda x. \rec \langle \0 ,\:\lambda 
    \_v.\newop{ite}\langle v,\0, \boldsymbol 1\rangle ,\:\ x \rangle\:.
\end{align*}
The byproduct of this relatively complex encoding is the fact that whatever
distribution one is able to compute in $\SystPl$ can also be computed back
in $\SystT$, and that this scales to first-order functions:
\begin{thm}\label{Th:finiteRep}
  Distributions in $\SystPl$ are finitely parameterically representable by $\SystT$-definable functions, {\em i.e.} 
  for any $\pmTermM:\NAT\rta\NAT$ in $\SystPl$ there are $F: \NAT\rta\NAT\rta\BIN$ and $Q:\NAT\rta\NAT$ in $\SystT$ 
  such that for all $n$:
  $$ 
  \Eval{\pmTermM\ \boldsymbol n}\ =\ \left\{\boldsymbol k\mapsto \NF{F\ 
  \boldsymbol n\ \boldsymbol k}\right\}\:; \quad\quad\quad \forall k\ge \NF{Q\ 
  \boldsymbol n}.\ \NF{F\ \boldsymbol n\ \boldsymbol k} =\0.
  $$
\end{thm}
%\todof{Do we have to do the full proof?}

\section{Countable Probabilistic Choice}
    \label{sec:fragmentRand}
      \newcommand{\tptstype}{\tpttype}
  \newcommand{\tptsvals}{\tptvals}

%%%%%%%%%%%%%%%%%%%%%%%%%%%%%%%%%%%%%%%%%%%%%%%%%%%%%%%%%
\subsection{Multistep Semantics}\label{sec:multStepRand}
%%%%%%%%%%%%%%%%%%%%%%%%%%%%%%%%%%%%%%%%%%%%%%%%%%%%%%%%%

We have seen that none of Theorem~\ref{thm:BSSred},
Theorem~\ref{thm:BSS-ASN} and Corollay~\ref{thm:BSS-PASN} hold in
$\SystFix$. Indeed Theorem~\ref{thm:BSSred}, which is a prologue to the
other two, does not hold on terms like, e.g., $\fixran\langle\0,\S\rangle$ that
will never $\Rta$-reduce to a value distribution.
The fragment $\SystRand$ is more interesting, as both
Theorem~\ref{thm:BSSred} and Theorem~\ref{thm:BSS-ASN} hold. However,
as we have seen in Theorem~\ref{thm:cePASN}, positive almost sure
normalization (and Theorem~\ref{thm:BSS-PASN}) do not hold. This is
because we are manipulating infinitely supported distributions (due to
the reduction rule of $\rand$).

\begin{exa}
  Recall that $\fixran_\rand\ :=\ \lambda x.\rec\ \langle \pi_2
  x,\lambda z.\pi_1 x,\rand\rangle$ is the encoding of $\fixran$ into
  $\SystRand$.  We have $\fixran:\rand\ \langle \S\S,\0\rangle \Rta
  \{\boldsymbol{2n}\mapsto\frac 1{2^{n+1}} \mid n\ge 0\}$; indeed, if
  we fix $U_n=\rec\ \langle \pi_2 \langle \S\S,\0\rangle,\lambda
  z.\pi_1 \langle \S\S,\0\rangle,\boldsymbol n\rangle$:
  \begin{center}
  	\tiny
  	\hspace{-20em}
  \begin{small}
    \AxiomC{\hspace{-2em}}
    \AxiomC{\hspace{-5.5em}}
    \AxiomC{\hspace{-1em}}
    \AxiomC{\hspace{3em}}
    \AxiomC{$\phantom {\pmDistM}$}
    %\RightLabel{{\scriptsize \rlap{$(R\dash refl)$}}}
    \UnaryInfC{$\0\Rta\{\0\}$}
    \RightLabel{{\scriptsize \rlap{$(R\dash{\in})$}}}
    \UnaryInfC{\llap{$\pi_2 \langle \S\S,\0\rangle\rta\ $}$\{\0\}\Rta\{\0\}$}
    \RightLabel{{\scriptsize \rlap{$(R\dash \tran)$}}}
    \BinaryInfC{$\pi_2 \langle \S\S,\0\rangle\Rta \{\0\}$}
    \RightLabel{{\scriptsize \rlap{$(R\dash{\in})$}}}
    \UnaryInfC{\llap{$U_0\rta\ $}$\{\pi_2 \langle \S\S,\0\rangle\}\Rta \{\0\} $}
    \RightLabel{{\scriptsize $(R\dash \tran)\hspace{6em}$}}
    \BinaryInfC{$U_0\Rta \{\0\}$}
    \AxiomC{$\xi$}
    \AxiomC{$\cdots$}
    \RightLabel{{\scriptsize \rlap{$(R\dash{\in})$}}}
    \TrinaryInfC{\llap{$U_\rand \rta\ $}$\{U_n\mapsto\frac 1 {2^{n+1}}\mid n\ge 0\}\Rta \{\boldsymbol {2n} \mapsto\frac 1{2^{n+1}}\mid n\ge 0\} $}
    \RightLabel{{\scriptsize \rlap{$(R\dash \tran)$}}}
    \BinaryInfC{$U_\rand \Rta\pmDistN$}
    \RightLabel{{\scriptsize \rlap{$(R\dash{\in})$}}}
    \UnaryInfC{\llap{$\fixran_\rand \langle \S\S,\0\rangle \rta$}$\{U_\rand\}\Rta\{\boldsymbol {2n} \mapsto\frac 1{2^{n+1}}\mid n\ge 0\}$}
    \LeftLabel{\hspace{12em}}
    \RightLabel{{\scriptsize \rlap{$(R\dash \tran)$}}}
    \BinaryInfC{$\fixran_\rand \langle \S\S,\0\rangle \Rta\ \{\boldsymbol{2n}\mapsto\frac 1{2^{n+1}} \mid n\ge 0\}$}
    \DisplayProof
  \end{small}
  \end{center}
  where $\xi$ is the following derivation:
  \begin{center}
  \begin{small}
  \AxiomC{\hspace{-7em}}
  \AxiomC{\hspace{3em}}
  \AxiomC{\hspace{7em}}
  \AxiomC{\hspace{3em}}
  \AxiomC{$\phantom {\pmDistM}$}
  %\RightLabel{{\scriptsize \rlap{$(R\dash refl)$}}}
  \UnaryInfC{$\S\S\ \0\Rta\{\boldsymbol2\}$}
  \RightLabel{{\scriptsize \rlap{$(R\dash{\in})$}}}
  \UnaryInfC{\llap{$\pi_1 \langle \S\S,\0\rangle\:\0\rta\ $}  $\{\S\S\ 
  \0\}\Rta\{\boldsymbol2\}$}
  \RightLabel{{\scriptsize \rlap{$(R\dash \tran)$}}}
  \BinaryInfC{$\pi_1 \langle \S\S,\0\rangle\:\0 \Rta\{\boldsymbol2\}$}
  \RightLabel{{\scriptsize \rlap{$(R\dash{\in})$}}}
  \UnaryInfC{\llap{$\pi_1 \langle \S\S,\0\rangle\:(\pi_2 \langle 
  \S\S,\0\rangle)\rta\ $}$\{\pi_1 \langle \S\S,\0\rangle\:\0\}\Rta\{\boldsymbol 
  2\}$}
  \RightLabel{{\scriptsize \rlap{$(R\dash \tran)$}}}
  \BinaryInfC{$\pi_1 \langle \S\S,\0\rangle\:(\pi_2 \langle \S\S,\0\rangle)\Rta 
  \{\boldsymbol 2\}$}
  \RightLabel{{\scriptsize \rlap{$(R\dash{\in})$}}}
  \UnaryInfC{\llap{$\pi_1 \langle \S\S,\0\rangle\:U_0\rta\ $}$\{\pi_1 \langle 
  \S\S,\0\rangle\:(\pi_2 \langle \S\S,\0\rangle)\}\Rta \{\pi_1 \langle 
  \S\S,\0\rangle\:\0\} $}
  \RightLabel{{\scriptsize \rlap{$(R\dash \tran)$}}}
  \BinaryInfC{$\pi_1 \langle \S\S,\0\rangle\:U_0\Rta \{\boldsymbol2\}$}
  \RightLabel{{\scriptsize \rlap{$(R\dash{\in})$}}}
  \UnaryInfC{\llap{$U_1\rta\ $}$\{\pi_1 \langle \S\S,\0\rangle\:U_0\}\Rta 
  \{\boldsymbol2\} $}
  \RightLabel{{\scriptsize \rlap{$(R\dash \tran)$}}}
  \BinaryInfC{$U_1\Rta \{\boldsymbol2\}$}
  \DisplayProof
  \end{small}
  \end{center}
  we can see that the tree is infinite due to the second application
  of $(R\dash{\in})$; but, despite being infinite, and with an
  infinite height, each subtree above the second application of
  $(R\dash{\in})$ is finite, making the derivation correct.  However,
  $\fixran \langle \S\S,\0\rangle \not\Rta
  \{\boldsymbol{2n}\mapsto\frac 1{2^{n+1}} \mid n\ge 0\}$.  We can
  approach this distribution, but it is impossible to $\Rta$-reduce to
  a value distribution.\EOE
\end{exa}

Remember that $\SystRand$ and $\SystFix$ are equivalent, so why 
such a difference? This is due to the discrepancy in nature 
between their execution trees. Indeed, we have seen that the 
execution trees are finitely branching in $\SystFix$, but with 
infinite paths, while those of $\SystRand$ are infinitely 
branching, but with finite paths. Since multistep reduction 
somehow reflects those execution trees,we can see that we only 
need derivations with infinite arity to get a correct multistep 
semantics for $\SystRand$. The whole point is that we can 
perform transfinite structural induction over these trees. 
Indeed, by considering the reduction trees themselves with the 
inclusion (or subtree) order gives you a well-founded poset, 
recalling that there is no infinite path. If one want to unfold 
this well-founded poset into an ordinal, then it should be the 
smallest ordinal $o$ such that $o = 1 + \omega o$, {\em 
i.e.}, $o=\omega^\omega$. This is unusual in operational 
semantics, where finitary induction suffices in most cases. 
Remark that, due to the encoding of $\bc$ and $\fixran$ 
into $\SystRand$, Theorem~\ref{thm:BSS-ASN} subsumes 
Theorem~\ref{thm:ASTofFullCalc}. Remark, moreover, that we did 
not have to go through the definition of approximants. 
Nonetheless, those approximations exists and point out that 
$\SystPl$ should be approximating $\SystRand$ in some way or 
another. This is precisely what we are going to do in the next 
section.
%%%%%%%%%%%%%%%%%%%%%%%%%%%%%%%%%%%%%%%%%%%%%%%%%%%%%%%%%
\subsection{The Approximants: State-Bounded Random Integers}
%%%%%%%%%%%%%%%%%%%%%%%%%%%%%%%%%%%%%%%%%%%%%%%%%%%%%%%%%
In this section, we show that $\SystPl$ approximates $\SystRand$: for
any term $\pmTermM\in\SystRand(\NAT)$, there is a term
$\pmTermN\in\SystPl(\NAT\rta\NAT)$ that represents a sequence
approximating $\pmTermM$ uniformly.  We will here make strong use of
the fact that $\pmTermM$ has type $\NAT$. This is a natural drawback
when we understand that the encoding $(\cdot)^\dagger$ on which the
result above is based is not direct, but goes through another
state-passing transformation.

A naive idea would be to use $\SystPl$
and to stop the evaluation after a given reduction time as schematized
in Figure~\ref{fig:TruncHor}.  Despite the encoding to be a nightmare,
this should be encodable in $\SystPl$. However, for the convergence
time to be independent from the term and uniform, there is virtually no
hope. That is why we have switched to $\SystRand$, which carries much
nicer properties as seen in the previous chapter.
The basic idea behind the embedding $(\cdot)^\dagger$ is to mimic any
instance of the $\rand$ operator in the source term by some term
$\0\bc(\boldsymbol 1\bc(\cdots (\boldsymbol n\bc \bot)\cdots\!)$, where
$n$ is \emph{sufficiently large}, and~$\bot$ is an arbitrary value of type
$\NAT$. Of course, the semantics of this term is \emph{not} the same
as that of $\rand$, due to the presence of $\bot$; however, $n$ will be
chosen sufficiently large for the difference to be
negligible. Notice, moreover, that this term can be generalized into
the following parametric form $\rand^\ddag:=\lambda
n.\rec\ \langle\bot,(\lambda x. \S \bc (\lambda y.\0))\rangle\ n$.
\begin{center}
  \begin{tikzpicture}
    \node (v) at (2.1,1.2) {$\rand^\ddag\ \boldsymbol n$};
    \node (v1) at (0,0) {$\boldsymbol0$};
    \node (v2) at (0.7,0) {$\boldsymbol1$};
    \node (v3) at (1.4,0) {$\boldsymbol2$};
    \node (v4) at (2.1,0) {$\boldsymbol3$};
    \node (v51) at (2.4,0) {$\vphantom 1$};
    \node (v52) at (2.8,0) {$\cdots\vphantom 1\!\cdots$};
    \node (v53) at (3.15,0) {$\vphantom 1$};
    \node (v6) at (3.5,0) {$\boldsymbol n$};
    \node (v7) at (4.1,0) {$\bot$};
    \node (v7b) at (4.7,0) {$\bot$};
    \node (v8) at (5.3,0) {$\bot$};
    \node (v91) at (5.6,0) {$\vphantom 1$};
    \node (v92) at (6.1,0) {$\cdots\vphantom 1\!\cdots$};
    \node (v93) at (6.5,0.1) {$\vphantom 1$};
    \node (v94) at (6.5,0.3) {$\vphantom 1$};
    \node (v95) at (6.5,0.5) {$\vphantom 1$};
    \node (v96) at (6.5,0.6) {$\vphantom 1$};
    \draw[->] ($(v.south)+(-0.2,0)$) -- node [left] {{\scriptsize $\frac 1 2\ \ \ \phantom.$}} (v1.north);
    \draw[->] ($(v.south)+(-0.1,0)$) -- node [below left] {{\scriptsize $\frac 1 {4_{\vphantom 1}}$}} (v2.north);
    \draw[->] ($(v.south)+(0,0)$) -- node [below] {{\scriptsize $\frac 1 8$}} (v3.north);
    \draw[->] ($(v.south)+(0.05,0)$) -- node [below] {} (v4.north);
    \draw[dashed,->] ($(v.south)+(0.05,0)$) -- node [below] {} (v51.north);
    \draw[dashed,->] ($(v.south)+(0.05,0)$) -- node [below] {} (v52.north);
    \draw[dashed,->] ($(v.south)+(0.05,0)$) -- node [below] {} (v53.north);
    \draw[->] ($(v.south)+(0.1,0)$) -- node [right] {} (v6.north);
    \draw[->] ($(v.south)+(0.1,0)$) -- node [right] {} (v7.north west);
    \draw[->] ($(v.south)+(0.1,0)$) -- node [right] {} (v7b.north west);
    \draw[->] ($(v.south)+(0.1,0)$) -- node [right] {} (v8.north west);
    \draw[dashed,->] ($(v.south)+(0.15,0)$) -- node [right] {{}} (v91.north);
    \draw[dashed,->] ($(v.south)+(0.2,0)$) -- node [right] {{}} (v92.north);
    \draw[dashed,->] ($(v.south)+(0.23,0)$) -- node [right] {{}} (v93.north);
    \draw[dashed,->] ($(v.south)+(0.26,0)$) -- node [right] {{}} (v94.north);
    \draw[dashed,->] ($(v.south)+(0.3,0)$) -- node [right] {{}} (v95.north);
    \draw[dashed,->] ($(v.south)+(0.3,0)$) -- node [right] {{}} (v96.north);
  \end{tikzpicture}
\end{center}

Once $\rand^\ddag$ is available, a natural candidate for the
encoding $(\cdot)^\dagger$ would be to consider something like
$\pmTermM^\ddag:=\lambda z.\pmTermM[(\rand^\ddag\ z)/\rand]
$. In the underlying execution tree, $(\pmTermM^\ddag\ \boldsymbol n)$
correctly simulates the first $n$ branches of each $\rand$ (which had infinite-arity),
but truncates the rest with garbage terms $\bot$. As schematized
in Figure~\ref{fig:TruncVer}, by increasing $n$, we can hope to
obtain the $\pmTermM$ at the limit.

The question is whether the remaining non-truncated tree has a
``sufficient weight'', {i.e.}, that there is a minimal bound to the
probability to stay in this non-truncated tree. However, in general 
$(\cdot)^\ddag$ 
fails on this point, not achieving to approximate $\pmTermM$ uniformly. 
In fact, this probability
is basically $(1-\frac 1 {2^n})^d$ where $d$ is its depth. Since in
general the depth of the non-truncated tree can grow very rapidly with respect 
to $n$
in a powerful system like $\npT$ , there is no hope for this
transformation to perform a uniform approximation. 
\begin{figure}
  \begin{center}
    \fbox{
      \begin{minipage}{.97\textwidth}
        \begin{minipage}[c]{.325\linewidth}
          \begin{subfigure}[c]{.99\linewidth}
          \begin{center}
          \begin{tikzpicture}
            \node (u) at (1.75,2.8) {};
            \node (u1) at (1.5,2.4) {};
            \node (u2) at (2,2.4) {};
            \node (u11) at (1.25,2) {};
            \node (u21) at (1.75,2) {};
            \node (u22) at (2.25,2) {};
            \node (u111) at (1,1.6) {};
            \node (u221) at (2,1.6) {};
            \node (u222) at (2.5,1.6) {};
            \node (u1111) at (0.75,1.2) {};
            \node (u1112) at (1.25,1.2) {};
            \node (u11111) at (0.5,0.8) {};
            \node (u11121) at (1,0.8) {};
            \node (u11122) at (1.5,0.8) {};
            \node (u111111) at (0.25,0.4) {};
            \node (u111211) at (0.75,0.4) {};
            \node (u111221) at (1.25,0.4) {};
            \node (u111222) at (1.75,0.4) {};
            \node (u2221) at (2.25,1.2) {};
            \node (u2222) at (2.75,1.2) {};
            \node (u22211) at (2,0.8) {};
            \node (u22221) at (2.5,0.8) {};
            \node (u22222) at (3,0.8) {};
            \node (u222221) at (2.75,0.4) {};
            \node (u222222) at (3.25,0.4) {};
            \node (dots) at (2,0.2) {\LARGE ...};
            \draw[->] (u.north) -- node [left] {{ }} (u1.north);
            \draw[->] (u.north) -- node [right] {{  }} (u2.north);
            \draw[->] (u1.north) -- node [left] {{  }} (u11.north);
            \draw[->] (u2.north) -- node [left] {{  }} (u21.north);
            \draw[->] (u2.north) -- node [right] {{  }} (u22.north);
            \draw[->] (u11.north) -- node [left] {{  }} (u111.north);
            \draw[->] (u22.north) -- node [left] {{}} (u221.north);
            \draw[->] (u22.north) -- node [right] {{  }} (u222.north);
            \draw[->] (u111.north) -- node [left] {{ }} (u1111.north);
            \draw[->] (u111.north) -- node [right] {{  }} (u1112.north);
            \draw[->] (u1111.north) -- node [left] {{  }} (u11111.north);
            \draw[->] (u1112.north) -- node [left] {{  }} (u11121.north);
            \draw[->] (u1112.north) -- node [right] {{  }} (u11122.north);
            \draw[->] (u11111.north) -- node [left] {{  }} (u111111.north);
            \draw[->] (u11121.north) -- node [left] {{  }} (u111211.north);
            \draw[->] (u11122.north) -- node [left] {{}} (u111221.north);
            \draw[->] (u11122.north) -- node [right] {{  }} (u111222.north);
            \draw[->] (u222.north) -- node [left] {{ }} (u2221.north);
            \draw[->] (u222.north) -- node [right] {{  }} (u2222.north);
            \draw[->] (u2221.north) -- node [left] {{  }} (u22211.north);
            \draw[->] (u2222.north) -- node [left] {{  }} (u22221.north);
            \draw[->] (u2222.north) -- node [right] {{  }} (u22222.north);
            \draw[->] (u22222.north) -- node [left] {{}} (u222221.north);
            \draw[->] (u22222.north) -- node [right] {{  }} (u222222.north);
            \draw[-,thick] (0,1.6) -- node [right] {} (3.5,1.6);
            \draw[-,thick] (0,0.8) -- node [right] {} (3.5,0.8);
            \draw[-,thick] (0,0) -- node [right] {} (3.5,0);
          \end{tikzpicture}
          \end{center}
          \caption{Horizontal}\label{fig:TruncHor}
          \end{subfigure}
        \end{minipage}
        \begin{minipage}[c]{.325\linewidth}
          \begin{subfigure}[c]{.99\linewidth}
          \begin{tikzpicture}
            \node (v) at (1.2,3) {};
            \node (v1) at (0,2) {};
            \node (v1') at (0,1.5) {};
            \node (v2) at (0.6,2) {};
            \node (v2') at (0.6,1.5) {};
            \node (v3) at (1.2,2) {};
            \node (v3') at (1.2,1.5) {};
            \node (v4) at (1.8,2) {};
            \node (v4') at (1.8,1.5) {};
            \node (v5) at (2.4,2) {};
            \node (v5') at (2.4,1.5) {};
            \node (v6) at (3,2) {};
            \node (v7) at (3.6,2) {};
            \node (v8) at (4,2.2) {};
            \node (vv1) at (0,0.5) {};
            \node (vv1') at (0,0) {};
            \node (vv2) at (0.6,0.5) {};
            \node (vv2') at (0.6,0) {};
            \node (vv3) at (1.2,0.5) {};
            \node (vv3') at (1.2,0) {};
            \node (vv4) at (1.8,0.5) {};
            \node (vv4') at (1.8,0) {};
            \node (vv5) at (2.4,0.5) {};
            \node (vv5') at (2.4,0) {};
            \node (vv6) at (3,0.5) {};
            \node (vv7) at (3.6,0.5) {};
            \node (vv8) at (4,0.7) {};
            \draw[->] (v.north) -- node [left] {} (v1.north);
            \draw[->] (v.north) -- node [below left] {} (v2.north);
            \draw[->] (v.north) -- node [below] {} (v3.north);
            \draw[->] (v.north) -- node [below] {} (v4.north);
            \draw[->] (v.north) -- node [right] {} (v5.north);
            \draw[dashed,->] (v.north) -- node [right] {{}} (v6.north);
            \draw[dashed,->] (v.north) -- node [right] {{}} (v7.north);
            \draw[dashed,->] (v.north) -- node [right] {{}} (v8.north);
            \draw[dashed,->] (v1.north) -- node [left] {} (v1'.north);
            \draw[dashed,->] (v2.north) -- node [left] {} (v2'.north);
            \draw[->] (v3.north) -- node [left] {} (v3'.north);
            \draw[dashed,->] (v4.north) -- node [left] {} (v4'.north);
            \draw[dashed,->] (v5.north) -- node [left] {} (v5'.north);
            \draw[->] (v3'.north) -- node [left] {} (vv1.north);
            \draw[->] (v3'.north) -- node [below left] {} (vv2.north);
            \draw[->] (v3'.north) -- node [below] {} (vv3.north);
            \draw[->] (v3'.north) -- node [below] {} (vv4.north);
            \draw[->] (v3'.north) -- node [right] {} (vv5.north);
            \draw[dashed,->] (v3'.north) -- node [right] {{}} (vv6.north);
            \draw[dashed,->] (v3'.north) -- node [right] {{}} (vv7.north);
            \draw[dashed,->] (v3'.north) -- node [right] {{}} (vv8.north);
            \draw[dashed,->] (vv1.north) -- node [left] {} (vv1'.north);
            \draw[dashed,->] (vv2.north) -- node [left] {} (vv2'.north);
            \draw[dashed,->] (vv3.north) -- node [left] {} (vv3'.north);
            \draw[dashed,->] (vv4.north) -- node [left] {} (vv4'.north);
            \draw[dashed,->] (vv5.north) -- node [left] {} (vv5'.north);
            {
              \draw[-,thick] (2,3.2) -- node [right] {} (2,0);
              \draw[-,thick] (3,3.2) -- node [right] {} (3,0);
              \draw[-,thick] (4,3.2) -- node [right] {} (4,0);
            }
          \end{tikzpicture}
      	  \caption{Vertical}\label{fig:TruncVer}
          \end{subfigure}
        \end{minipage}
       	\begin{minipage}[c]{.325\linewidth}
        \begin{subfigure}[c]{.99\linewidth}
          \begin{center}
          \begin{tikzpicture}
            \node (v) at (1.2,3) {};
            \node (v1) at (0,2) {};
            \node (v1') at (0,1.5) {};
            \node (v2) at (0.6,2) {};
            \node (v2') at (0.6,1.5) {};
            \node (v3) at (1.2,2) {};
            \node (v3') at (1.2,1.5) {};
            \node (v4) at (1.8,2) {};
            \node (v4') at (1.8,1.5) {};
            \node (v5) at (2.4,2) {};
            \node (v5') at (2.4,1.5) {};
            \node (v6) at (3,2) {};
            \node (v7) at (3.6,2) {};
            \node (v8) at (4,2.2) {};
            \node (vv1) at (0,0.5) {};
            \node (vv1') at (0,0) {};
            \node (vv2) at (0.6,0.5) {};
            \node (vv2') at (0.6,0) {};
            \node (vv3) at (1.2,0.5) {};
            \node (vv3') at (1.2,0) {};
            \node (vv4) at (1.8,0.5) {};
            \node (vv4') at (1.8,0) {};
            \node (vv5) at (2.4,0.5) {};
            \node (vv5') at (2.4,0) {};
            \node (vv6) at (3,0.5) {};
            \node (vv7) at (3.6,0.5) {};
            \node (vv8) at (4,0.7) {};
            \draw[->] (v.north) -- node [left] {} (v1.north);
            \draw[->] (v.north) -- node [below left] {} (v2.north);
            \draw[->] (v.north) -- node [below] {} (v3.north);
            \draw[->] (v.north) -- node [below] {} (v4.north);
            \draw[->] (v.north) -- node [right] {} (v5.north);
            \draw[dashed,->] (v.north) -- node [right] {{}} (v6.north);
            \draw[dashed,->] (v.north) -- node [right] {{}} (v7.north);
            \draw[dashed,->] (v.north) -- node [right] {{}} (v8.north);
            \draw[dashed,->] (v1.north) -- node [left] {} (v1'.north);
            \draw[dashed,->] (v2.north) -- node [left] {} (v2'.north);
            \draw[->] (v3.north) -- node [left] {} (v3'.north);
            \draw[dashed,->] (v4.north) -- node [left] {} (v4'.north);
            \draw[dashed,->] (v5.north) -- node [left] {} (v5'.north);
            \draw[->] (v3'.north) -- node [left] {} (vv1.north);
            \draw[->] (v3'.north) -- node [below left] {} (vv2.north);
            \draw[->] (v3'.north) -- node [below] {} (vv3.north);
            \draw[->] (v3'.north) -- node [below] {} (vv4.north);
            \draw[->] (v3'.north) -- node [right] {} (vv5.north);
            \draw[dashed,->] (v3'.north) -- node [right] {{}} (vv6.north);
            \draw[dashed,->] (v3'.north) -- node [right] {{}} (vv7.north);
            \draw[dashed,->] (v3'.north) -- node [right] {{}} (vv8.north);
            \draw[dashed,->] (vv1.north) -- node [left] {} (vv1'.north);
            \draw[dashed,->] (vv2.north) -- node [left] {} (vv2'.north);
            \draw[dashed,->] (vv3.north) -- node [left] {} (vv3'.north);
            \draw[dashed,->] (vv4.north) -- node [left] {} (vv4'.north);
            \draw[dashed,->] (vv5.north) -- node [left] {} (vv5'.north);
            {
              \draw[-,thick] (1.2,3.2) -- node [right] {} (2.2,0);
              \draw[-,thick] (1.7,3.2) -- node [right] {} (3.4,0);
              \draw[-,thick] (2.2,3.2) -- node [right] {} (4.4,0.4);
            }
          \end{tikzpicture}
          \end{center}
          \caption{Biaised}\label{fig:TruncBia}
        \end{subfigure}
        \end{minipage}
      \end{minipage}}
    \caption{Various Forms of Approximation}
  \end{center}
\end{figure}
It might well be possible to perform a complex monadic transformation in 
the style of Section~\ref{sec:coding_T+}, that computes a function 
relating the size $n$ to the depth $d$ of the execution 
tree. But there is a much easier solution.

The solution we are
using is to have the precision $m$ of 
$\0\bc (\boldsymbol{1}\bc(\cdots
(\boldsymbol{m}\bc \bot)\cdots))$ to dynamically grow along the
computation, as schematized in Figure~\ref{fig:TruncBia}. 
More specifically, in the approximants $\pmTermM^\dagger\boldsymbol n$,
the growing speed of $m$ will increase with~$n$: in the $n$-th
approximation $\pmTermM^\dagger\boldsymbol n$, 
$\rand$ will be simulated as $\0\bc\boldsymbol (1\bc(\cdots (\boldsymbol m\bc \bot)\cdots))$
\emph{and}, somehow,~$m$ will be updated to $m+n$. Why does it work?
Simply because even for an (hypothetical) infinite and complete execution tree of $\pmTermM$,
we would stay inside the $n^{th}$ non-truncated tree with probability 
$\prod_{k\ge 0} (1-\frac 1 {2^{k*n}})$
which is asymptotically above $(1-\frac 1 n)$. %(for $n\ge 4$).

Implementing this scheme in $\pT{\bc}$ requires a feature which is not
available (but which can be encoded), namely ground-type references.
We then prefer to show that the just described scheme can be realized
in an intermediate language called $\SystSRand$, whose operational
semantics is formulated not on \emph{terms}, but rather on triples
in the form $(\pmTermM,m,n)$, where $\pmTermM$ is the term 
currently being evaluated, $m$ is the current approximation threshold value,
and $n$ is the value of which $m$ is
incremented whenever $\rand$ is simulated. The operational semantics
is standard, except for the following rule:\vspace{-0.1em}
\begin{center}
  \AxiomC{}
  \RightLabel{{\scriptsize$(r\protect\dash \Srand)$}}
  \UnaryInfC{$(\Srand, m,n) \rta \Bigl\{(k,m{+}n,n) \mapsto \frac 1 {2^{k+1}} \mid k< m\Bigr\}$}
  \DisplayProof\vspace{-0.5em}
\end{center}
Notice how this operator behaves similarly to $\rand$ with the
exception that it fails when drawing too big of a number ({\em i.e.},
bigger that the fist state $m$). Notice that the failure is
represented by the fact that the resulting distribution does not necessarily sum
to $1$. The intermediate language $\SystSRand$ is able to approximate
$\SystRand$ at every order (Theorem~\ref{th:TSRapproxTR} below). Moreover,
the two memory cells can be shown to be expressible in $\pT{\bc}$, again by
way of a continuation-passing transformation. Crucially, the initial
value of $n$ can be passed as an argument to the encoded term.

\begin{defi}
  For any $\pmTermM\in\SystRand$, the expression $\pmTermM^*$ stands
  for $\pmTermM[\Srand/\rand]$.  We say that~$(\pmTermM, m,n)\in
  \SystSRand$ if~$m,n\in\Nat$ and~$\pmTermM=\pmTermN^*$ for some
  $\pmTermN\in\SystRand$. Similarly, $\Dist\SystSRand$ is the set of
  probabilistic distributions over $\SystSRand_C\times\Nat^2$, {\em
    i.e.}, over the terms plus states. The reduction rules of
  system $\SystT$ with state-bounded random integers are given by Figure
  \ref{fig:OSTSrand}.
\end{defi}
\begin{figure*}
  \fbox{
  \begin{minipage}{.97\textwidth}
  \begin{small}
  \begin{center}
    \AxiomC{\vphantom{M}}
    \RightLabel{{\scriptsize $(s\protect\dash\beta)$}}
    \UnaryInfC{$(\lambda x.\pmTermM,m,n)\ \pmValV\rta \Bigl\{(\pmTermM[\pmValV/x],m,n)\Bigr\}$}
    \DisplayProof\hskip 10pt
    \AxiomC{$(\pmTermM,m,n)\rta \pmDistM$}
    \RightLabel{{\scriptsize $(s\protect\dash c\protect\at L)$}}
    \UnaryInfC{$(\pmTermM\ \pmValV,m,n) \rta \pmDistM\ \pmValV \vphantom{\Bigl\{}$}
    \DisplayProof\vspace{0.5em}\\
    \AxiomC{$(\pmTermN,m,n)\rta \pmDistN$}
    \RightLabel{{\scriptsize $(s\protect\dash c\protect\at R)$}}
    \UnaryInfC{$(\pmTermM\ \pmTermN,m,n) \rta \pmTermM\ \pmDistN \vphantom{\Bigl\{}$}
    \DisplayProof \hskip 15pt 
    \AxiomC{$\vphantom{M}$}
    \RightLabel{{\scriptsize $(s\protect\dash \protect\rec\protect \0)$}}
    \UnaryInfC{$(\rec\ \pmValU\ \pmValV,m,n)\ \0 \rta \Bigl\{(\pmValU,m,n)\Bigr\}$}
    \DisplayProof\vspace{0.3em}\\
    \AxiomC{$\vphantom{M}$}
    \RightLabel{{\scriptsize $(s\protect\dash \protect\rec\protect \S)$}}
    \UnaryInfC{$(\rec\ \pmValU\ \pmValV\ (\S\ \boldsymbol k),\ m,n) \rta \Bigl\{(\pmValV\ \boldsymbol k\ (\rec\ \pmValU\ \pmValV\ \boldsymbol k),m,n)\Bigr\}$}
    \DisplayProof\vspace{0.3em}\\
    \AxiomC{$\vphantom{M}$}
    \RightLabel{{\scriptsize $(s\protect\dash \protect\piO)$}}
    \UnaryInfC{$\!(\piO\ \langle\pmTermM,\pmTermN\rangle, m,n) \rta \Bigl\{(\pmTermM,m,n)\Bigr\}\!$}
    \DisplayProof\hskip 2pt
    \AxiomC{$\vphantom{M}$}
    \RightLabel{{\scriptsize $(s\protect\dash \protect\piT)$}}
    \UnaryInfC{$\!(\piT\ \langle\pmTermM,\pmTermN\rangle,m,n) \rta \Bigl\{(\pmTermN,m,n)\Bigr\}\!$}
    \DisplayProof\vspace{0.3em}\\
    \AxiomC{$\vphantom{M}$}
    \RightLabel{{\scriptsize $(s\protect\dash \Srand)$}}
    \UnaryInfC{$(\Srand,m,n) \rta 
      \left\{
      \begin{matrix} 
        (k,m+n,n) \mapsto \frac{1}{2^{k+1}},\ \text{if }k< m\quad \\ 
        (\pmTermN,m',n')\mapsto 0,\quad \text{otherwise} 
      \end{matrix}
      \right\}$}
    \DisplayProof  \vspace{0.7em}\\
    \AxiomC{$\forall (\pmTermM,m,n)\in \supp\pmDistM,\ (\pmTermM,m,n)\rta^? \pmDistN_{(\pmTermM,m,n)}$}
    \RightLabel{{\scriptsize $(s\protect\dash{\in})$}}
    \UnaryInfC{$\pmDistM \rta  \sumd{(\pmTermM,m,n)}{\pmDistM}{\pmDistN_{\pmTermM,m,n}}$}
    \DisplayProof
  \end{center}
  \end{small}
  \end{minipage}}
  \caption{Operational semantics of $\SystSRand$}\label{fig:OSTSrand}
\end{figure*}

For any $m$ and $n$, the behavior of $\pmTermM$ and~$(\pmTermM^*,m,n)$ 
are similar, except that $(\pmTermM^*,m,n)$ will
``fail'' more often. In other words, any~$(\pmTermM^*,m,n)$ approximates the 
behavior of $\pmTermM$ from below:

\begin{lem}\label{lm:M>M*}
  For any $\pmTermM\in\SystRand$ and any $m,n\in\Nat$, $\Eval{\pmTermM}\succeq \Evall{\pmTermM^*,m,n}$, i.e., for every~$\pmValV\in\SystRand_V$, we have \vspace{-0.1em}
  $$ \Eval{\pmTermM}(\pmValV) \ge \sum_{l,p} 
  \Evall{\pmTermM^*,m,n}(\pmValV^*,l,p). $$
\end{lem}
\begin{proof}
  By an easy induction, one can show that for any
  $\pmDistM\in\Dist\SystRand$ and $\pmDistN\in\Dist\SystSRand$ if
  $\pmDistM\succeq \pmDistN$, $\pmDistM\rta\pmDistL$ and
  $\pmDistN\rta \pmDistP$, then $\pmDistL\succeq\pmDistP$.  This
  ordering is then preserved at the limit so that we get our result.
\end{proof}

In fact, the probability of ``failure'' of any $(M,m,n)_{m,n\in \Nat}$
can be upper-bounded explicitly. More precisely, we can find an
infinite product underapproximating the success rate of $(M,m,n)$ by
reasoning inductively over $(\pmTermM,m,n)\Rta\Eval{(\pmTermM,m,n)}$.
\begin{lem}\label{lm:succGe}
  For any $M\in\SystSRand$ and any $m,n\ge 1$
  $$ \Succ(M,m,n) \ge \prod_{k\ge 0}\Bigl(1-\frac 1 {2^{m+kn}}\Bigr) . $$
\end{lem}
  \begin{proof}
    We use the following notation:
    $$
    \#(m,n) := \prod_{k\ge 0}\Bigl(1-\frac 1 {2^{m+kn}}\Bigr)\:; 
    \qquad\qquad 
    \#\pmDistM := \sumd{(\pmTermM,m,n)}{\pmDistM}{\#(m,n)}\:.
    $$
    By induction on $\Rta$, we can show that if $(\pmTermM,m,n)\Rta\pmDistM$ 
    then $\#\pmDistM = \#(m,n)$ and that if $\pmDistN\Rta\pmDistM$ then  
    $\#\pmDistM = \#\pmDistN$:
    \begin{varitemize}
    \item If $(\pmTermM,m,n)\rta\pmDistN\Rta \pmDistM$ then $\pmDistN$ is 
    either of the form $\{(\pmTermN,m,n)\}$ or $\{(\pmTermN_i,m+n,n)\mapsto 
    \frac1 {2^{i+1}} \mid i< m\}$ for some $\pmTermN$ of $(\pmTermN_i)_{i\le 
    m}$. In the first case it is clear that $\#\pmDistN=\#(m,n)$, but the 
    equality holds also in the second: 
      $$ \#\pmDistN \quad =\quad  \sum_{i\le m}\frac 1{2^{i+1}}\#(m+n,n)
      \quad =\quad (1-\frac 1{2^m})\#(m+n,n)
      \quad = \quad \#(m,n)\:.
      $$
      By IH, we conclude that $\#\pmDistM=\#\pmDistN=\#(m,n)$.
    \item The other cases are immediate.  
    \end{varitemize}
    In particular, we have that:
    \begin{align*}
      \Succ(M,m,n) &= \sumd{(\pmTermM,m,n)}{\Evall{\pmTermM,m,n}}{1}\\
      &\ge \#\Eval{\pmTermM}   &&\text{since } \forall m,n,\ \ 1\ge \#(m,n) \\
      &= \#(m,n)  &&\text{since } (\pmTermM,m,n)\Rta \Eval{\pmTermM,m,n}.
                     \tag*{\qedhere}
    \end{align*}
  \end{proof}

This gives us an analytic lower bound to the success rate of $(\pmTermM,m,n)$. 
However, it is not obvious that this infinite product is an interesting bound: 
it is not even clear that it can be different from $0$. This is why we will
further underapproximate this infinite product to get a simpler expression whenever $m=n$:
\begin{lem}
  For any $M\in\SystSRand$ and any $n\ge 4$, we have that
  $$ \Succ(M,n,n)\quad \ge\quad 1 - \frac 1 n . $$
\end{lem}
\begin{proof}
  By Lemma \ref{lm:succGe} we have that $\Succ(M,n,n) \ge \prod_{k\ge
    1}\bigl(1-\frac 1 {2^{k*n}}\bigr)$ which is above the product~$\prod_{k\ge
    1}\bigl(1-\frac {1} {n^2k^2}\bigr)$ whenever $n\ge 4$. This
  infinite product  has been shown by Euler to be
  equal to $\frac {sin(\frac \pi n)} {\frac \pi n}$. By an easy
  numerical analysis we then obtain that $\frac {sin(\frac \pi n)}
  {\frac \pi n}\ge 1- \frac 1 n$.
\end{proof}
This lemma can be restated by saying that the probability of ``failure'' of 
$(\pmTermM^*,n,n)$, {\em i.e.} the difference between $\Eval{\pmTermM^*,n,n}$ and $\Eval\pmTermM$, is bounded by $\frac 1 n$.
With this we then get our first theorem, which is the uniform approximation of
elements of $\SystRand$ by those of~$\SystSRand$:
\begin{thm}\label{th:TSRapproxTR}
  For any $\pmTermM\in\SystRand$ and any $n\in\Nat$,
  $$ \sum_{\pmValV}\ \Bigl|\ \Eval{\pmTermM}(\pmValV)\ -\ 
  \Sigma_{l,p}\Evall{\pmTermM^*, n,n}(\pmValV^*,l,p)\ \Bigr|\quad \le\quad 
  \frac 1 n .$$
\end{thm}
\begin{proof}
  By Lemma~\ref{lm:M>M*}, for each $\pmValV$ the difference is positive, thus 
  we can remove the absolute value and distribute the sum. We conclude by using 
  the fact that $\Succ(M)=1$ and $\Succ(\pmTermM^*,n,n)\ge 1-\frac 1 n$. 
\end{proof}
The second theorem, {\em i.e.}, the uniform approximation of ground
elements of $\SystRand$ by those of~$\SystPl$, follows immediately:
\begin{thm}\label{th:T+approxTR}
  Distributions in $\SystRand(\NAT)$ can be approximated by
  $\SystPl$-distributions (which are finitely $\SystT$-representable),
  i.e., for any $\pmTermM\in\SystRand(\NAT)$, there is
  $\pmTermM^\dagger\in\SystPl(\NAT)$ such that for every natural number
  $n$, it holds that
  $$
  \sum_{k}\ \Bigl|\ \Eval{\pmTermM}(\boldsymbol
  {k})\ -\ \Evall{\pmTermM^\dagger\ \boldsymbol{n}}(\boldsymbol
  k)\ \Bigr|\quad \le\quad \frac 1 n.
  $$
  Moreover:
  \begin{varitemize}
  \item 
    the encoding is parametric, {\em i.e.}, for all
    $\pmTermM\in\SystRand(\NAT\rta\NAT)$, there is
    $\pmTermM^\dagger\in\SystPl(\NAT\rta\NAT)$ such that
    $(\pmTermM\ \boldsymbol n)^\dagger=\pmTermM^\dagger\ \boldsymbol
    n$ for all $n\in\Nat$;
  \item 
    the encoding is such
    that $\Eval{\pmTermM}(\boldsymbol
    {k})\le\Evall{\pmTermM^\dagger\ \boldsymbol{n}}(\boldsymbol k)$
    only when $k=0$.
  \end{varitemize}
\end{thm}
\begin{proof}
  It is clear that in an extension of $\SystPl$ with two global memory
  cells $m, n$ and with exceptions, the $\Srand$ operator can be
  encoded by 
  $$
  \Srand:=\rec\langle \lambda u.\bot,\: \lambda xyu.\: \0 \bc \S (y\ u),\ 
  m:=!m+!n\rangle\ \0,$$
  where $\bot$ is raising an error/exception and $m:=!m+!n$ is
  returning the value of $m$ before changing the memory cell to $m+n$.
  Remark that the only objective of the dummy abstraction over $u$ and 
  of the dummy application to $\0$, preventing $\bot$ from being evaluated.
  We can conclude by referring to the usual state passing style
  encoding of exceptions and state-monads into $\SystT$ (and thus into
  $\SystPl$).
  In fact, we do not have any requirement over $\bot$, i.e.,
  we can replace $\bot$ by any value $\bot_\pmTypa$ of the
  correct type $\pmTypa$ (which is possible since every type is
  inhabited). In other words, we do not need to implement the exception monad,
  but only the state monad which we can present easily here:
  \begin{align*}
    \tptstype \pmTypa &:= \NAT^3\rta(\tptsvals\pmTypa\times \NAT^3)\:; &
    \tptsvals\NAT &:= \NAT\:; \\
    \tptsvals{\pmTypa\rta\pmTypb} &:=\tptsvals\pmTypa\rta\tptstype\pmTypb\:; &
    \tptsvals{\pmTypa\times\pmTypb} 
    &:=\tptstype\pmTypa\times\tptstype\pmTypb\:.  
  \end{align*}
  The state here is an element of $\NAT^3$, the first natural number
  monitoring the presence of an error along the reduction, 
  the second represents the state $m$ and the third represents the state $n$. 
  The encoding $\tptsvals -$ of extended values is the same as for the encoding of Section~\ref{sec:coding_T+}:
  \begin{align*}
    \tptvals\S&:= \lambda y. \tptvals{\S\ y}\:;& 
    \tptvals{\S\ \pmValV}&:= \S\ \tptvals{\pmValV}\:; &
    \tptvals\0&:= \0\:; \\
    \tptvals{\langle\pmTermM,\pmTermN\rangle} &:=  
    \langle\tptvals\pmTermM,\tptvals\pmTermN\rangle\:;&
    \tptvals{\lambda x.M}&:= \lambda x.\tpttype \pmTermM\:;&
    \tptvals x &:= x\:;\\
    \tptvals{\pi_i} &:= \lambda x. \ret (\pi_i x)\:;\\
    \tptvals\rec &:= \lambda \langle u,v,w\rangle.\rec\:\langle \ret\: u ,\: 
    \rlap{$\lambda xy. (v x)\bindo y ,\: w\rangle$\;.}
  \end{align*}
  Where $\ret$ and $\bindo$ are the return and the bind of the considered monad:
  \begin{align*}
    \ret &: \tptvals\pmTypa\rta\tpttype\pmTypa\:;  &
    \ret &:= \lambda xs.\langle x , s\rangle\:;\\
    (\cdot)\bindo(\cdot)  & : 
    \tpttype{\pmTypa\rta\pmTypb}\times\tpttype\pmTypa\rta\tpttype\pmTypb\:;&
    \pmTermM\bindo \pmTermN & := \lambda s.(\lambda\langle x,t\rangle. 
    (\lambda\langle y,u\rangle.x y u)\ (\pmTermN t))\ (\pmTermM s)\:.
  \end{align*}
  What $\pmTermM\bindo \pmTermN$ does is looking at the current state
  $s$, evaluating $\pmTermM$ under the state $s$ which results to
  $\langle x,t\rangle$, then evaluating $\pmTermN$ under the state
  $s_2$ which results to $\langle y, u\rangle$, and, finally,
  evaluating $(x\ y):\tptstype\pmTypb$ under the state $u$.
  The encoding $\tpttype\cdot$ is given by the return and the bind
  operations as well as the encoding of effectful operations:
  \begin{align*}
    \tpttype{\pmValV} &:= \ret\ \tptvals\pmValV\:;&
    \tpttype{\pmTermM\ \pmTermN} &:= \tpttype\pmTermM\bindo\tpttype\pmTermN\:;\\
    \tptstype{\pmTermM\oplus\pmTermN} &:= \lambda s.\; \left(\tptstype\pmTermM\ 
    s\right)\;\oplus\;\left( \tptstype\pmTermN\ s\right)\:;  &
    \tptstype{m:=!m+!n} &:=  \lambda \langle e,m,n\rangle.\; \langle\; m\;,\; 
    \langle e,m+n,n\rangle\;\rangle\:;\\
    \tptstype{\newop{\bot}} &:=  \lambda \langle e,m,n\rangle.\langle *,\langle 
    \boldsymbol 1,\langle m,n\rangle\rangle\rangle\:.
    \end{align*}
    where $*$ is any term of correct type. In the end, we set
    $$
    \pmTermM^\dagger:= \lambda x.(\lambda\langle y, e,
    m,n\rangle. \rec\ y\ (\lambda
    uv.\0)\ e)\ (\tptstype{\pmTermM}\ \langle x,x\rangle)
    $$ 
    for
    $\pmTermM\in\SystRand(\NAT)$. The parametrization is obtained
    by using the equality $\tptvals{\boldsymbol n}=\boldsymbol n$.
\end{proof}  

\begin{cor}\label{Th:functionalRep}
  Distributions in $\SystRand$ are functionally parametrically representable by 
  $\SystT$-definable functions, {\em i.e.} 
  for any $\pmTermM:\NAT\rta\NAT$ in $\SystRand$ there is $F: \NAT\rta\NAT\rta\NAT\rta\BIN$ and $Q:\NAT\rta\NAT\rta\NAT$ in $\SystT$ 
  such that for all $m$ and $n$:
  $$ 
  \sum_{k\in \Nat}\Bigl|\:\Eval{\pmTermM\ \boldsymbol m}(\boldsymbol k) - 
  \NF{F\ \boldsymbol m\ \boldsymbol n\ \boldsymbol k}\:\Bigr|\ \le\ 
  \frac{1}{n}\:;
  \quad\quad\quad \forall k\ge \NF{Q\ \boldsymbol m\ \boldsymbol n}.\ \NF{F\ 
  \boldsymbol m\ \boldsymbol n\ \boldsymbol k} =\0.
  $$
\end{cor}

\section{Subrecursion}
    \label{sec:subrec}
    Up to now, $\SystTot$ and its fragments' expressiveness has been evaluated
by considering programs of type $\NAT\rta\NAT$ as representing functions
from $\Nat$ to $\Dist{\Nat}$. Probabilistic computational models, however,
are often treated as representing ordinary functions, like in probabilistic
complexity theory~\cite{AroraBarak2009,Papadimitriou1994}, where $\BPP$ or $\ZPP$ are classes
of decision problems. If one wishes to define $\SystPl$-definable or $\SystRand$-definable
functions as a set of ordinary functions (say from $\Nat$ to $\Nat$),
it is necessary to somehow collapse the probabilistic output into a deterministic
one. As already acknowledged by the complexity community, there are at
least two reasonable ways to do so: by using a either Monte Carlo (like
in $\BPP$) or Las Vegas observations (like in $\ZPP$).

We recall that a function $f:\Nat\rta\Nat$ is $\SystT$-definable if
there is a $\SystT$ program $\vdash\pmTermM:\NAT\rta\NAT$ in $\SystT$
such that $(\pmTermM\ \boldsymbol{n})\rta^*\boldsymbol{f(n)}$ for all
$n$. We denote the set of $\SystT$-definable functions as $\DT$.

%%%%%%%%%%%%%%%%%%%%%%%%%%%%%%%%%%%%%
\subsection{Monte Carlo Observations}
%%%%%%%%%%%%%%%%%%%%%%%%%%%%%%%%%%%%%
In Monte Carlo observations, what one observes is the output with
the highest probability, which must be sufficiently high to rule
out any ambiguity. The class of {\em Monte Carlo representable functions}
on $\SystPl$ (respectively, $\SystRand$)
is the class $\BP^\oplus$ (respectively, $\BP^\rand$) of functions $f$ 
definable by a
$\SystPl$ program  (respectively,
a $\SystRand$ program) $\vdash \pmTermM:\NAT\rta\NAT$ which computes
$f$ with at least probability $\frac{2}{3}$ of outputting the correct result
$p\ge\frac 2 3$. Formally:
\begin{align*}
f\in \BP^\oplus \quad\text{iff}\quad\exists\pmTermM\in\SystPl(\NAT\rta\NAT),\ \ 
 \forall n\in\Nat,\ \ \Eval{\pmTermM \boldsymbol n} (\boldsymbol{f(n)}) \ge \frac 2 3\\
f\in \BP^\rand \quad\text{iff}\quad\exists\pmTermM\in\SystRand(\NAT\rta\NAT),\ 
\ 
 \forall n\in\Nat,\ \ \Eval{\pmTermM \boldsymbol n} (\boldsymbol{f(n)}) \ge \frac 2 3
\end{align*}
As is well known in complexity theory, the bound $\frac{2}{3}$ is
arbitrary and we could have used equivalently any bound \emph{strictly}
above~$\frac 1 2$. It is also natural to consider $\frac{1}{2}$
as a bound, but force the probability of error be \emph{strictly}
below it. We can then obtain the following classes of {\em probabilistically
representable functions}:
\begin{align*}
f\in \PrC^\oplus \quad\text{iff}\quad\exists\pmTermM\in\SystPl(\NAT\rta\NAT),\ \ 
 \forall n\in\Nat,\ \ \Eval{\pmTermM \boldsymbol n} (\boldsymbol{f(n)}) > \frac 1 2\\
f\in \PrC^\rand \quad\text{iff}\quad\exists\pmTermM\in\SystRand(\NAT\rta\NAT),\ 
\ 
 \forall n\in\Nat,\ \ \Eval{\pmTermM \boldsymbol n} (\boldsymbol{f(n)}) > \frac 1 2
\end{align*}
The pertinence of these classes is however controversial. Indeed, it can well be that $(\pmTermM\ 
\boldsymbol n)$ evaluates into $\boldsymbol{f(n)}$ with probability at least $\frac 1 2 +h(n)$
where $h$ is an uncomputable function.

Due to the functional aspect of the considered objects\footnote{in
contrast to what happened for the polynomial classes.},  we can
nonetheless consider subclasses of $\PrC^\oplus$ (rep. $\PrC^\rand$) for a
reasonable \emph{dynamic} bound $h$, namely one which
can itself be computed in $\SystT$. We obtain
this way the following {\em dynamic Monte Carlo classes}
\begin{align*}
f\in \BPf^\oplus \quad\text{iff}\quad\exists h\in\DT,\ \exists\pmTermM\in\SystPl(\NAT\rta\NAT),\ \ 
 \forall n\in\Nat,\ \ \Eval{\pmTermM \boldsymbol n} (\boldsymbol{f(n)}) > \frac 1{h(n)}\\
f\in \BPf^\rand \quad\text{iff}\quad\exists h\in\DT,\ 
\exists\pmTermM\in\SystRand(\NAT\rta\NAT),\ \ 
 \forall n\in\Nat,\ \ \Eval{\pmTermM \boldsymbol n} (\boldsymbol{f(n)}) > \frac 1{h(n)}
\end{align*}
There are easy inclusions between the just introduced classes of functions,
since probabilistic observations are more liberal than dynamic Monte Carlo, themselves
more liberal than Monte Carlo:
$$
\DT\subseteq\BP^\oplus\subseteq\BPf^\oplus\subseteq\PrC^\oplus\qquad\qquad
\DT\subseteq\BP^\rand\subseteq\BPf^\rand\subseteq\PrC^\rand
$$
Is any of the above inclusions \emph{strict}? In presence of
binary probabilistic choice, the answer is negative:
\begin{thm}\label{th:PRCPl}
  $\PrC^\oplus=\BP^\oplus=\BPf^\oplus=\DT$.
\end{thm}
\begin{proof}
  Let $f\in \PrC^\oplus$. 
  There is $ \pmTermM\in\SystPl(\NAT\rta\NAT)$ such that $\Eval{\pmTermM\ \boldsymbol m}(f(m))>\frac 1 2$.
    By Theorem~\ref{Th:finiteRep}, there are $F\in\SystT(\NAT\rta\NAT\rta\BIN)$ and $G\in\SystT(\NAT\rta\NAT)$ such that
  $$ \forall k\le \NF{G\:n},\quad\quad
  \NF{F\:\boldsymbol n\: \boldsymbol k} > \frac 1 2 \quad \Lra \quad 
  k=f(n)\ . $$
  In this case can we set $\pmTermM'\in\SystT(\NAT\rta\NAT)$ such that $\NF{\pmTermM'\ \boldsymbol n} = f(n)$, by:
  \begin{align*}
    \pmTermM' &:= \lambda n. \rec\langle F n \0,\: \lambda k y. \newop{ite}\langle \newop{sup}_{\frac 1 2} (F\: n\: k),\: k,\: y\rangle ,\: G\:n\rangle
  \end{align*}
  where $\newop{sup}_{\frac 1 2}$ is testing whether the input is above $\frac 1 2$:
  \begin{align*}
    \newop{sup}_{\frac 1 2} &:\NAT\times\NAT\rta\NAT & 
    \newop{sup}_{\frac 1 2} &:= \lambda \langle m,n\rangle. (m+m) > 2^n\\
    \_>\_&:\NAT\times\NAT\rta\NAT &
    \pmTermM >\pmTermN&:= \rec\langle \lambda u.\boldsymbol 0 ,\: \lambda xyu.\rec\langle \boldsymbol 1, \lambda a \_. y a , u\rangle ,\ \pmTermM \rangle\ \pmTermN
  \tag*{\qedhere}
  \end{align*}
\end{proof}
When considering countable probabilistic choice, we cannot quite
get to the same result, but close to that:
\begin{thm}
  $\BPf^\rand=\BP^\rand=\DT$.
\end{thm}
\begin{proof}
  Let $f\in \BPf^\rand$.
  There is $\pmTermM\in\SystRand(\NAT\rta\NAT)$ and $H\in\SystT(\NAT\rta\NAT)$ such that 
  $$\Eval{\pmTermM\ \boldsymbol m}(f(m))\ge \frac 1 2 +\frac 1 {\NF{H\ \boldsymbol m}}.$$
  By Theorem~\ref{Th:functionalRep}, there exists  $F: \NAT{\rta}\NAT{\rta}\NAT{\rta}\BIN$ and $Q:\NAT{\rta}\NAT{\rta}\NAT$ in $\SystT$ 
  such that:
  $$ 
  \sum_{k\in \Nat}\Bigl|\:\Eval{\pmTermM\ \boldsymbol m}(\boldsymbol k) - \NF{F\ \boldsymbol m\ \boldsymbol n\ \boldsymbol k}\:\Bigr|\ \le\ \frac{1}{n}
  \quad\quad\quad \forall k\ge \NF{Q\ \boldsymbol m\ \boldsymbol n},\ \NF{F\ \boldsymbol m\ \boldsymbol n\ \boldsymbol k} =\0.
  $$
  In particular, $f(m)$ is the only $k\le \NF{Q\ \boldsymbol m\ (H\ \boldsymbol m)}$ such that:
  $$ 
  \NF{F\ \boldsymbol m\ (H\ \boldsymbol m)\ \boldsymbol k} > \frac 1 2
  $$
  In this case can we define $\pmTermN$ such that $\NF{\pmTermN\ \boldsymbol n} = f(n)$ by:
  \begin{align*}
    \pmTermN &:= \lambda m. \rec\langle F\: m\: (H\: m)\: \0,\: \lambda k y. \newop{ite}\langle \newop{sup}_{\frac 1 2} (F\: m\: (H\: m)\: k), k, y\rangle ,\: G\:m\: (H\: m)\rangle
  \tag*{\qedhere}
  \end{align*}
\end{proof}
%%%%%%%%%%%%%%%%%%%%%%%%%%%%%%%%%%%
\subsection{Las Vegas Observations}
%%%%%%%%%%%%%%%%%%%%%%%%%%%%%%%%%%%
In Las Vegas observations, one requires the underlying program to either
return a special value (by convention, we take $\boldsymbol{0}$ here)
representing failure, or to return the correct value of the function,
the latter with at least a certain probability of success. Mimicking
what we have done for Monte Carlo observations in the previous
section, we can thus define six classes of functions as follows:
\begingroup
\allowdisplaybreaks
\begin{align*}
f\in \LV^\oplus \quad&\text{iff}\quad\exists\pmTermM\in\SystPl(\NAT\rta\NAT).\  
 \forall n\in\Nat.\ \Eval{\pmTermM \boldsymbol n}=\left\{\begin{matrix} 
 \S\boldsymbol{f(n)}&\mapsto &p \\ \0 & \mapsto & (1-p)\end{matrix}\right\}\ 
 \text{ with }\ p\ge\frac 1 3\\
f\in \LV^\rand \quad&\text{iff}\quad\exists\pmTermM\in\SystRand(\NAT\rta\NAT).\ 
 \forall n\in\Nat.\ \Eval{\pmTermM \boldsymbol n}=\left\{\begin{matrix} 
 \S\boldsymbol{f(n)}&\mapsto &p \\ \0 & \mapsto & (1-p)\end{matrix}\right\}\ 
 \text{ with }\ p\ge\frac 1 3\\
f\in \NDC^\oplus 
\quad&\text{iff}\quad\exists\pmTermM\in\SystPl(\NAT\rta\NAT).\  
 \forall n\in\Nat.\ \Eval{\pmTermM \boldsymbol n}=\left\{\begin{matrix} 
 \S\boldsymbol{f(n)}&\mapsto &p \\ \0 & \mapsto & (1-p)\end{matrix}\right\}\ 
 \text{ with }\ p>0\\
f\in \NDC^\rand 
\quad&\text{iff}\quad\exists\pmTermM\in\SystRand(\NAT\rta\NAT).\  
 \forall n\in\Nat.\ \Eval{\pmTermM \boldsymbol n}=\left\{\begin{matrix} 
 \S\boldsymbol{f(n)}&\mapsto &p \\ \0 & \mapsto & (1-p)\end{matrix}\right\}\ 
 \text{ with }\ p>0\\
f\in \LVf^\oplus \quad&\text{iff}\quad\exists h\in\DT\ 
\exists\pmTermM\in\SystPl(\NAT\rta\NAT).\  
 \forall n\in\Nat.\ \Eval{\pmTermM \boldsymbol n}=\left\{\begin{matrix} 
 \S\boldsymbol{f(n)}&\mapsto &p \\ \0 & \mapsto & (1-p)\end{matrix}\right\}\\ 
 &\text{ with }\ p\ge \frac 1 {h(n)}\\
\end{align*}
\endgroup
\begin{align*}
f\in \LVf^\rand \quad&\text{iff}\quad\exists h\in\DT\ 
\exists\pmTermM\in\SystRand(\NAT\rta\NAT).\ 
 \forall n\in\Nat.\  \Eval{\pmTermM \boldsymbol n}=\left\{\begin{matrix} 
 \S\boldsymbol{f(n)}&\mapsto &p \\ \0 & \mapsto & (1-p)\end{matrix}\right\}\\
 &\text{ with }\ p\ge \frac 1 {h(n)}\\
\end{align*}
The polynomial-time equivalent to, say, $\LV^\oplus$ or $\LV^\rand$
are classes in the style of $\mathbf{ZPP}$, whose
name comes from an equivalent presentation using zero-error probabilistic 
programs running in average-case polynomial time. Notice that the equivalence does not 
hold here. As usual the bound $\frac 1 3$ is arbitrary. The
classes $\NDC^\oplus$ and $\NDC^\rand$ in fact model a form of of
\emph{nondeterministic} observation: the actual value $\boldsymbol{f(n)}$ can
be obtained with \emph{any} probability, making the underlying
notion of computation to collapse to may-convergence.

The same trivial inclusions between the introduced classes hold here:
$$
\DT\subseteq\LV^\oplus\subseteq\LVf^\oplus\subseteq\NDC^\oplus\:;\qquad\qquad
\DT\subseteq\LV^\rand\subseteq\LVf^\rand\subseteq\NDC^\rand.
$$
As for the reverse inclusions, a picture very similar to the one we had in
Monte Carlo observations can be given here:
\begin{thm}\label{th:NDCPl}
  $\NDC^\oplus=\LV^\oplus=\LVf^\oplus=\DT$.
\end{thm}
\begin{proof}
  Let $f\in \NDC^\oplus$. 
  There is $\pmTermM\in\SystPl(\NAT\rta\NAT)$ such that $f(m)$ is the only $k\in\Nat$ such that $\Eval{\pmTermM\ \boldsymbol m}(\S {\boldsymbol k})>0$.
  By Theorem~\ref{Th:finiteRep}, there is $F\in\SystT(\NAT\rta\NAT\rta\BIN)$ and $G\in\SystT(\NAT\rta\NAT)$ such that
  $$ \forall k\le \NF{G\:n},\quad\quad
  \NF{F\:\boldsymbol n\: (\S \boldsymbol k)} > 0 \quad \Lra \quad 
  k=f(n)\ . $$
  In this case can we set $\pmTermN\in\SystT(\NAT\rta\NAT)$ such that $\NF{\pmTermN\ \boldsymbol n} = f(n)$, by:
  \begin{align*}
    \pmTermN &:= \lambda n. \rec\langle F n \0,\: \lambda k y. \newop{ite}\langle \newop{sup}_{0} (F\: n\: (\S k)),\: k,\: y\rangle ,\: G\:n\rangle
  \end{align*}
  where $\newop{sup}_{0}$ is testing whether the input is above $\frac 1 2$:
  \begin{align*}
    \newop{sup}_{0} &:\NAT\times\NAT\rta\NAT & 
    \newop{sup}_{0} &:= \lambda \langle m,n\rangle. m > \0
  \tag*{\qedhere}
  \end{align*}
  %\begin{alignat*}6
  %  \pmTermN := \lambda u.
  %  \rec\ \boldsymbol 0\ 
  %  &\Bigl(\lambda x. \newop{ifte}\ &&(\tptbranch{\pmTermM}\ x\ u)\ (\newop{Pred} (\tptbranch{\pmTermM}\ x\ u))\Bigr)\ (\newop{exp} (\tpttype{\pmTermM}\at_\#\langle u,0\rangle))
  %\end{alignat*}
\end{proof}
\begin{thm}
  $\LVf^\rand=\LV^\rand=\DT$.
\end{thm}
\begin{proof}
  Let $f\in \LVf^\rand$.
  There is $\pmTermM\in\SystRand(\NAT\rta\NAT)$ and $H\in\SystT(\NAT\rta\NAT)$ such that 
  $$ \Eval{\pmTermM\ \boldsymbol m}(\S\ \boldsymbol k) > \0 \quad \Lra \quad \Eval{\pmTermM\ \boldsymbol m}(\S\ \boldsymbol k) > \frac 1 {\NF{H\ \boldsymbol m}} \quad\Lra\quad k= f(m) $$
  By Theorem~\ref{Th:functionalRep}, there exists $F: \NAT{\rta}\NAT{\rta}\NAT{\rta}\BIN$ and $Q:\NAT{\rta}\NAT{\rta}\NAT$ in $\SystT$ 
  such that:
  $$ 
  \sum_{k\in \Nat}\Bigl|\:\Eval{\pmTermM\ \boldsymbol m}(\boldsymbol k) - \NF{F\ \boldsymbol m\ \boldsymbol n\ \boldsymbol k}\:\Bigr|\ \le\ \frac{1}{n}
  \quad\quad\quad \forall k\ge \NF{Q\ \boldsymbol m\ \boldsymbol n},\ \NF{F\ \boldsymbol m\ \boldsymbol n\ \boldsymbol k} =\0.
  $$
  In particular, for $n= 2*\NF{H\ \boldsymbol m}$, we get that $f(m)$ is the only $k\le \NF{Q\ \boldsymbol m\ (\boldsymbol2{*}(H\ \boldsymbol m))}$ such that:
  $$ \NF{F\ \boldsymbol m\ (2*(H\ \boldsymbol m))\ (\S\ \boldsymbol k)} > \frac 1 {2*\NF{H\ \boldsymbol m}}$$
  In this case can we set $\pmTermN$ such that $\NF{\pmTermN\ \boldsymbol n} = f(n)$ by:
  \begin{align*}
    \pmTermN := \lambda m. \rec\langle &F\: m\: (\boldsymbol2{*}(H\: m))\: 
    \boldsymbol 1,\\
    &\lambda k y. \newop{ite}\langle\: \newop{sup}_{\frac 1 2} ((H\ \boldsymbol m) *_b (F\: m\: (\boldsymbol2{*}(H\: m))\: (\S k))) \:,\ k,\ y\rangle, \\
    & G\:m\: (\boldsymbol2{*}(H\: m))\rangle\:;
  \end{align*}
  where
  \begin{align*}
    (\cdot)*(\cdot) &:\NAT\times\NAT\rta\NAT\:; & 
    M * N &:= \rec\langle \0, \lambda\_y. M{+}y , N \rangle\:;\\
    (\cdot)*_b(\cdot) &:\NAT\times\BIN\rta\BIN\:;& 
    (M *_b) &:= \lambda \langle m,n\rangle. \langle (M*m,n) \rangle\:.
  \tag*{\qedhere}
  \end{align*}
\end{proof}

%%%%%%%%%%%%%%%%%%%%%%%%%%%%%%%%%%%%%%%%%%%%%%%%%%%%%%%%%%%%
\subsection{On Probabilistic and Nondeterministic Observations}
%%%%%%%%%%%%%%%%%%%%%%%%%%%%%%%%%%%%%%%%%%%%%%%%%%%%%%%%%%%%
In the last two sections, we have not been able to 
precisely delineate the status of $\PrC^\rand$ and
$\NDC^\rand$. As we previously mentioned, the practical pertinence of
these classes is questionable, in the sense that the result will
be obtained after an unbounded number of tries and the proof that the
algorithm is correct is given as an oracle.

\begin{figure}
\fbox{
\begin{minipage}{.97\textwidth}
\begin{center}
\includegraphics[scale=1.2]{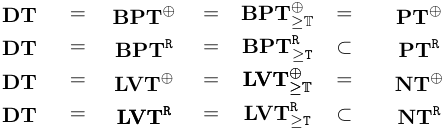}
\end{center}
\end{minipage}}	
\caption{Subrecursive Classes --- A Summary of the Obtained Results}\label{fig:summaryobtainedresult}
\end{figure}

In this section, we exploit this intuition, by proving that both of
them contain functions which are recursive bot not definable in
$\SystT$. More precisely, we show that $\NDC^\rand$, the
nondeterministic class over $\SystRand$, exactly captures (total)
recursive functions, while $\PrC^\rand$ has a bit more complex
structure and corresponds to a recursive choice over two
$\SystT$-definable possible results.  Before giving these two results,
a remark is in order: contrary to the polynomial case where
$\mathbf{NP}\subseteq\mathbf{PP}$,
we have $\PrC^\rand\subseteq\NDC^\rand$ here. In fact, in the realm of
decision problems, the two classes collapse to the one of recursive
decision problems.  The difference between them can only be observed
when considering proper functions, which are neglected in
probabilistic complexity theory. For any subset $X$ of $\Nat$, the
class $\mathbf{Rec}^{X}$ stands for the class of recursive total
function whose range is included in $X$.
\begin{thm}\label{Th:NDC=Rec}
  $\NDC^\rand \ =\ \mathbf{Rec}^\Nat$
\end{thm}
\begin{proof}
  \begin{varitemize}
  \item Let us first consider the
    inclusion $ \NDC^\rand \subseteq \mathbf{Rec}^\Nat $.
    Let $f\in \NDC^\rand=\NDC^\fixran$; there is $\pmTermM\in\SystFix(\NAT\rta\NAT)$ such that $f(m)$ is the only $k\in\Nat$ such that $\Eval{\pmTermM\ \boldsymbol m}(\S {\boldsymbol k})>0$. This means that there is a finite execution of $\pmTermM$ converging to $(\S {\boldsymbol k})$. Thus we only have to perform a simple Breadth-first search in the binary execution tree of $\pmTermM$.
  \item Now,
    consider the inclusion $ \NDC^\rand \supseteq \mathbf{Rec}^\Nat $.
    Let $f\in \mathbf{Rec}^\Nat$; then $f$ is computed by a program 
    $\pmTermM:\NAT\rta\NAT$ that makes use of the operators of System $\SystT$ 
    and of unguarded recursion $\newop{Y}:(\pmTypa\rta\pmTypa)\rta\pmTypa$; 
    since the execution of $\pmTermM \boldsymbol n$ is finite, there exists an 
    error-free execution of $\pmTermM[\newop Y:=\lambda x.\fixran\langle 
    x,\bot\rangle]\: \boldsymbol n\in \SystFix_\bot$ that gives the same 
    result; using an encoding of the error monad, we can easily get a term 
    $N\in\SystFix(\Nat\rta\Nat)$ such that $f(m)$ is the only $k\in\Nat$ such 
    that $\Eval{\pmTermN\ \boldsymbol m}(\S {\boldsymbol k})>0$. We conclude by 
    $\NDC^\rand = \NDC^\fixran$
    \qedhere
  \end{varitemize}
\end{proof}
\begin{thm}
   $f\in\PrC^\rand$ iff there are $g_1,g_2\in\DT$ and
   $h\in\mathbf{Rec}^{\{1,2\}}$ such that $f(n)=g_{h(n)}(n)$.
\end{thm}
\begin{proof}
  Let $\DT\circ\mathbf{Rec}^{\{1,2\}}$ be the class of all those functions $f$ 
  such
  that $f(n)=g_{h(n)}(n)$, where $g_1,g_2\in\DT$ and
   $h\in\mathbf{Rec}^{\{1,2\}}$. We prove the equality between $\PrC^\rand$ and
   $\DT\circ\mathbf{Rec}^{\{1,2\}}$ as follows:
  \begin{varitemize}
  \item First of all, $ \PrC^\rand \subseteq \DT\circ\mathbf{Rec}^{\{1,2\}}$.
    Let $f\in\PrC^\rand$; there is $ \pmTermM\in\SystRand(\NAT\rta\NAT)$ such that $\Eval{\pmTermM\ \boldsymbol m}(f(m))>\frac 1 2$.
    By Theorem~\ref{Th:functionalRep}, there exists  $F: \NAT{\rta}\NAT{\rta}\NAT{\rta}\BIN$ and $Q:\NAT{\rta}\NAT{\rta}\NAT$ in $\SystT$ 
    such that:
    $$ 
    \sum_{k\in \Nat}\Bigl|\:\Eval{\pmTermM\ \boldsymbol m}(\boldsymbol k) - 
    \NF{F\ \boldsymbol m\ \boldsymbol n\ \boldsymbol k}\:\Bigr|\ \le\ 
    \frac{1}{n}\:;
    \quad\quad\quad \forall k\ge \NF{Q\ \boldsymbol m\ \boldsymbol n},\ \NF{F\ \boldsymbol m\ \boldsymbol n\ \boldsymbol k} =\0.
    $$
    Then for $n=8$, we get that $\NF{F\ \boldsymbol m\ \boldsymbol n\ \boldsymbol k}>\frac 3 8$ for $k=f(m)$ and for at most one other value (since the total has to be bellow $\frac 9 8$), both bellow $\NF{Q\ \boldsymbol m\ \boldsymbol n}$. We can thus construct two terms $\pmTermN_1,\pmTermN_2:\Nat\rta\Nat$ in $\SystT$ such that 
    \begin{varitemize}
    \item\: $\pmTermN_1 \boldsymbol m$ gives the smaller of those $k$ such that $\NF{F\ \boldsymbol m\ \boldsymbol n\ \boldsymbol k}>\frac 3 8$
    \item\: and $\pmTermN_2\boldsymbol m$ the bigger.
    \end{varitemize}
    A recursive procedure can then easily choose which one between the two failures is the correct one.
  \item Moreover, $ \PrC^\rand \supseteq \DT\circ\mathbf{Rec}^{\{1,2\}}$.
    Let $g_1,g_2\in\DT$ and a recursive function $h:\Nat\rta\{1,2\}$. 
    Trivially, we can write $G:\NAT\rta\NAT\rta\NAT$ in $\SystT\subseteq\SystRand$ such that $\NF{G\boldsymbol1\boldsymbol n}=g_1(n)$ and $\NF{G\boldsymbol2\boldsymbol n}=g_2(n)$.
    As we have seen in Theorem~\ref{Th:NDC=Rec}, $h\in \NDC^\rand$ and thus there is $\pmTermM\in\SystRand(\NAT\rta\NAT)$ such that $f(m)$ is the only $k\in\Nat$ such that $\Eval{\pmTermM\ \boldsymbol m}(\S {\boldsymbol k})>0$. We thus set:
    \[ \pmTermN := \lambda n. \newop{ite}\langle \pmTermM n,\: G\: (\pmTermM
      n)\: n,\: (G\: \boldsymbol1\: n){\oplus}(G\: \boldsymbol2\: n)\rangle
      \tag*{\qedhere}
    \]
   \end{varitemize}
\end{proof}
\noindent A summary of the introduced subrecursive classes and the obtained 
results is
in Figure~\ref{fig:summaryobtainedresult}

\section{Conclusions}
    This paper is concerned with the impact of adding various forms of
probabilistic choice operators to a higher-order subrecursive calculus
in the style of G\"odel's $\npT$.  The three probabilistic choice
operators we analyze in this paper are equivalent if employed in the
context of untyped or Turing-powerful
$\lambda$-calculi \cite{dallagozorzi2012}. As an example, $\xc$ can be
easily expressed by way of $\bc$, thanks to fixpoints. Moreover, there
is no hope to get termination in any of those settings. We give
evidence that this is \emph{not} the case in a subrecursive setting.

We claim that all we have said in this paper could
have been spelled out in a probabilistic variation of Kleene's 
primitive recursive functions,
e.g.~\cite{dallagozuppiroli2014}. Going higher-order makes our
results, and in particular the termination results from
Sections~\ref{sec:fullCalculus} and~\ref{sec:fragmentBinChoice},
significantly stronger. This is one of the reasons why we have
proceeded this way. Classically, subrecursion refers to the study of relatively small
classes of computable functions lying strictly below the partially
recursive ones, and typically consisting of \emph{total} functions. In
this paper, we have initiated a study of the corresponding notion of
subrecursive computability in presence of probabilistic choice
operators, where computation itself becomes a stochastic process.

However, we barely scratched the tip of the iceberg, since the kinds
of probabilistic choice operators we consider here are just examples
of the possible ways one can turn a deterministic calculus like $\npT$
into a probabilistic model of computation. The expressiveness of
$\SystTot$ is sufficient to encode most reasonable probabilistic
operators, but what can we say about their own expressive power?  For
example, what about a ternary operator in which either of the first
two operators is chosen with a probability \emph{which depends} on the
value of the third operator? This ternary operator would have the type
$\newop{Ter}:
\pmTypa{\rta}\pmTypa{\rta} (\NAT{\rta}\NAT){\rta} \pmTypa$, where 
the third argument $z:\NAT{\rta}\NAT$ is seen as a probability $p\in[0,1]$ (whose $n^{th}$ 
binary component is given by $(z\ \boldsymbol n)$). The expressivity of
$\SystRand$ is sufficient to encode
$\newop{Ter}:=\lambda xyz.\rec\ x$ $(\lambda uv.y)\ (z\ \rand)$. The
expressivity of $\SystT^{\newop{Ter}}$, however, strictly lies
between that of $\SystPl$ and of $\SystRand$: $\SystT^{\newop{Ter}}$
can construct non binomial distributions\footnote{Such as $\newop{Ter}\
\0\ \boldsymbol 1\ (\rec\ \0\ (\lambda x.\rec\ \boldsymbol1 (\lambda 
yz.\0)))$.} while enforcing PAST.  A general theory of probabilistic
choice operators and of their expressive power is still lacking, and
is an intriguing topic for future work.

Another research direction to which this paper hints at consists in
studying the logical and proof-theoretical implications of endowing a
calculus like $\npT$ with probabilistic choice operators. The calculus
$\npT$ was born as a language of realizers for arithmetical formulas,
and indeed the class of first-order functions $\npT$ can express
precisely corresponds to the ones which are provably total in Peano's
arithmetic. But how about, e.g., $\pT{\rand}$? Is there a way to
characterize the functions (from natural numbers
to \emph{distributions} of natural numbers) which can be represented
in it? Or even better: to which extent do \emph{real} numbers in
the codomain of a distribution in the form $\Eval{\pmTermM}$
(where $\pmTermM$ is, say, a $\pT{\rand}$ term of type $\NAT$)
are computable? They are of course computable in the sense
of Turing computability, but how about subrecursive notions of
real-number computability?

What is even more exciting, however, is the application of the ideas
presented here to polynomial time computation. This would allow to
go towards a characterization of expected polynomial time computation,
thus greatly improving on the existing works on the implicit complexity
of probabilistic systems \cite{dallagoparisentoldin2012,dallagozuppiroli2014},
which only deals with worst-case execution time. The authors are
currently engaged in that.

\subsection*{Acknowledgements:}
We would like to thank Martin Avanzini and Charles Grellois for their precise 
comments and for their careful proofreading. The second author is partially 
supported by the ERC CoG DIAPASoN, GA 818616.%\vspace{-0.3em}

\bibliographystyle{alpha}
\bibliography{biblio}

\end{document}